%% file: main.tex
\documentclass{article}
\usepackage{arxiv}

\usepackage[T1]{fontenc}
\usepackage[latin9]{inputenc}
\usepackage{geometry}
\usepackage{framed,multirow}
\usepackage{latexsym}
\usepackage{graphicx}
\usepackage{amsfonts,amssymb,amsmath}
\usepackage{hyperref}

\def\BibTeX{{\rm B\kern-.05em{\sc i\kern-.025em b}\kern-.08em T\kern-.1667em\lower.7ex\hbox{E}\kern-.125emX}}
\usepackage{color}
\usepackage{subcaption}
\usepackage{hyperref}
\input{boxes}
\usepackage{babel}
\usepackage{cite}
\usepackage{booktabs} 
\usepackage{xcolor} 
\usepackage{fancyvrb} 
\usepackage[nolist]{acronym} 
\usepackage{booktabs} % To thicken table lines

\begin{document}

\title{Single image deep defocus estimation and its applications}

\author{ \href{https://orcid.org/
0000-0002-7456-201X}{\includegraphics[scale=0.06]{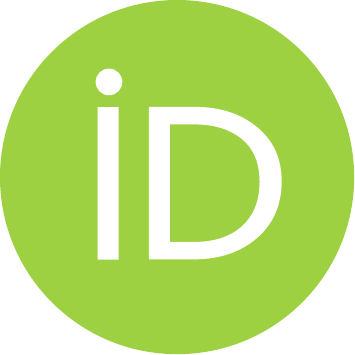}\hspace{1mm}Fernando J.~Galetto}\thanks{Corresponding author.} \\
  Department of  Engineering \\
  La Trobe University\\
  Bundoora, VIC 3086, Australia\\
	\texttt{f.galetto@latrobe.edu.au} \\
	%% examples of more authors
	\And
    \href{https://orcid.org/
0000-0003-1803-4578}{\includegraphics[scale=0.06]{orcid.pdf}\hspace{1mm}Guang ~Deng}\\
    Department of  Engineering \\
    La Trobe University\\
    Bundoora, VIC 3086, Australia\\
    \texttt{d.deng@latrobe.edu.au} \\
}
\maketitle
\begin{abstract}

     Depth information is useful in many image processing applications. However, since taking a picture is a process of projection of a 3D scene onto a 2D imaging sensor, the depth information is embedded in the image.  Extracting the depth information from the image is a challenging task. A guiding principle is that the level of blurriness due to defocus is related to the distance between the object and the focal plane. Based on this principle and the widely used assumption that Gaussian blur is a good model for defocus blur, we formulate the problem of estimating the spatially varying defocus blurriness as a Gaussian blur classification problem. We solved the problem by training a deep neural network to classify image patches into one of the 20 levels of blurriness. We have created a dataset of more than 500000 image patches of size $32\times32$ which are used to train and test several well-known network models. We find that MobileNetV2 is suitable for this application due to its low memory requirement and high accuracy.  The trained model is used to determine the patch blurriness which is then refined by applying an iterative weighted guided filter. The result is a defocus map that carries the information of the degree of blurriness for each pixel. We compare the proposed method with state-of-the-art techniques and we demonstrate its successful applications in adaptive image enhancement, defocus magnification, and multi-focus image fusion.
\end{abstract}

\keywords{ Defocus blur estimation \and CNN \and adaptive image enhancement \and shallow depth of field \and  multi focus fusion}

\begin{acronym}[AAAA ] % Give the longest label here so that the list is nicely aligned

\acro{IFM}{Image Formation Model}
\acro{DCP}{Dark Channel Prior}
\acro{IBM}{Image Blur Map}
\acro{CNN}{Convolutional Neural Network}
\acro{GIF}{Guided Image Filter}
\acro{DoF}{Depth of Field}
\acro{SDoF}{Shallow Depth of Field}
\acro{CV}{Computer Vision}
\acro{LTI}{Linear translation-invariant}
\acro{GF}{Guided Filter}
\acro{WGIF}{Weighted Guided Image Filter}
\acro{NLT}{Non Linear Transformation}
\acro{NN} {Neural Network}
\acro{CNNs}{Convolutional Neural Networks}
\acro{PSF}{Point Spread Function}
\acro{CoC}{Circle of Confusion} 
\acro{KNN}{K-Nearest Neighbour}
\acro{UM}{Unsharp Masking}

\end{acronym}

\input{01_introduction2}

\input{02_Method}
\input{02b_Refinement}

\input{03a1_Blurmap_comparison}

\section{Applications}\label{sec:applications}
In this section, we demonstrate applications of the proposed defocus map estimation in adaptive image enhancement, creating shallow depth of field, and multi-focus fusion.  

\input{03a_Adaptive_enhancement}

\input{03b_SDoF}

\input{03c_Fusion}

\section{Conclusion}\label{sec:conclusion}

In this paper,  we address a challenge problem of estimating the degree of defocus blur from a single image. This work is based on the assumption that the defocus blur can be modelled as spatially varying Gaussian blur. As such the variance of the Gaussian kernel can be used as a proxy for the degree of blurriness. We train a CNN to classify the blurriness of an image patch as one of the 20 classes. Each class label is the variance of a Gaussian kernel. We create a dataset  by (1) choosing sharply focused patches of size ($32\times32$) from a high resolution image dataset, and (2) filtering those patches by using Gaussian filters of different standard deviations $\sigma$ ranging from 1 to 19 to produce 20 classes of blurriness (the original patch is regarded as being filtered by a kernel with $\sigma=0$). The compiled dataset is used to train a deep CNN model called MobileNetV2 to classify patches into 20 different levels of blurriness. Using a sliding window algorithm and a weighted guided filter-based refinement, a pixel level defocus map is obtained.

We have tested and validated the proposed method by comparing it with classical and state-of-the-art methods. We have also demonstrated applications of the blur map estimation in  adaptive image enhancement, producing shallow of depth of field effects, and multi-focus image fusion. Results have shown that the proposed algorithms lead to competitive performance and are thus effective tools in single image blurriness estimation and a range of applications.


\end{document}

%% file: boxes.tex
\usepackage{tikz} 

\usepackage{pgfplots}
\pgfplotsset{compat=1.16}
\usetikzlibrary{spy,calc}

\newif\ifblackandwhitecycle
\gdef\patternnumber{0}
\pgfkeys{/tikz/.cd,
    zoombox paths/.style={
        draw=orange,
        very thick
    },
    black and white/.is choice,
    black and white/.default=static,
    black and white/static/.style={ 
        draw=white,   
        zoombox paths/.append style={
            draw=white,
            postaction={
                draw=black,
                loosely dashed
            }
        }
    },
    black and white/static/.code={
        \gdef\patternnumber{1}
    },
    black and white/cycle/.code={
        \blackandwhitecycletrue
        \gdef\patternnumber{1}
    },
    black and white pattern/.is choice,
    black and white pattern/0/.style={},
    black and white pattern/1/.style={    
            draw=white,
            postaction={
                draw=black,
                dash pattern=on 2pt off 2pt
            }
    },
    black and white pattern/2/.style={    
            draw=white,
            postaction={
                draw=black,
                dash pattern=on 4pt off 4pt
            }
    },
    black and white pattern/3/.style={    
            draw=white,
            postaction={
                draw=black,
                dash pattern=on 4pt off 4pt on 1pt off 4pt
            }
    },
    black and white pattern/4/.style={    
            draw=white,
            postaction={
                draw=black,
                dash pattern=on 4pt off 2pt on 2 pt off 2pt on 2 pt off 2pt
            }
    },
    zoomboxarray inner gap/.initial=5pt,
    zoomboxarray columns/.initial=2,
    zoomboxarray rows/.initial=2,
    subfigurename/.initial={},
    figurename/.initial={zoombox},
    zoomboxarray/.style={
        execute at begin picture={
            \begin{scope}[
                spy using outlines={%
                    zoombox paths,
                    width=\imagewidth / \pgfkeysvalueof{/tikz/zoomboxarray columns} - (\pgfkeysvalueof{/tikz/zoomboxarray columns} - 1) / \pgfkeysvalueof{/tikz/zoomboxarray columns} * \pgfkeysvalueof{/tikz/zoomboxarray inner gap} -\pgflinewidth,
                    height=\imageheight / \pgfkeysvalueof{/tikz/zoomboxarray rows} - (\pgfkeysvalueof{/tikz/zoomboxarray rows} - 1) / \pgfkeysvalueof{/tikz/zoomboxarray rows} * \pgfkeysvalueof{/tikz/zoomboxarray inner gap}-\pgflinewidth,
                    magnification=3,
                    every spy on node/.style={
                        zoombox paths
                    },
                    every spy in node/.style={
                        zoombox paths
                    }
                }
            ]
        },
        execute at end picture={
            \end{scope}
        },
        spymargin/.initial=0.5em,
        zoomboxes xshift/.initial=1,
        zoomboxes right/.code=\pgfkeys{/tikz/zoomboxes xshift=1},
        zoomboxes left/.code=\pgfkeys{/tikz/zoomboxes xshift=-1},
        zoomboxes yshift/.initial=0,
        zoomboxes above/.code={
            \pgfkeys{/tikz/zoomboxes yshift=1},
            \pgfkeys{/tikz/zoomboxes xshift=0}
        },
        zoomboxes below/.code={
            \pgfkeys{/tikz/zoomboxes yshift=-1},
            \pgfkeys{/tikz/zoomboxes xshift=0}
        },
        caption margin/.initial=0.01ex,
    },
    adjust caption spacing/.code={},
    image container/.style={
        inner sep=0pt,
        at=(image.north),
        anchor=north,
        adjust caption spacing
    },
    zoomboxes container/.style={
        inner sep=0pt,
        at=(image.north),
        anchor=north,
        name=zoomboxes container,
        xshift=\pgfkeysvalueof{/tikz/zoomboxes xshift}*(\imagewidth+\pgfkeysvalueof{/tikz/spymargin}),
        yshift=\pgfkeysvalueof{/tikz/zoomboxes yshift}*(\imageheight+\pgfkeysvalueof{/tikz/spymargin}+\pgfkeysvalueof{/tikz/caption margin}),
        adjust caption spacing
    },
    calculate dimensions/.code={
        \pgfpointdiff{\pgfpointanchor{image}{south west} }{\pgfpointanchor{image}{north east} }
        \pgfgetlastxy{\imagewidth}{\imageheight}
        \global\let\imagewidth=\imagewidth
        \global\let\imageheight=\imageheight
        \gdef\columncount{1}
        \gdef\rowcount{1}
        
    },
    image node/.style={
        inner sep=0pt,
        name=image,
        anchor=south west,
        append after command={
            [calculate dimensions]
            node [image container,subfigurename=\pgfkeysvalueof{/tikz/figurename}-image] {\phantomimage}
            node [zoomboxes container,subfigurename=\pgfkeysvalueof{/tikz/figurename}-zoom] {\phantomimage}
        }
    },
    color code/.style={
        zoombox paths/.append style={draw=#1}
    },
    connect zoomboxes/.style={
    spy connection path={\draw[draw=none,zoombox paths] (tikzspyonnode) -- (tikzspyinnode);}
    },
    help grid code/.code={
        \begin{scope}[
                x={(image.south east)},
                y={(image.north west)},
                font=\footnotesize,
                help lines,
                overlay
            ]
            \foreach \x in {0,1,...,9} { 
                \draw(\x/10,0) -- (\x/10,1);
                \node [anchor=north] at (\x/10,0) {0.\x};
            }
            \foreach \y in {0,1,...,9} {
                \draw(0,\y/10) -- (1,\y/10);                        \node [anchor=east] at (0,\y/10) {0.\y};
            }
        \end{scope}    
    },
    help grid/.style={
        append after command={
            [help grid code]
        }
    },
}

\newcommand\phantomimage{%
    \phantom{%
        \rule{\imagewidth}{\imageheight}%
    }%
}
\newcommand\zoombox[2][]{
    \begin{scope}[zoombox paths]
        \pgfmathsetmacro\xpos{
            (\columncount-1)*(\imagewidth / \pgfkeysvalueof{/tikz/zoomboxarray columns} + \pgfkeysvalueof{/tikz/zoomboxarray inner gap} / \pgfkeysvalueof{/tikz/zoomboxarray columns} ) + \pgflinewidth
        }
        \pgfmathsetmacro\ypos{
            (\rowcount-1)*( \imageheight / \pgfkeysvalueof{/tikz/zoomboxarray rows} + \pgfkeysvalueof{/tikz/zoomboxarray inner gap} / \pgfkeysvalueof{/tikz/zoomboxarray rows} ) + 0.5*\pgflinewidth
        }
        \edef\dospy{\noexpand\spy [
            #1,
            zoombox paths/.append style={
                black and white pattern=\patternnumber
            },
            every spy on node/.append style={#1},
            x=\imagewidth,
            y=\imageheight
        ] on (#2) in node [anchor=north west] at ($(zoomboxes container.north west)+(\xpos pt,-\ypos pt)$);}
        \dospy
        \pgfmathtruncatemacro\pgfmathresult{ifthenelse(\columncount==\pgfkeysvalueof{/tikz/zoomboxarray columns},\rowcount+1,\rowcount)}
        \global\let\rowcount=\pgfmathresult
        \pgfmathtruncatemacro\pgfmathresult{ifthenelse(\columncount==\pgfkeysvalueof{/tikz/zoomboxarray columns},1,\columncount+1)}
        \global\let\columncount=\pgfmathresult
        \ifblackandwhitecycle
            \pgfmathtruncatemacro{\newpatternnumber}{\patternnumber+1}
            \global\edef\patternnumber{\newpatternnumber}
        \fi
    \end{scope}
}

%% file: 01_introduction2.tex
\section{Introduction}

An image captured by a camera is the projection of a 3-D scene onto a 2-D plane. When an object in a scene is outside the focal plane, it is blurred due to defocus. A larger distance from the focal plane leads to a higher level of defocus resulting in more blurriness of the object \cite{Pentland1987}. The defocus blur is one of the simplest and efficient depth cues used in photography. It allows the viewer to have a rich appreciation of the 3D space \cite{Mather1996}. In computer vision, the defocus blur is used in a wide range of applications such as deblurring \cite{xue2014novel, chan2011single, cheong2015fast}, blur magnification \cite{Bae2007}, image quality assessment \cite{van2007robust, al2015image}, image sharpening \cite{Nasonov2019,ye2018blurriness} and depth estimation \cite{ziou2001depth, lin2013absolute, shi2015break, Tang_2017_CVPR}.

Defocus blur estimation has been actively studied in the literature. Some estimation methods used multiple images or special hardware \cite{Levin2007, Vu2014, Zhou2009} while others methods rely on a single image. In this paper, we focus on single image defocus blur estimation methods. A common approach is to model the defocus blur in the spatial domain as a linear filter with the impulse response of a Gaussian function. The defocus estimation problem becomes finding the standard deviation $\sigma$ of the Gaussian kernel for each pixel in the image. The gradient information is usually employed to estimate $\sigma$. Tai et al.  \cite{tai2009single} uses the relationship between the gradient and the local contrast to estimate an initial defocus map which is refined using a Markov Random Field. Others generate a coarse map using a ratio of gradient magnitudes at the edges \cite{Zhang2016, zhuo2011defocus, Hu2006}. The map is refined using a matting algorithm \cite{levin2007closed, Chen2013} or guided filter \cite{he2012guided}. Although these methods have demonstrated successful applications for some images, they can not differentiate between close edges \cite{zhuo2011defocus}. A statistical method \cite{xue2014novel} was proposed to overcome such limitation by formulating the filter kernel estimation problem as a SURE minimization problem. This method does not rely on edge information. 

Another common approach is to determine the defocus map by analyzing the image in the frequency domain \cite{Chakrabarti2010} and \cite{Zhu2013a}. The idea is to calculate the likelihood of a group of pixels being blurred with a certain kernel. Shi et al. \cite{shi2015just, shi2015break} proposed two methods to estimate the just noticeable blur level present in natural images by using dictionaries. However, these methods can not handle high degrees of spatially varying defocus blur. 

In recent years, we have witnessed more and more successful applications of deep neural networks (DNN) such as image classification \cite{AlexNet, sandler2018mobilenetv2}, object detection \cite{ren2016faster}, semantic segmentation \cite{lateef2019survey}, image restoration \cite{jin2019flexible, anwar2020diving} and image super-resolution \cite{yang2019deep} to just name a few. The use of DNNs for defocus blur estimation or detection has also been studied. In \cite{Nasonov2019}, a \ac{CNN} is used to estimate the optimum values for the grid wrapping algorithm (GWIS) \cite{Nasonova2015} for image sharpening. The CNN is trained using blurred patches to estimate the optimum parameter based on blurriness to achieve the best sharpening result. In \cite{Ma2018a, zhao2019defocus, jiang2020multianet, tang2020br, zhai2021global} a \ac{CNN} architecture is proposed for an end-to-end defocus map estimation. These networks were trained using either natural or synthetic images labelled at a pixel level to segment focused regions. Li et al. \cite{lee2019deep} on the other hand, proposed a CNN-based method to estimate spatially varying defocus blur on a single image, using synthetic data to train the network and using domain transfer to bridge the gap between real and synthetic data. Two recently published methods \cite{bohra2020texturetomtf} and \cite{ying2020patches},  generate a patch-based quality map, which is similar to the blur map when the distortion on the image is defocus blur only. 

% http://ice.dlut.edu.cn/ZhaoWenda/BTBCRLNet.html

Inspired by these works, the motivation of this study is to explore the application of deep neural networks in solving the challenging problem of estimating the defocus map from a single image. The main contribution of this work is that unlike previous works \cite{Ma2018a, zhao2019defocus, jiang2020multianet, tang2020br, zhai2021global}, we treat the blurriness estimation as a self-supervised multi-class classification problem which is solved by training a CNN to classify a patch of the input image into one of the 20 levels of blurriness. The output of the CNN is a patch-based estimation of blurriness. To obtain an estimate of a pixel-based blurriness, we use an iterative weighted guided filter to perform the refinement which generates the defocus map. 

Another main contribution of this work is the development of three algorithms based on the defocus map which carries the blurriness information for each pixel.  Since the blurriness is related to the distance between an object and the focal plane, the blur map provides useful information about the depth which is used in the following applications: 
\begin{itemize}
    \item \textbf{Adaptive image enhancement:} The defocus map is used to control the level of enhancement in the image to reduce halos and artifacts due to over enhancement of high contrast regions and the enhancement of heavily blur regions. 
    \item \textbf{Defocus magnification or shallow depth of field:} The defocus map is non-linearly transformed to estimate the weights for the combination of the sharpened and blurred images to create a shallow depth of field effect. 
    \item \textbf{Multi focus image fusion:} The defocus map is used to generate a decision map to blend partially in-focus images to form an all-in-focus image. 
\end{itemize}

The organization of this paper is summarized as follows. In section \ref{sec:background and related work} we briefly review the modelling of defocus blur as a Gaussian blur process. We then formulate the problem of defocus estimation as a classification of spatially varying Gaussian blur. In section \ref{sec:map estimation} we present the details of the proposed method including dataset creation, model testing and selection, the training process to classify the blurriness of a patch, a sliding window algorithm for transferring patch-based blurriness information to the pixel-based format, and a refinement process that produces final pixel-based defocus map. In section IV, we present a comparison of the proposed method with 7 other methods for defocus estimation. In section \ref{sec:applications} we present 3 algorithms to demonstrate the usefulness of the defocus map on 3 typical computational photography applications: adaptive image enhancement, generating effect of shallow depth of field, and fusion of images which are focused on different parts of the scene. We present experimental results and comparisons for qualitative and quantitative analysis. We summarize the main ideas and contributions of this work in Section \ref{sec:conclusion}.

\section{Modelling of defocus and problem formulation}\label{sec:background and related work}

\begin{figure}[h!]
    \centering
    \includegraphics[width = 0.7\linewidth]{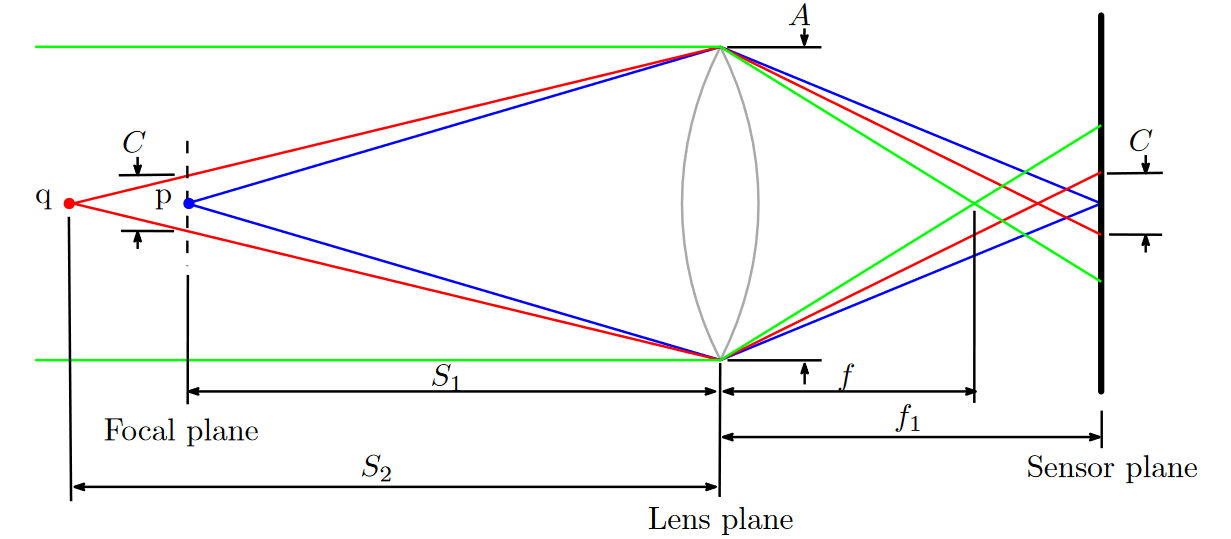}
    \caption{Optical model for defocus blur. }
    \label{fig:lens_image}
\end{figure}

In this section, we briefly review the optical model of out-of-focus and show how it can be approximated by a Gaussian filter \cite{Sakamoto1984, Tai2009a, Favaro2005, Kubota2005}. Although a practical optical system usually has more than one moving lens element to correct and fix the focal plane, Fig. \ref{fig:lens_image} allows us to understand the two main reasons for the presence of defocus blur in the observed image: (1) limited depth of field and  (2) lens aberrations \cite{Tai2009a}. To model the defocus by a Gaussian filter, we limit our discussion on the limited depth of field property. As shown in Fig. \ref{fig:lens_image}, the point $p$ is in perfect focus because it is placed right in the focal plane. The projection of $p$ on the sensor plane is a single point.

%Lens aberrations can cause that other points in the focal plane are not represented as a single point in the sensor %plane. i.e., the projection of the focal point is not exactly placed in the sensor plane. 

According to the thin lens law, the relationship between the distances is given by: 
\begin{equation}
\frac{1}{S_1}+\frac{1}{f_1}=\frac{1}{f}
\end{equation}
where $f$ is the focal length and $f_1$ is the distance between the sensor plane and the lens plane. We can see that since the point $q$ is placed in a different plane than the focal plane at a distance $S_2$ from the lens plane, the projection of point $q$ in the sensor plane is not a single point but a region (or a blurred spot) called circle of confusion. The diameter of the circle of confusion for the point $q$ can be calculated as:
\begin{equation}
    d = A \frac{f|S_2-S_1|}{S_2|S_1-f|} 
\end{equation}
where $A$  is the diameter of the aperture. The diameter $d$ is directly proportional to $A$ and increases when the difference $|S_2-S_1|$ increases. From a system point of view, we can approximate the lens as a linear shift-invariant system. The point source can be regarded as the impulse input signal and the resulting blurred spot due to defocus blur is thus the impulse response. The shape of the impulse response (the blurred spot) can be modelled as a 2D Gaussian 
\begin{equation}
    k(x,y;\sigma) \propto \exp\left(-\frac{x^2+y^2}{2\sigma^2}\right)
\end{equation}
where $\sigma$ is the scale parameter proportional to $d$, and $x,y$ are spatial variables. Because a scene usually has many objects placed at different distances from the focal plane, these objects will appear to have different degrees of blurriness. Using the linear system model, we can thus model the observed image as
\begin{equation}
    J(x,y) = k(x,y;\sigma_{x,y})*I(x,y)+Z(x,y)
\end{equation}
where $I(x,y)$ is the desired image where all objects are in focus, $k(x,y;\sigma_{x,y})$ is a spatially varying Gaussian kernel, "*" represents the convolution operations, and $Z(x,y)$ models the possible sensor noise \cite{Sakamoto1984, Tai2009a, Favaro2005, Kubota2005}. The spatial varying-parameter $\sigma_{x,y}$ is theoretically determined by the distance between the object at location $(x,y)$ and the focal plane.

From the above brief discussion of the Gaussian model for the defocus blur, We can see that (1) the blurriness is caused by the distance of an object from the focal plane, (2) the defocus blur can be modelled as applying a spatial varying Gaussian filter to an image, and (3) the degree of blurriness can implicitly measure by the scale parameter $\sigma_{x,y}$ of the Gaussian smoothing kernel. In other words, if we know the parameter $\sigma_{x,y}$ for an image patch, we can infer its blurriness and reveal its relative distance from the focal plane. Thus, the problem of inferring depth information or the spatially varying blurriness from a single image reduces to determining the parameter $\sigma_{x,y}$ of the Gaussian smoothing kernel. In this paper, we proposed a CNN-based approach to first classify an image patch into one of the 20 classes. Each class represents a Gaussian smoothing model of a particular parameter $\sigma_{x,y}$. The patch-based blurriness map is refined to obtain a pixel-based blurriness map.

%% file: 02_Method.tex
\begin{figure}[h!]
    \centering
    \includegraphics[width = 0.5\linewidth]{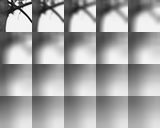}
    \caption{Sample of a training patch with different levels of blurriness.}
    \label{fig:patches}
\end{figure}

\section{Defocus map estimation via CNN based blurriness classification}\label{sec:map estimation}

The proposed method for defocus map estimation has two main building blocks. The first block is a \ac{CNN}  based classifier which takes a patch of $32\times32$ pixels as input and outputs the blur level of the patch. The classifier is applied to pixels in an image using the sliding window algorithm. The result is an image of patch blurriness. The second block refines the image obtained from the first block to produce an image of pixel blurriness. An additional threshold operation can be applied to produce a binary map for applications that require it. 

In this section, we first present the details of the dataset creation, the training of a \ac{CNN}, and how to use the \ac{CNN} to estimate the blurriness at a patch level. We then describe the refinement method for producing the pixel-level blur map.

\subsection{Blurriness classification}

As mentioned earlier, out-of-focus blur is usually modelled as the result of applying a Gaussian low pass filter on an image. The problem of blur estimation is reduced to finding the kernel of the filter \cite{Hu2006} which is parameterized by the standard deviation $\sigma$ of the Gaussian function. In this work, we follow this model. We take a different approach by casting the estimation problem as a classification problem that can be efficiently solved by a \ac{CNN}.  

More specifically, we aim to train a CNN which can classify 20 different levels of blurriness. To train the model, we need to prepare the dataset. The dataset consists of image patches of different levels of blurriness. A blurred patch is produced by filtering a sharp/unblurred image patch by a Gaussian filter with a specific $\sigma$. Different levels of blurriness are generated by using different values of $\sigma$. For an image patch, we generate 19 blurred patches of different blur levels resulting in 20 classes (including the unblurred patch). The 19 levels of blurriness are generated by Gaussian filters of $\sigma=1,2,...,19$. The unblurred patch can be regarded as being filtered by the filter with $\sigma=0$ whose impulse response is a unit sample sequence. An example of patches of 20 levels of blurriness can be seen in Fig.\ref{fig:patches}, from top to bottom and left to right the standard deviation $\sigma$ ranges from 0 to 19. We use the value of $\sigma$ as a label for the patch. The justification of why 20 levels of blurriness are selected can be seen in Fig. \ref{fig:patches}, the difference in blurriness for patches where $\sigma>15$ is almost imperceptible for the human eye so 20 levels of blurriness is the optimal equilibrium between efficiency and complexity.

The size of the dataset depends on the number of unblurred/sharp patches. To find as many as possible sharp patches to generate the dataset for training the CNN, we use the DIV2K \cite{Agustsson_2017_CVPR_Workshops} database which has 1000 high-resolution images of diverse contents. To find sharp image patches from the database, we convert each image in the database to gray-scale and divide them into patches of $32\times32$ pixels,  denoted $I_p$. We first filter each patch $I_p$ by a $3 \times 3$ Gaussian kernel to produce a slightly smoothed patch $B_p$. The Gaussian filter is used to remove noise in the patch which could lead to a wrong measurement of the sharpness. We use the variance of the Laplacian of a patch as a measure of the sharpness for the $p$th patch, which is calculated as
\begin{equation}
    \phi_p = \frac{1}{N}\sum_{q} (\nabla B_p(q)- \overline{\nabla B_p})^2
    \label{eq:variance}
\end{equation}
where $\nabla$ is the Laplacian operator, $ \overline {\nabla B_p}$ is the mean value of  $\nabla B_p(q)$ over the all pixels in the patch, $q$ is the pixel index, and $N$ is the number of pixels in the patch. Patches satisfying $\phi_p > 1000$ have a high level of sharpness and are regarded as in sharp focus and are assigned the blur level 0. Using the above procedure for patch selection and generation, we create a gray-scale image dataset with more than $5\times10^5$ patches of 20 classes. The dataset is split into 3 sets for training, validation, and testing. Table \ref{tab:Dataset} details the number of patches of each set. 
\begin{table}[h!]
    \caption{Number of image patches in the data set.}
    \centering
    \begin{tabular}{ l l}
    \toprule
        \textbf{Data set} & \textbf{Patches} \\ 
        \midrule
         Training & 392240 \\
         Validation & 98040 \\
         Testing & 49920 \\
    \bottomrule
    \end{tabular}
    \label{tab:Dataset}
\end{table}

Using this dataset, we train a neural network to classify the blurriness of a $32\times32$ gray-scale image patch into one of the 20 classes. As such, the trained network has an input array of size ($32\times32$) and the network output is an integer ranging from 0 to 19 indicating the level of blurriness. The development framework is  TensorFlow Keras. Because our goal is to train a classifier for blurriness estimation, we do not attempt to develop new neural network models. Instead, we test some well-known models such as MobileNetV2, NASNet and DenseNet and ShufflenetV2 \cite{sandler2018mobilenetv2, NASNet,DenseNet, ma2018shufflenet} to find a model which has good accuracy and is of small size. We have trained these models and presented results in  Table \ref{tab:training_results}. We can see that the MobileNetV2 \cite{sandler2018mobilenetv2} has a relatively higher accuracy for the test data (0.97) and is computationally more efficient with low complexity and low memory requirements (the model size is 28MB). This indicates that it is a more efficient and accurate tool for the estimation of the defocus map. ShufflenetV2 \cite{ma2018shufflenet} presented the worst performance among the models, even after we tuned the hyper-parameters. The sub-optimal performance can be attributed to the small patch size for which the ShufflenetV2 is not designed to work with. It is expected that using a larger patch size will lead to an improvement in the accuracy.  However, this is at the increase of the computational complexity in the refinement process of the proposed algorithm because of excessive overlaps of pixels in neighbouring patches. 

\begin{table}[h!]
    \centering
\caption{Accuracy and loss results on training, validation, and test set after 100 epochs for different models.}

    \begin{tabular}{l c c c c c c c }
    \toprule
         \textbf{Model} & \textbf{ Train loss } & \textbf{ Train acc } & \textbf{ val loss } & \textbf{ val acc } & \textbf{ Test loss } & \textbf{ Test acc} & \textbf{ Model Size}\\
         \midrule 
         \textbf{MobileNetV2} \cite{sandler2018mobilenetv2}  & 0.0117  & 0.9958 & 0.1201  & 0.9699 & 0.1610 &  0.9704 & 28MB\\ %checked 3dec2020 /Projects/Blur_map_estimation/2020_Project/Notebooks/01%20-%20Blur%20Map%20estimation%20%20-%20Trainning-V2.ipynb
          \textbf{EfficientNet0} \cite{tan2019efficientnet}  & 0.0075 & 0.9975 & 0.0944 & 0.9793 & 0.0794 & 0.9798 & 49MB \\
   
         \textbf{NASNetMobile} \cite{NASNet} & 0.0073 & 0.9976 & 0.2717 & 0.9585 & 0.2243 & 0.9582  & 55MB \\

         \textbf{DenseNet121} \cite{DenseNet}  & 0.0124 & 0.9957 & 0.6755 & 0.8996 & 0.5359 & 0.9050 & 86MB \\
         
          \textbf{ShufflenetV2} \cite{ma2018shufflenet}  & 0.1140 & 0.9563 & 1.9462 & 0.6818 & 1.8099 & 0.6902 & 33MB \\
          
        %  \textbf{AlexNet} \cite{AlexNet}  & 0.0151 & 0.9947 & 0.3187 & 0.9365 & 0.2901 & 0.0498 & 337MB \\
 
        %  \textbf{VGG16} \cite{VGG16} & 0.0151 & 0.9947 & 0.3187 & 0.9365 & 0.2901 & 0.0498 & 400MB  \\
         \bottomrule  
    \end{tabular}
    
    \label{tab:training_results}
\end{table}

\begin{figure}[h!]
\centering
\begin{subfigure}{0.7\linewidth}
    \includegraphics[width=1\linewidth]{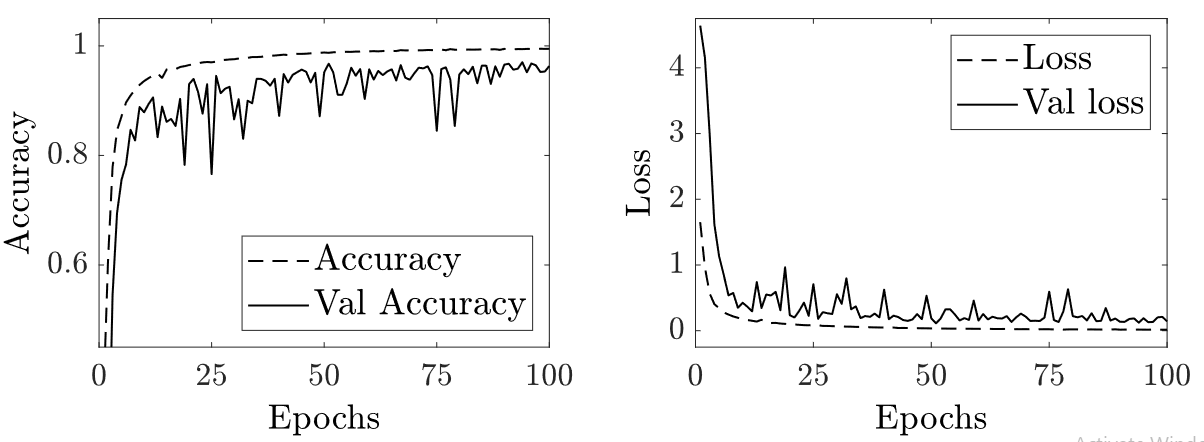}
\end{subfigure}
\caption{Accuracy and loss over epochs using MobileNetV2 on training and validation set.}
\label{fig:training_accuracy}
\end{figure}

  We train MobileNetV2 using the following parameters. The batch size is 128. The sparse categorical cross-entropy function is used to calculate the loss. Adam optimization method is used with a learning rate of  $\alpha = 0.001$, an exponential decay rates for the 1st and 2nd moment estimates are $\beta_1 = 0.9$ and $\beta_2 = 0.999 $ and a stability constant $\hat{\epsilon} = 10^{-8}$ \cite{Adam}.  After 100 epochs, classification accuracy is 97.04\% on the test set. The accuracy and loss for the validation and training set after each training epoch is shown in Fig. \ref{fig:training_accuracy}.

Table \ref{tab:evaluation_model} displays results of the MobileNetV2 including the precision, recall and F1-score for each class when evaluated using the testing set. The model can predict with high levels of accuracy on the testing dataset. However, we can identify two sources of error that are not considered here: 1) error due to noise in the image, which can lead to classify the patch as sharper than it is and 2) error because the patch being classified is not represented by those examples in the dataset, which could result on inaccurate predictions. We minimize the effect of 1 and 2 during the refinement stage.
\begin{table}[h!]
\centering
\caption{Model evaluation of MobileNetV2}
\label{tab:evaluation_model}
\begin{tabular}{ccccc}
\toprule
\textbf{Class} & \textbf{Precision} & \textbf{Recall} & \textbf{F1-score} & \textbf{Support} \\
\midrule
\textbf{0}            & 0.99               & 1.00            & 1.00              & 2475             \\
\textbf{1}            & 1.00               & 0.99            & 0.99              & 2513             \\
\textbf{2}            & 0.98               & 1.00            & 0.99              & 2457             \\
\textbf{3}            & 0.98               & 0.98            & 0.98              & 2513             \\
\textbf{4}            & 0.98               & 0.98            & 0.98              & 2490             \\
\textbf{5}            & 0.99               & 0.97            & 0.98              & 2530             \\
\textbf{6}            & 0.96               & 0.99            & 0.98              & 2437             \\
\textbf{7}            & 0.99               & 0.96            & 0.97              & 2587             \\
\textbf{8}            & 0.98               & 0.95            & 0.97              & 2561             \\
\textbf{9}            & 0.94               & 0.99            & 0.97              & 2371             \\
\textbf{10}           & 0.98               & 0.99            & 0.99              & 2474             \\
\textbf{11}           & 0.98               & 0.99            & 0.98              & 2461             \\
\textbf{12}           & 0.98               & 0.99            & 0.98              & 2479             \\
\textbf{13}           & 0.99               & 0.96            & 0.97              & 2566             \\
\textbf{14}           & 0.92               & 0.96            & 0.94              & 2405             \\
\textbf{15}           & 0.99               & 0.98            & 0.98              & 2523             \\
\textbf{16}           & 0.94               & 0.94            & 0.94              & 2495             \\
\textbf{17}           & 0.98               & 0.92            & 0.95              & 2666             \\
\textbf{18}           & 0.95               & 0.98            & 0.96              & 2429             \\
\textbf{19}           & 0.97               & 0.97            & 0.97              & 2488             \\
\midrule
\textbf{Accuracy}     &                    &                 & 0.97              & 49920            \\
\textbf{Macro avg}    & 0.97               & 0.97            & 0.97              & 49920            \\
\textbf{Weighted avg} & 0.97               & 0.97            & 0.97              & 49920           \\ \bottomrule
\end{tabular}
\end{table}

\subsection{Blur map estimation}

The trained neural network predicts the blur level of a patch of $32 \times 32$ pixels. This section explains how to obtain the blur map of the whole image at a pixel level using this patch estimation. The blur-map (denoted $M$) is an image of the same size as the input image (denoted $I$). A pixel at location $q$ within the $p$th patch is denoted $I_{p}(q)$. The image $I$ is divided into overlapping patches of ($32\times32$). The extraction of patches is implemented by using a sliding window method. The amount of overlap depends on the step-size of moving the widow from the current location to the next one. Results presented in this paper are produced with a step size of 16 pixels unless specified. Each patch, e.g., the $p$th patch is fed to the trained network. The network output, denoted $O_{p}$, is assigned to every pixel in the patch. Because of patch overlapping, one pixel belongs to multiple patches. Specifically, let $\Omega_{q}$ be the set of patches where the pixel $I(q)$ belongs to. The blurriness of the pixel at location $q$ is defined as the average of classification results due to all patches which contain the pixel $I(q)$:

\begin{equation}
M(q)=\frac{1}{|\Omega_{q}|}\sum_{p\in\Omega_{q}}O_{p}\label{eq:map}
\end{equation}
where $|\Omega_{q}|$ is the number of patches which have the pixel $I(q)$. The rationale for performing the average is this. A pixel $I(q)$ belongs to $|\Omega_q|$ patches. Each patch produces a prediction of the blurriness of this pixel. A simple way to aggregate these predictions is by taking the average. In addition,  since the classification result $O_{p}$ is in the interval $[0,19]$, the blur-map $M(q)$ is also in the same interval.

We now discuss the computational complexity. Assume the \ac{CNN} takes $T$ seconds to process a patch. The processing time for an image is given by: 
\begin{equation}
    t = T \times \frac{N}{s^2}
\end{equation}

where  $N$ is the number of pixels in the image and $s$ is the step-size. The optimal step-size depends on the image resolution and the complexity of the image. Generally, for images of large size the step size should be 32 pixels to avoid long processing times. For images of small size, better results are obtained by using a step size of 4 pixels such that so small objects with different blur levels are preserved. As a rule thumb we suggest to use a step size that can preserve the main structure of the image. 

%% file: 02b_Refinement.tex
\subsection{Refinement of the defocus map}\label{refinement}

The defocus map obtained using a patch and sliding window approach needs to be refined to get rid of the undesirable blocking effect and to make the edges follow those in the original image. A well-known tool for the refinement is the matting Laplacian \cite{levin2007closed} which produces the alpha matting matrix.  However, this method is computationally very expensive.  The \ac{GIF} \cite{he2012guided} has demonstrated a performance comparable to that of the matting Laplacian algorithm at a lower computational cost. The weighted version of the \ac{GIF} called \ac{WGIF} \cite{li2014weighted} improves the performance of the \ac{GIF} and produces fewer halo effects. However both \ac{GIF} and WGIF transfer details from the guidance image to the processed image, which is an undesirable feature for our applications because texture information from the original image will be transferred to the refined defocus map leading to wrong predictions in pixel blurriness. To overcome this limitation, we propose to use a \ac{WGIF} \cite{li2014weighted} based algorithm to refine the defocus map. A new feature in the proposed method is the use of a smoothed version of the original image (through edge-aware filtering) as a guide to producing a sharp defocus map without transferring the texture information.

\begin{figure}[h!]
    \centering
    \includegraphics[width = 0.55 \linewidth]{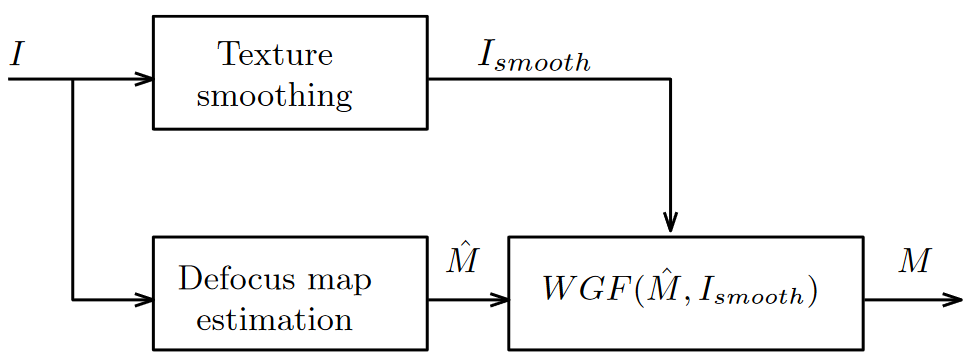}
    \caption{Defocus map refinement.}
    \label{fig:block_diagram_refinement}
\end{figure}

Figure \ref{fig:block_diagram_refinement} shows the block diagram of the proposed refinement process. First, we perform edge-aware filtering of the original image to smooth out texture information. The filtered image, which retains information of sharp edges, is used as a guidance image in the \ac{WGIF} algorithm to refine the defocus map. We aim to keep only silhouettes of elements in the image which should present an equal level of blurriness preserving large scale edges.  To produce the guidance image $I_{smooth}$ we use the \ac{WGIF}\cite{li2014weighted} iteratively.  
\begin{figure}[t]
\centering
\begin{subfigure}{0.24\linewidth}
    \includegraphics[width=1\linewidth]{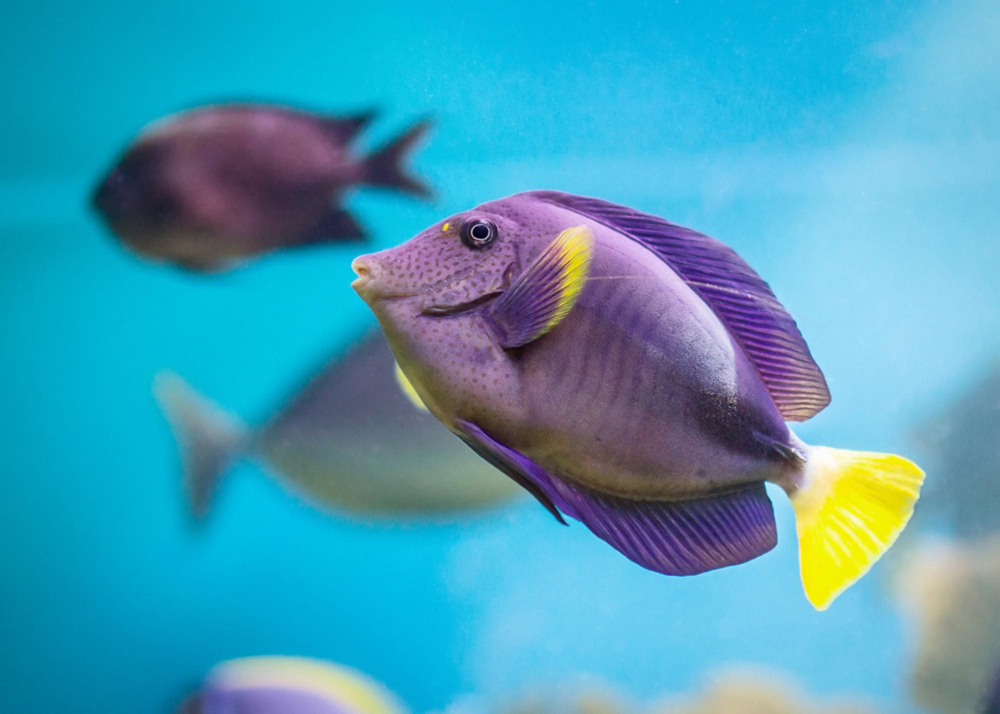}
    \caption{Input}
    \label{fig:fishin}
\end{subfigure}
\begin{subfigure}{0.24\linewidth}
    \includegraphics[width=1\linewidth]{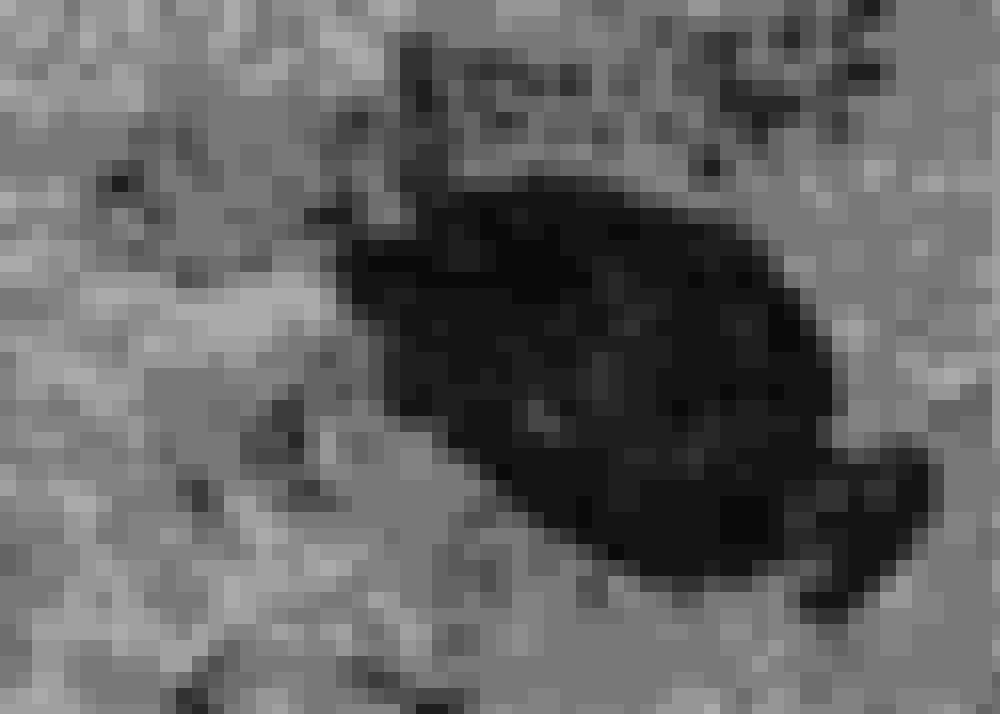}
    \caption{Blur map}
    \label{fig:fishmap}
\end{subfigure}
\begin{subfigure}{0.24\linewidth}
    \includegraphics[width=1\linewidth]{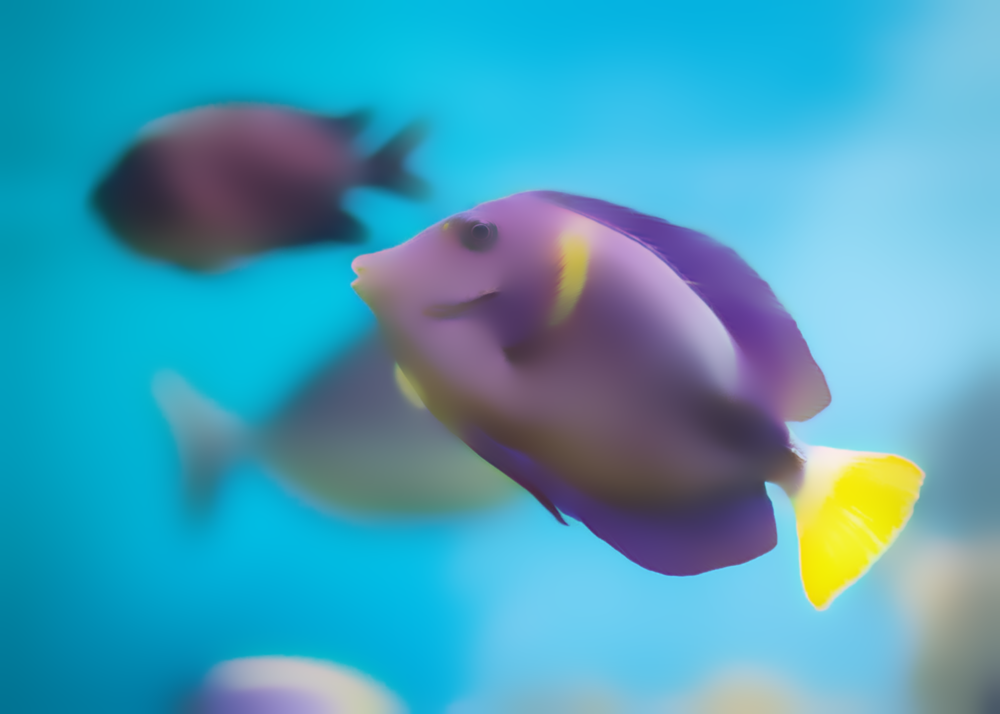}
    \caption{Guidance }
    \label{fig:fishmaprefguide}
\end{subfigure}
\begin{subfigure}{0.24\linewidth}
    \includegraphics[width=1\linewidth]{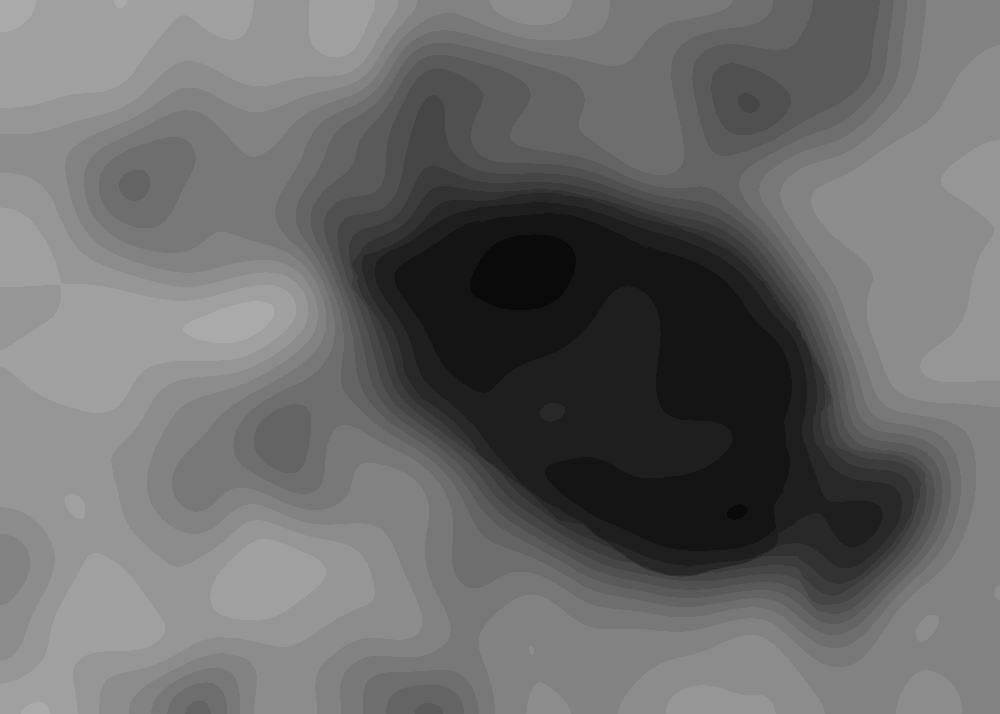}
        \caption{WFIG\cite{li2014weighted}}
    \label{fig:fishmaprefWFIG}
\end{subfigure}
\begin{subfigure}{0.24\linewidth}
    \includegraphics[width=1\linewidth]{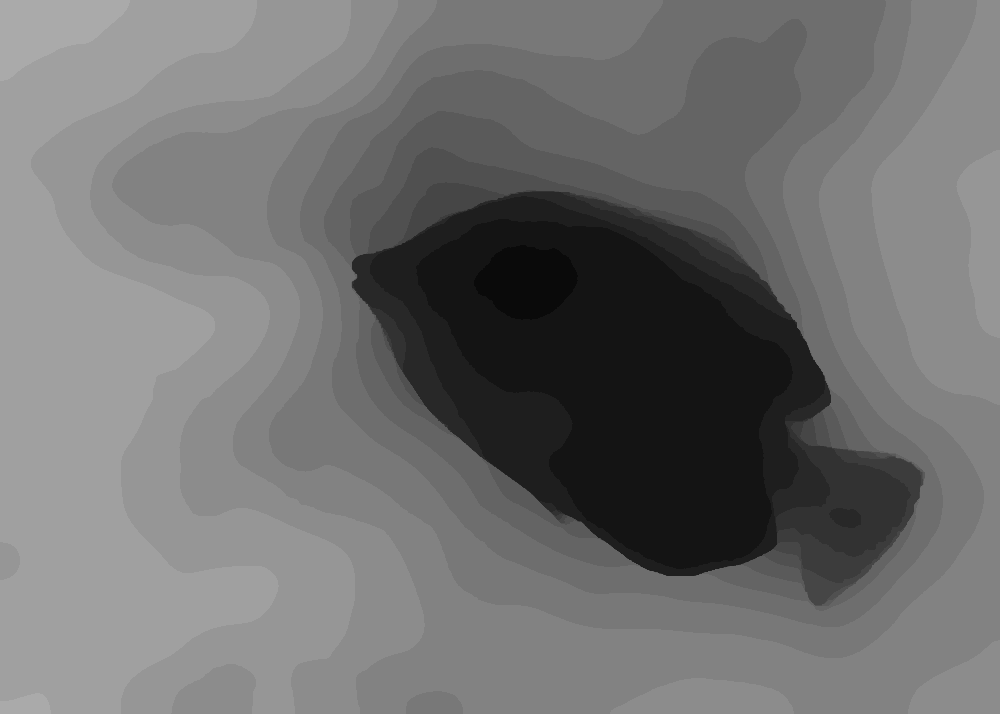}
        \caption{muGIF\cite{guo2018mutually}}
    \label{fig:fishmaprefmuGIF}
\end{subfigure}
\begin{subfigure}{0.24\linewidth}
    \includegraphics[width=1\linewidth]{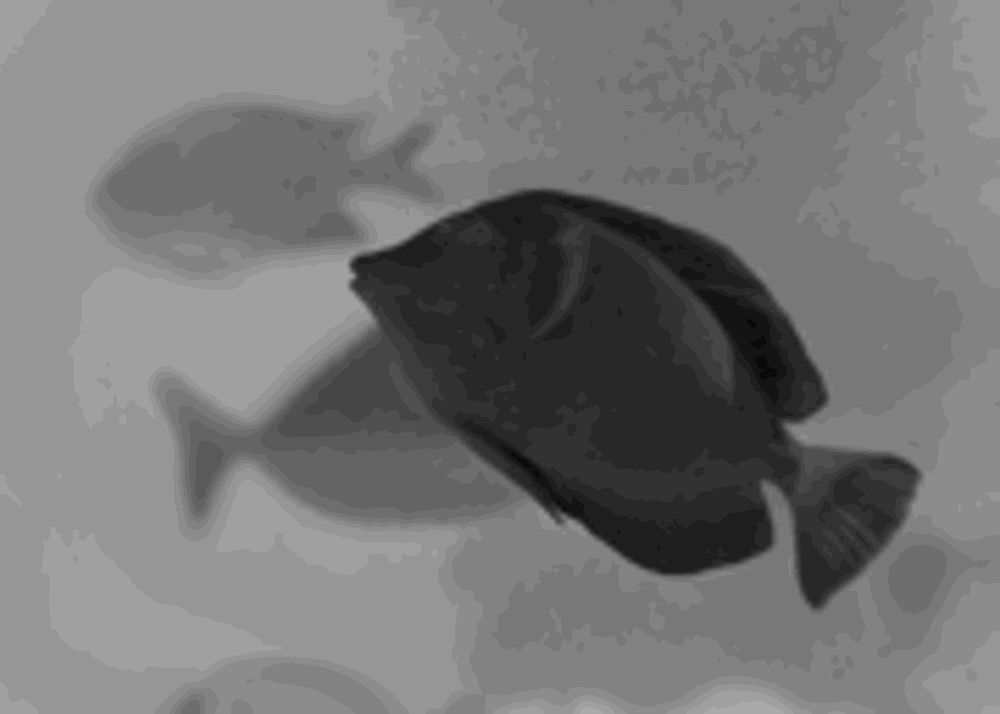}
        \caption{Matting\cite{levin2007closed}}
    \label{fig:fishmaprefMatting}
\end{subfigure}
\begin{subfigure}{0.24\linewidth}
    \includegraphics[width=1\linewidth]{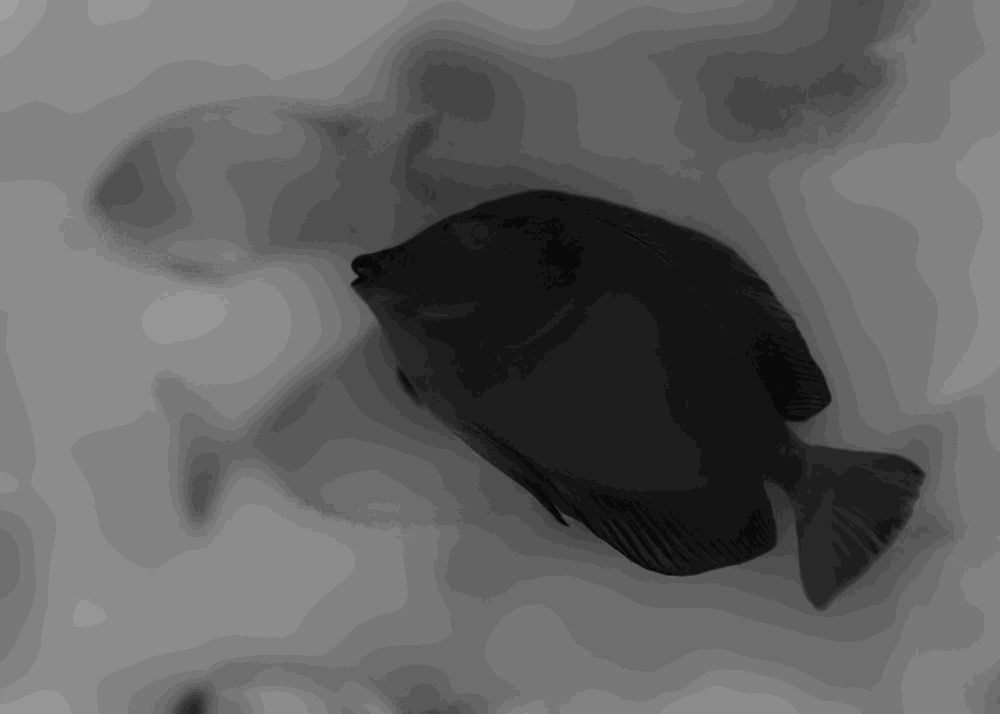}
        \caption{Proposed}
    \label{fig:fishmaprefref2}
\end{subfigure}

\caption{Defocus blur map refinement. a) Input image, b) Blur map, c) Guidance image, d) to g) Refined blur maps by WGIF, muGIF, Matting laplacian and the proposed method.}

\label{fig:frefinement2}
\end{figure}

Figure \ref{fig:frefinement2} shows an example of the refinement algorithm. The blur map shown in Fig. \ref{fig:frefinement2}b was estimated using the proposed CNN approach with a step size of 16 pixels. It was then refined by using the proposed algorithm with a patch radius  $r = 16$, regularization parameter $\epsilon = 0.005$, and the number of iterations $N_{iter}=7 $. The guidance image is shown in Fig \ref{fig:fishmaprefguide} and the refinement result is shown in Fig \ref{fig:fishmaprefref2}. Only the fish's silhouette has been transferred to the refined defocus map. The darker area represents the area in sharp focus, while the brighter area represents the area of more out of focus. We can see that the refined defocus map accurately represents objects in focus such as the fish in the foreground and blurred objects such as those fish in the background.

\begin{figure*}[h!]
\centering
\begin{subfigure}{0.19\linewidth}
    \includegraphics[width=1\linewidth]{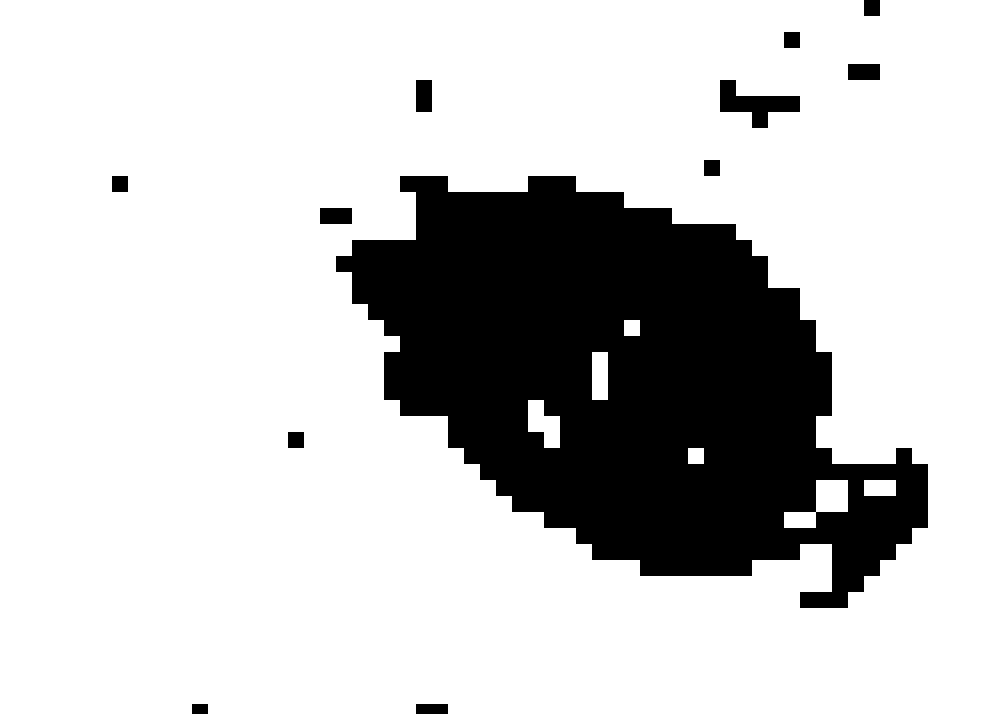}
    \caption{CNN prediction}
    \label{fig:binaryOriginal}
\end{subfigure}
\begin{subfigure}{0.19\linewidth}
    \includegraphics[width=1\linewidth]{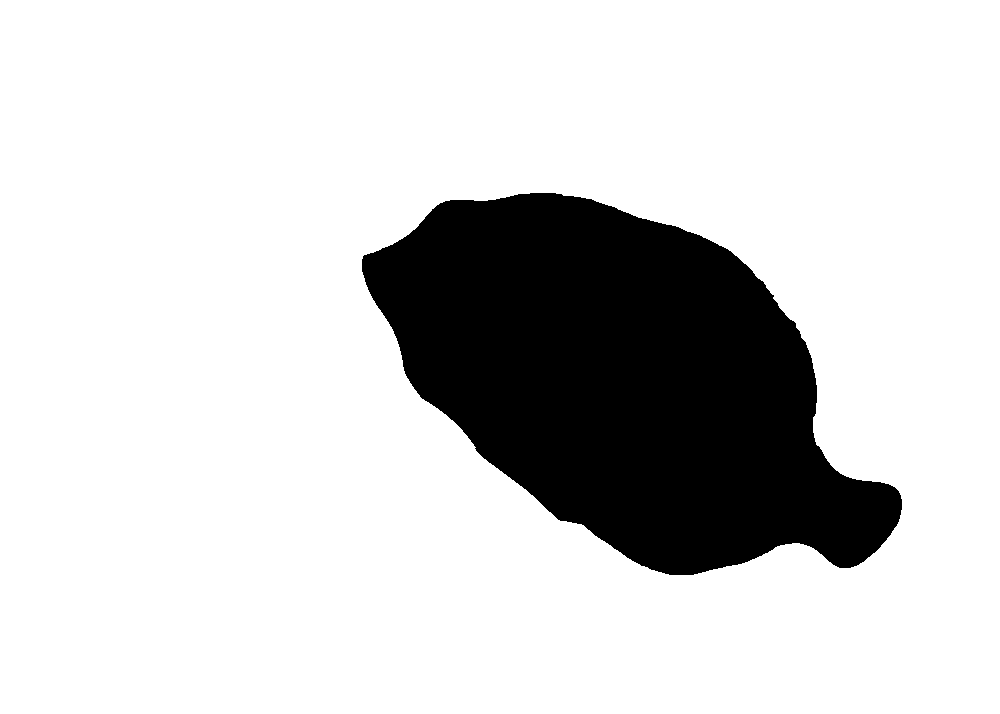}
    \caption{WFIG\cite{li2014weighted}}
    \label{fig:binaryWFIG}
\end{subfigure}
\begin{subfigure}{0.19\linewidth}
    \includegraphics[width=1\linewidth]{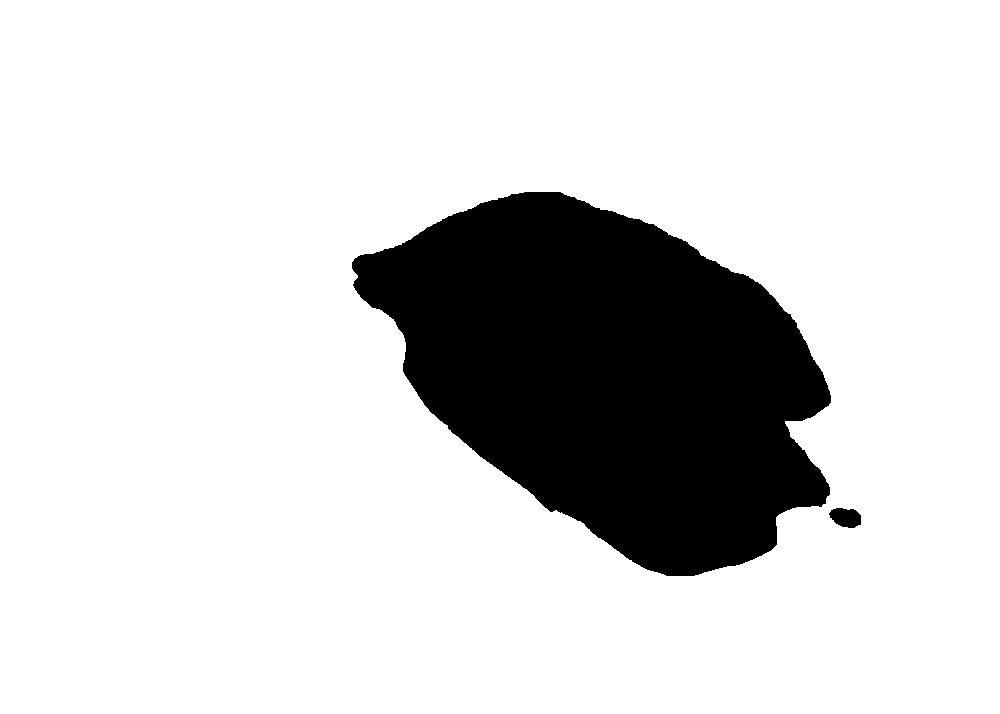}
    \caption{muGIF\cite{guo2018mutually}}
    \label{fig:binarymuGIF}
\end{subfigure}
\begin{subfigure}{0.19\linewidth}
    \includegraphics[width=1\linewidth]{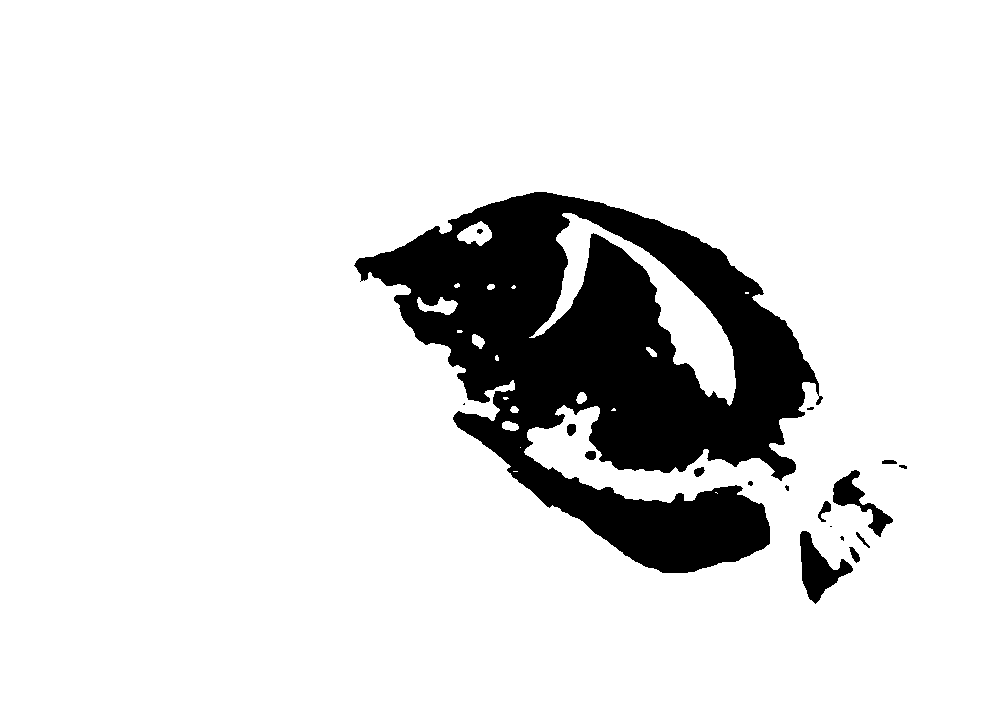}
    \caption{Matting\cite{levin2007closed}}
    \label{fig:binaryMatting}
\end{subfigure}
\begin{subfigure}{0.19\linewidth}
    \includegraphics[width=1\linewidth]{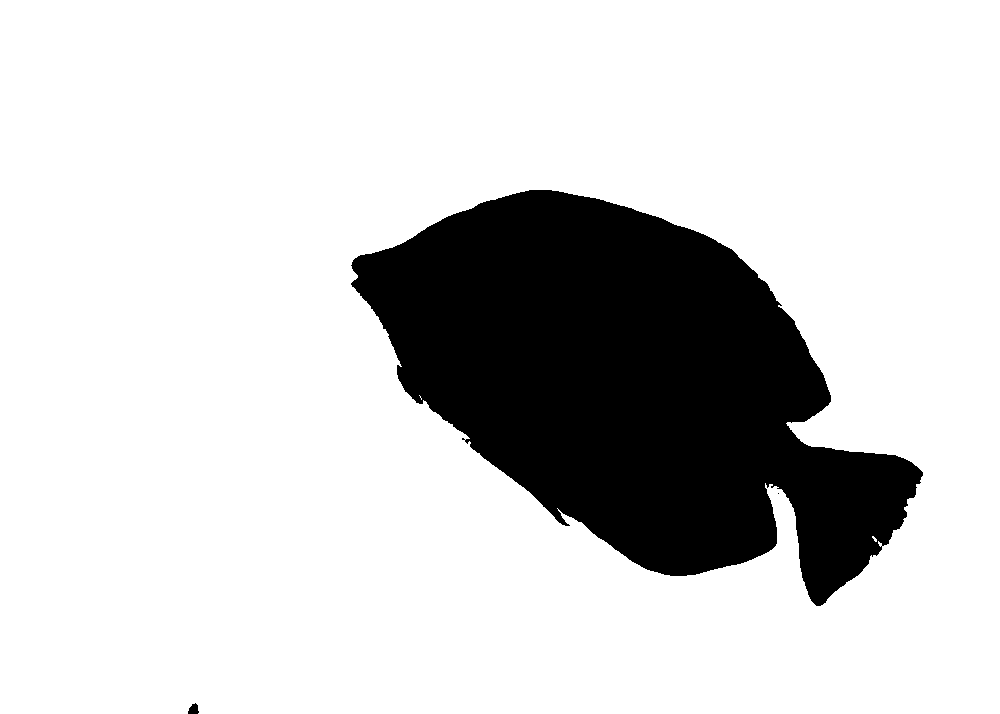}
    \caption{Proposed}
    \label{fig:binaryOurs}
\end{subfigure}

\caption{Binary defocus maps obtained by setting a threshold $\lambda=4$ and assigning a 0 to those pixels $q$ where ${M}_(q)\leq \lambda$ and 1 otherwise. a) Binary defocus map from original prediction, b) to e) Binary defocus map from the map refined by WGIF, muGIF, Matting Laplacian, and our method respectively. This image shows the effectiveness of our method in preserving the defocus level predicted by the CNN.}

\label{fig:frefinement3}
\end{figure*}

Figure \ref{fig:frefinement2} also shows the result of refining the blur map with other methods such as an iterative WGIF \cite{li2014weighted} using the original image as guidance, mutually guided image faltering muGIF \cite{guo2018mutually} and matting Laplacian \cite{Levin2007} to visually appreciate the effectiveness of our method. All four methods clearly remove the blocking effect and wrong predictions in the blur map. However, the WGIF does not preserve edge consistency. The smoothing effect of muGIF is very strong, it removes all the background objects and shrinks the main object's silhouette. Apart from being computationally too expensive, the matting Laplacian algorithm properly preserves the objects in the background but it does transfer some undesirable details from the original image to the defocus map (brightness on the upper side and tail of the closest fish). Our method does preserve defined edges and smooths objects at a lower computational cost. 

Another important characteristic of a our refinement method is that it accurately preserves the defocus level classified by the \ac{CNN}, i.e., after the refinement process is applied the effect of wrong predictions are mitigated and the edges are improved while the defocus prediction values are preserved. This can be better appreciated in Figure \ref{fig:frefinement3}a, where we created a binary version of the original defocus map, the binary image is obtained by assigning a value 0 to every pixel $q$ where $M(q)\leq \lambda$ and 1 otherwise. The parameter $\lambda$ is a user defined threshold ($\lambda=4$ in the figure). In the figure we can see all the pixels with a level of blurriness lower or equal than 4 are shown in black color. We can clearly see that most of those pixels belong to the biggest fish in the image and a small number of pixels belong to the wrong predictions. Figures \ref{fig:frefinement3}b to c show the same threshold ($\lambda = 4$) applied to the defocus maps refined using the WFIG, muGIF, Matting Laplacian and the proposed method.  Our method is the only one which can segment the main object in focus correctly, correct the edges and remove incorrect predictions.

% \begin{figure}[h]
% \centering
% \begin{subfigure}{\linewidth}
% \begin{subfigure}{0.49\linewidth}
%     \includegraphics[width=1\linewidth]{images/refinement2/tedy.png}
% \end{subfigure}
% \begin{subfigure}{0.49\linewidth}
%     \includegraphics[width=1\linewidth]{images/refinement2/RF_cellphoneman_original.png}
% \end{subfigure}
% \caption{Input image}
% \end{subfigure}
% \begin{subfigure}{\linewidth}
% \begin{subfigure}{0.49\linewidth}
%     \includegraphics[width=1\linewidth]{images/refinement2/tedy_ours_s16_bmap.png}
% \end{subfigure}
% \begin{subfigure}{0.49\linewidth}
%     \includegraphics[width=1\linewidth]{images/refinement2/RF_cellphoneman_bmap_8.png}
% \end{subfigure}
% \caption{Blur map}
% \end{subfigure}
% \begin{subfigure}{\linewidth}
% \begin{subfigure}{0.49\linewidth}
%     \includegraphics[width=1\linewidth]{images/refinement2/tedy_ours_s16_refined2.png}
% \end{subfigure}
% \begin{subfigure}{0.49\linewidth}
%     \includegraphics[width=1\linewidth]{images/refinement2/RF_cellphoneman_refbmap.png}
% \end{subfigure}
% \caption{Refined blur map}
% \end{subfigure}

% \caption{Defocus blur map refinement, a) Original image, b) Blur map estimation, c) Refined blur map. }
% \label{fig:refinement_comparison_2}
% \end{figure}

%% file: 03a1_Blurmap_comparison.tex
\section{Comparison of defocus estimation algorithms}

In this section, we compare the proposed method with  other methods in two categories: classical methods such as entropy, standard deviation and variance of the laplacian of the patch, and modern methods including three handcrafted features "Lbp-based segmentation of defocus blur" from Yi et al. \cite{yi2016lbp}, "Defocus map estimation from a single image" from Zhuo et al. \cite{zhuo2011defocus} and   "Fast defocus map estimation" (FDM) from Chen et al.\cite{chen2016fast} and a CNN based method from Lee et al. "Deep Defocus Map Estimation Using Domain Adaptation" (DMNet) \cite{lee2019deep}.

\begin{figure*}[t]
\centering
\begin{subfigure}{0.24\linewidth}
    \includegraphics[width=1\columnwidth]{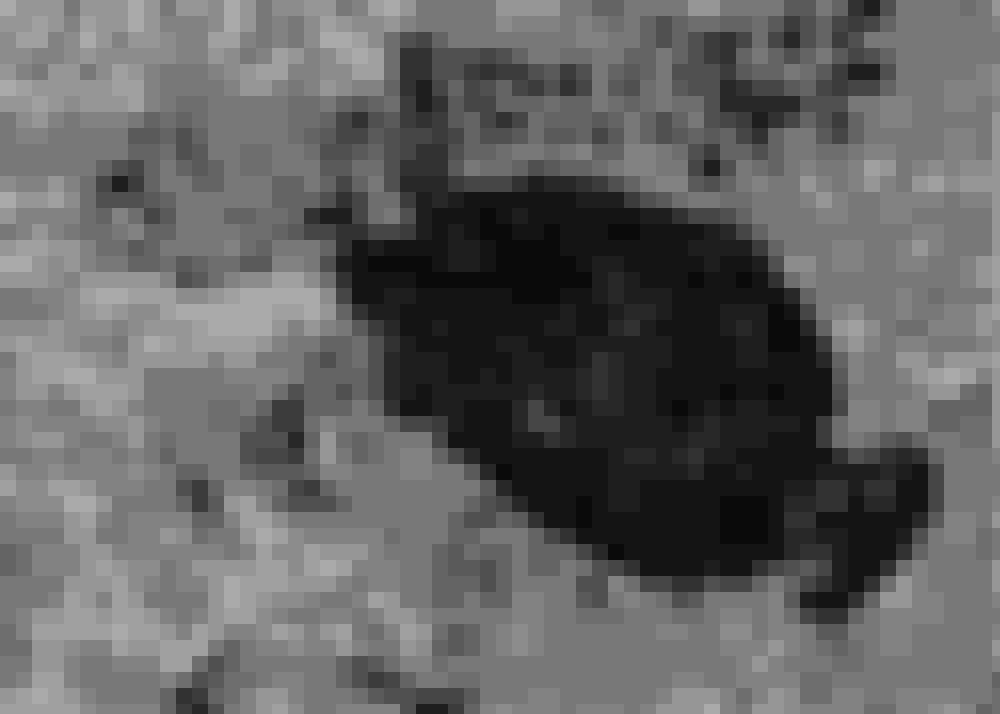}
\end{subfigure}
\hfill
\begin{subfigure}{0.24\linewidth}
    \includegraphics[width=1\columnwidth]{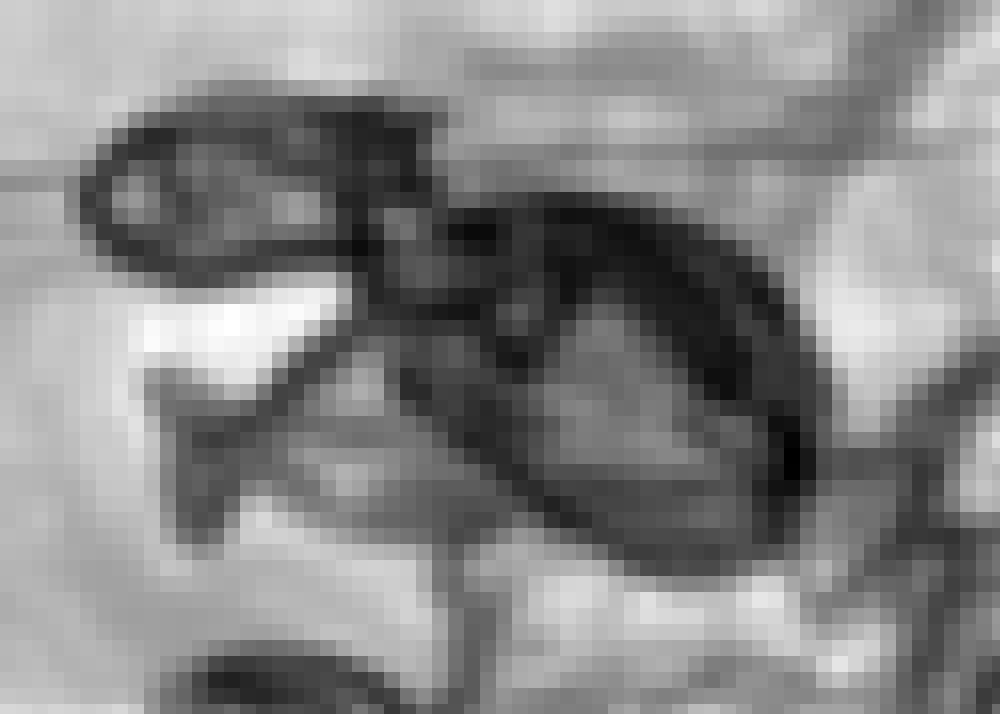}
\end{subfigure}
\hfill
\begin{subfigure}{0.24\linewidth}
    \includegraphics[width=1\columnwidth]{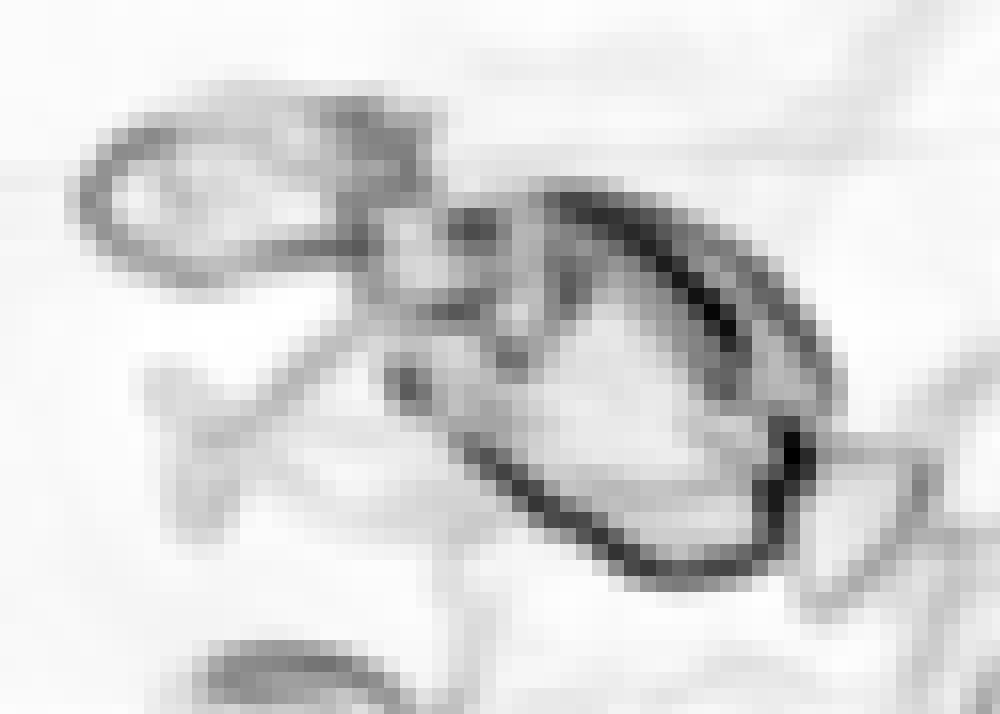}
\end{subfigure}
\hfill
\begin{subfigure}{0.24\linewidth}
    \includegraphics[width=1\columnwidth]{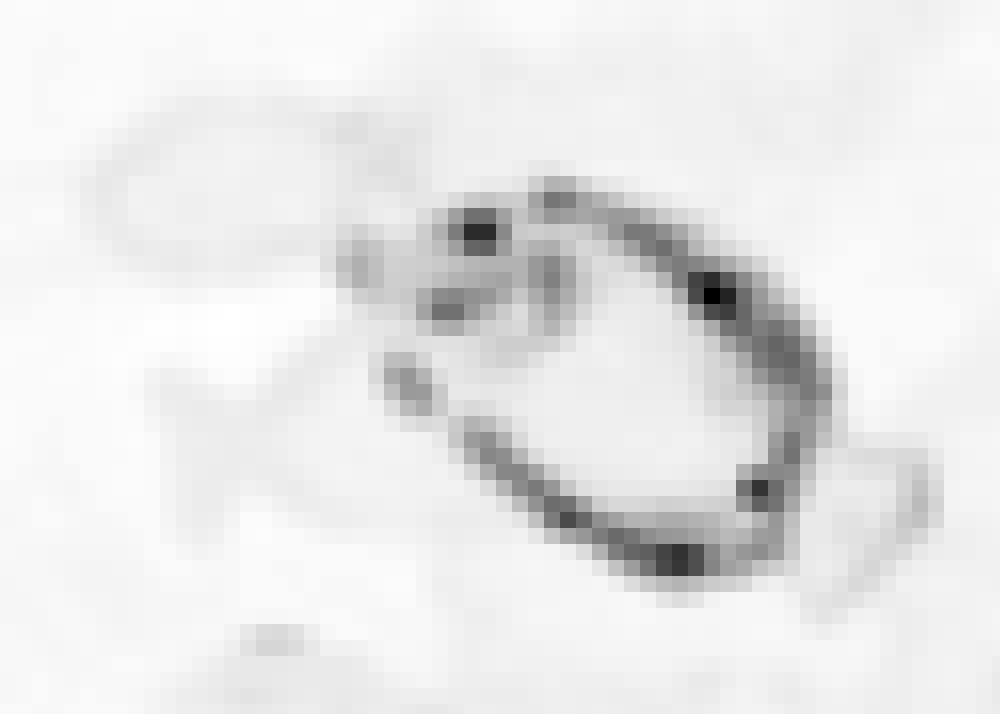}
\end{subfigure}
\hfill

\begin{subfigure}{0.24\linewidth}
    \includegraphics[width=1\columnwidth]{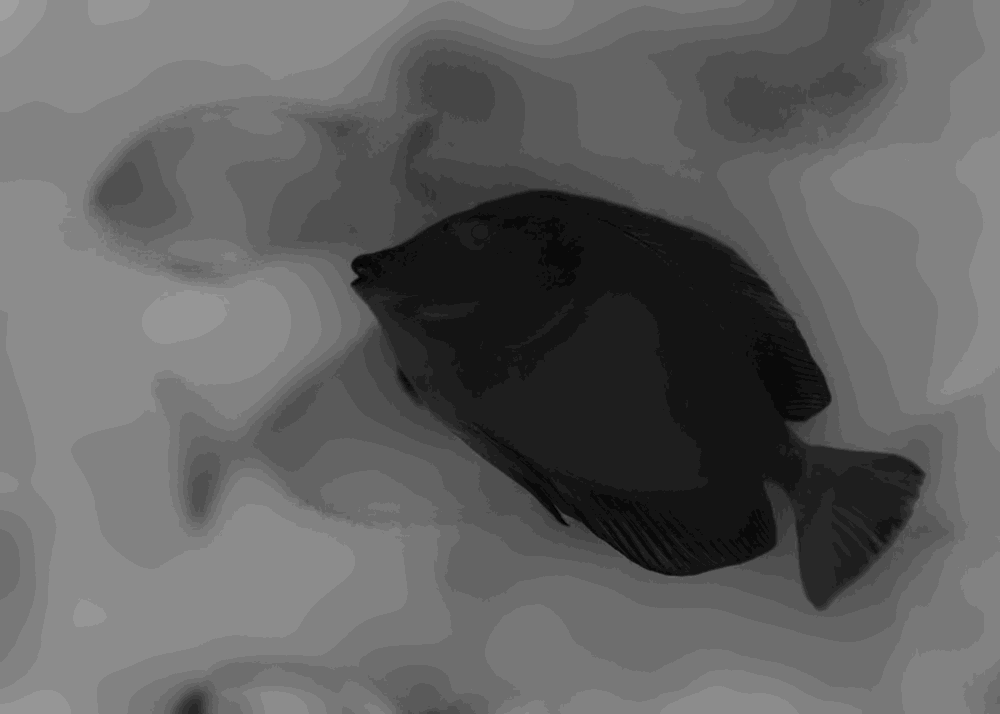}
    \caption{Proposed method}
\end{subfigure}
\hfill
\begin{subfigure}{0.24\linewidth}
    \includegraphics[width=1\columnwidth]{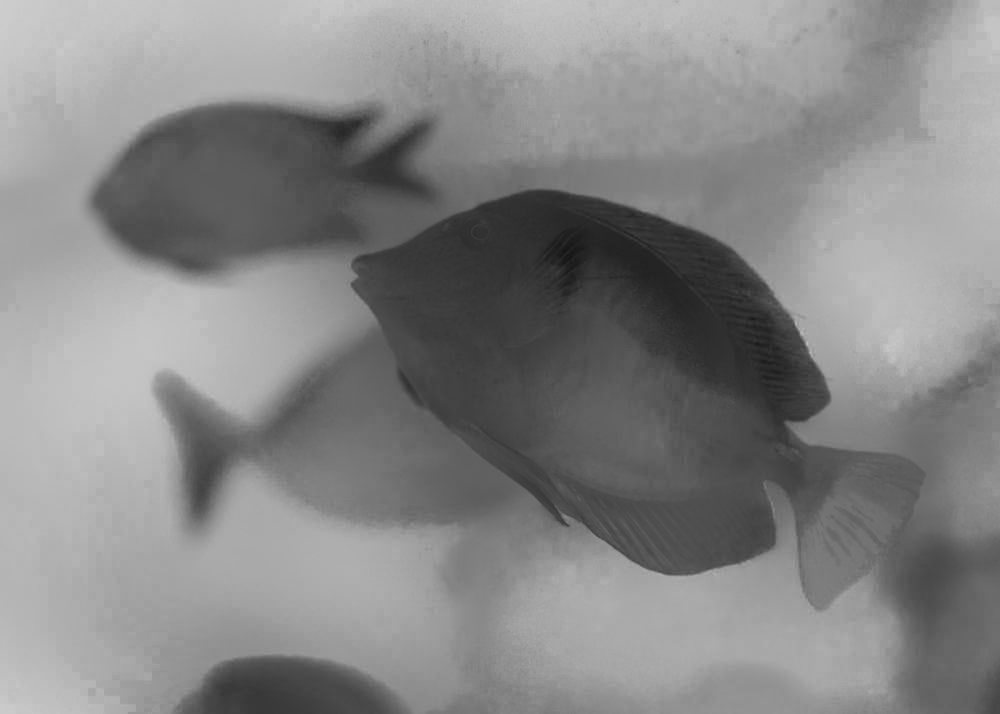}
    \caption{Entropy}
\end{subfigure}
\hfill
\begin{subfigure}{0.24\linewidth}
    \includegraphics[width=1\columnwidth]{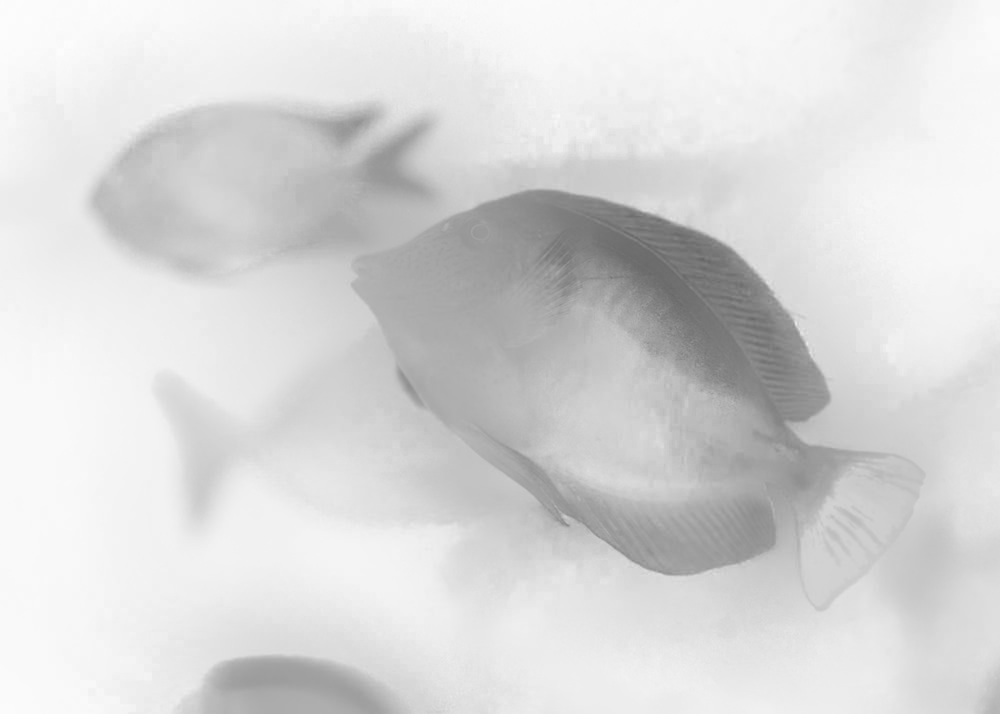}
    \caption{Standard deviation}
\end{subfigure}
\hfill
\begin{subfigure}{0.24\linewidth}
    \includegraphics[width=1\columnwidth]{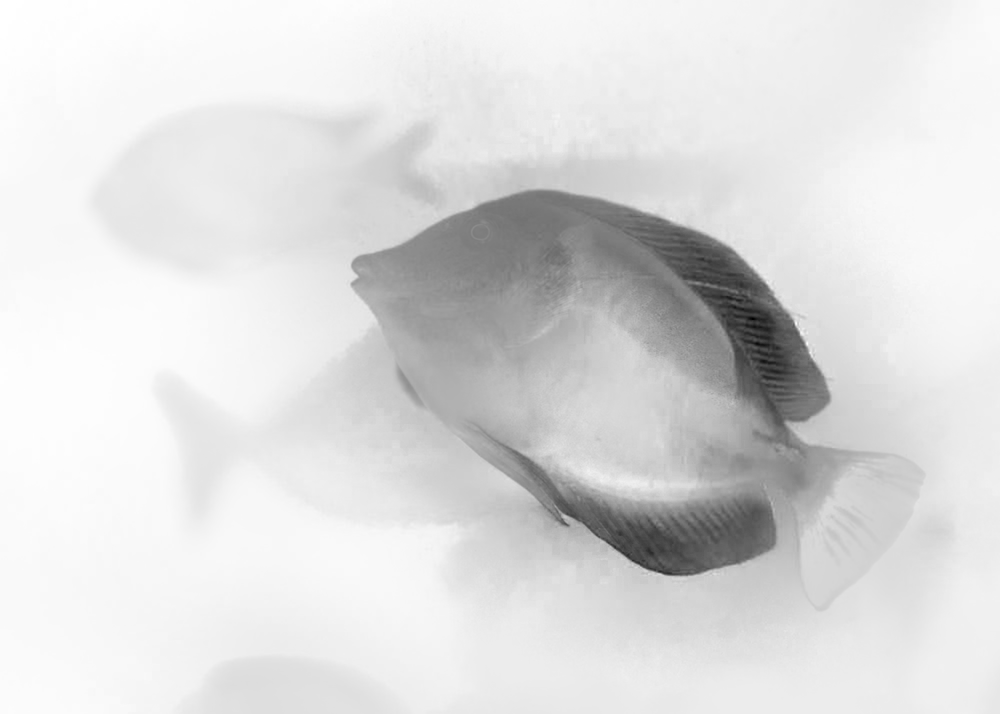}
    \caption{Var. of the Laplacian \cite{variance_lapla}}
\end{subfigure}
\hfill

\caption{Defocus blur map and refined map using different methods. The blur map obtained using different methods is displayed on the top row and the refined version using the algorithm introduced in section \ref{refinement} is shown in the bottom row. (a) Our method. (b) Entropy. (c) Standard deviation. (d) Variance of the Laplacian.  }

\label{fig:refinement2_comparison}
\end{figure*}

\subsection{Comparison with classical methods}
 We studied three methods: entropy, standard deviation, and variance of Laplacian  \cite{variance_lapla}. The ideas of these three methods are similar. The blurriness of a pixel is measured by a quantity such as the entropy or standard deviation of a patch of pixels centred at that pixel. For the variance of Laplacian, the image is filtered by using the Laplacian operator. The blurriness of a pixel is measured by the variance of a patch of pixels of the filtered image centred at that pixel.  For these three methods, we use a patch size of $16 \times 16$ to calculate the entropy, standard deviation and variance of the Laplacian.  Simulation results are shown in Fig. \ref{fig:refinement2_comparison}. The proposed CNN-based method classifies the level of defocus properly for both the edges of objects and the featureless patches. On the other hand, the entropy and standard deviation of the patch can only classify adequately the edges and classify all featureless patches as out of focus. Because both methods are trying to capture edge detection rather than blurriness estimation, they failed to capture information that the fish in the foreground is in focus, see e.g., the certain parts of the body of the fish which are in bright area indicating the area is wrongly classified as out of focus by these two methods. We made a similar observation when we compared the results of the CNN-based method with the variance of the Laplacian. For example, the tail of the fish in the foreground is in bright grayscale indicating it is wrongly classified as out of focus by this method.  

\begin{figure*}[h]
\centering
\begin{subfigure}{0.16\linewidth}
    \includegraphics[width=1\linewidth]{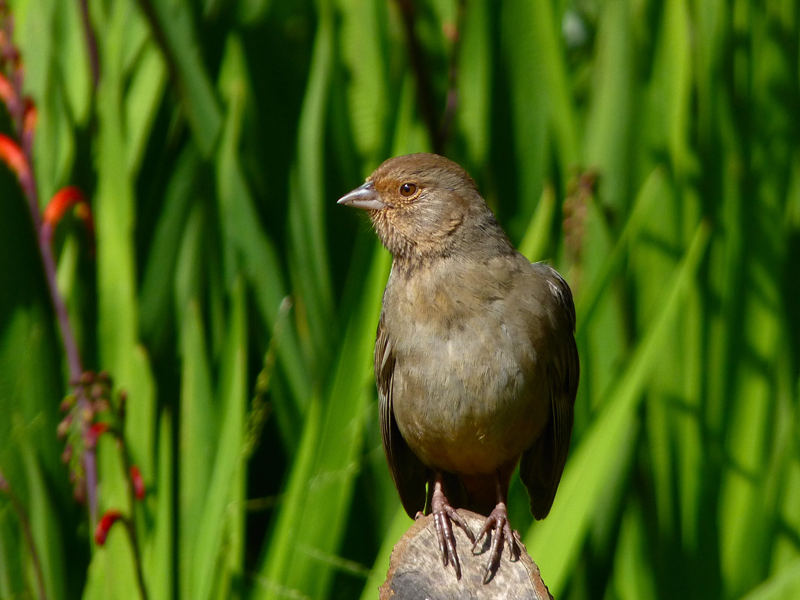}
    \label{fig:fishin}
\end{subfigure}
\begin{subfigure}{0.16\linewidth}
    \includegraphics[width=1\linewidth]{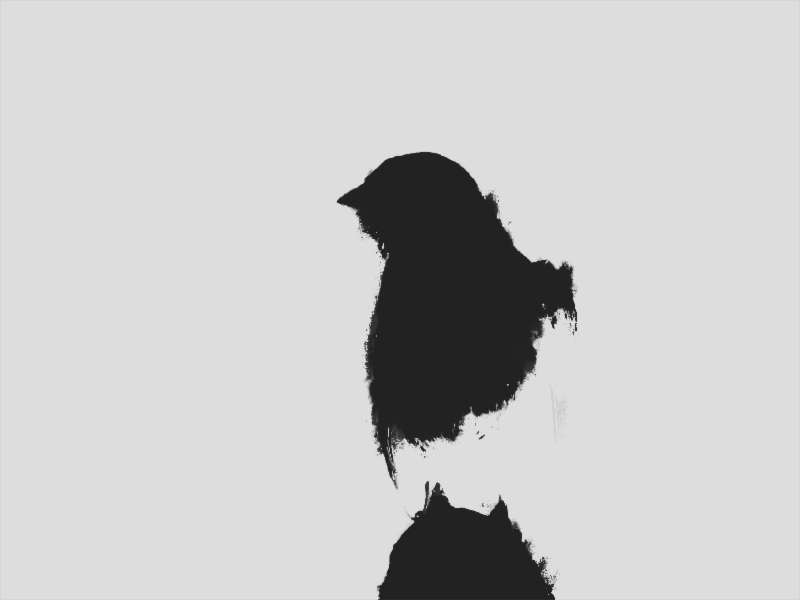}
    \label{fig:fishmap}
\end{subfigure}
\begin{subfigure}{0.16\linewidth}
    \includegraphics[width=1\linewidth]{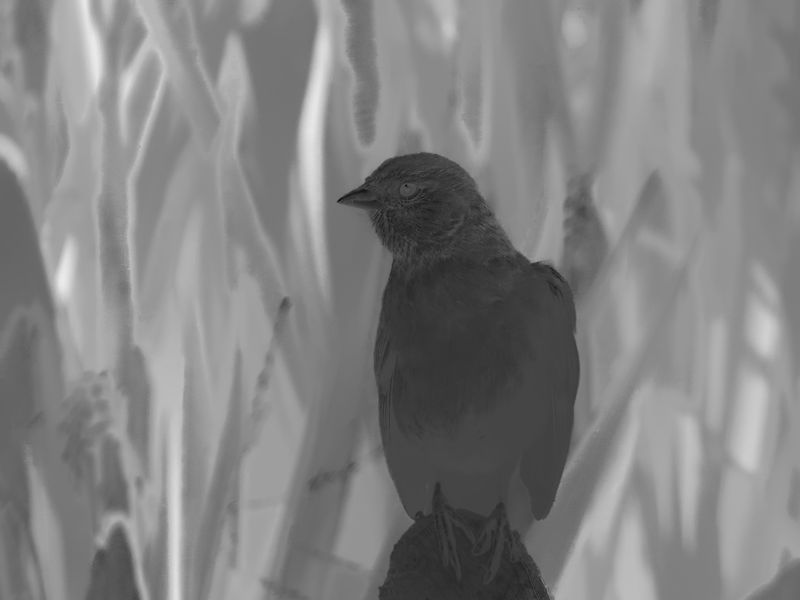}
    \label{fig:fishrefined}
\end{subfigure}
\begin{subfigure}{0.16\linewidth}
    \includegraphics[width=1\linewidth]{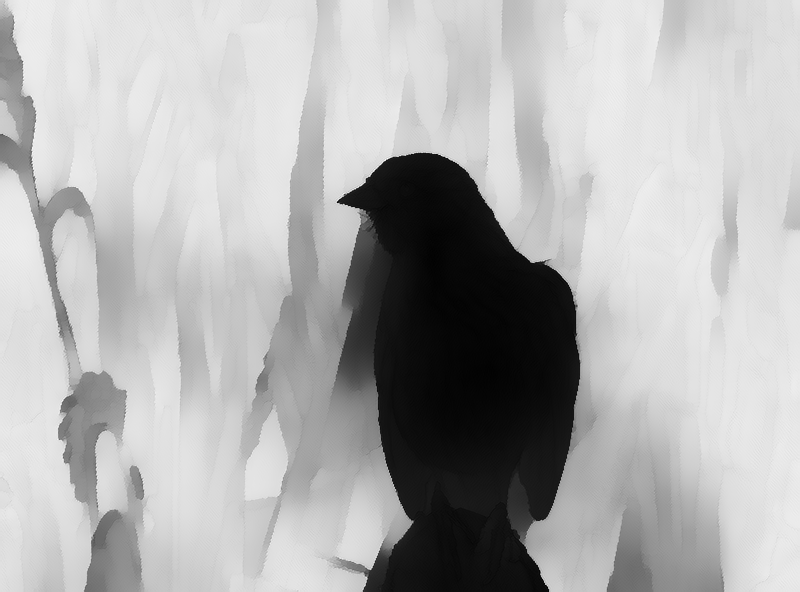}
    \label{fig:fishrefined}
\end{subfigure}
\begin{subfigure}{0.16\linewidth}
    \includegraphics[width=1\linewidth]{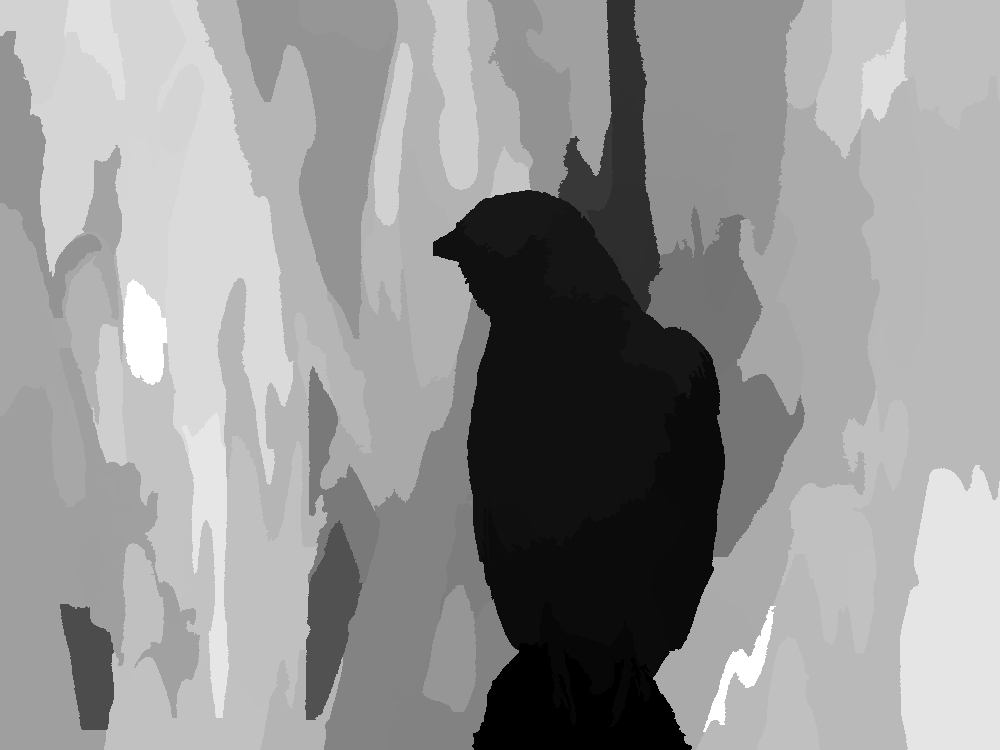}
    \label{fig:fishrefined}
\end{subfigure}
\begin{subfigure}{0.16\linewidth}
    \includegraphics[width=1\linewidth]{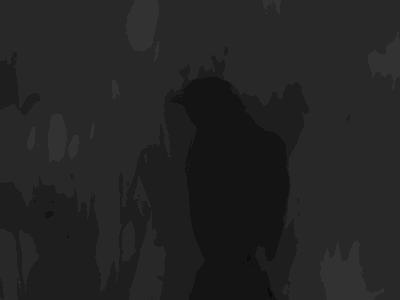}
    \label{fig:fishrefined}
\end{subfigure}
% \begin{subfigure}{0.33\linewidth}
%     \includegraphics[width=1\linewidth]{images/estimation/bird_W2_m0_2_m1_4_step_32.png}
%     \caption{Ours binary}
%     \label{fig:fishrefined}
% \end{subfigure}

\begin{subfigure}{0.16\linewidth}
    \includegraphics[trim=2cm 0 0 0, clip,width=1\linewidth]{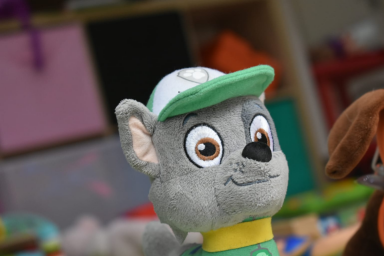}
    \caption{Input}
    \label{fig:fishin}
\end{subfigure}
\begin{subfigure}{0.16\linewidth}
    \includegraphics[trim=2cm 0 0 0, clip,width=1\linewidth]{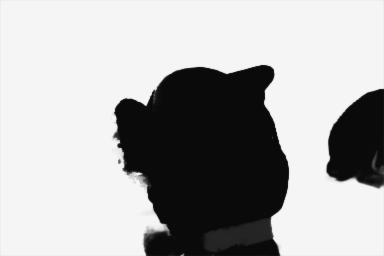}
    \caption{Yi \cite{yi2016lbp}}
    \label{fig:fishmap}
\end{subfigure}
\begin{subfigure}{0.16\linewidth}
    \includegraphics[trim=2cm 0 0 0, clip,width=1\linewidth]{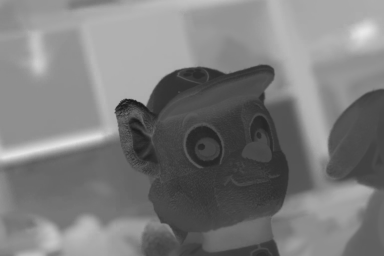}
    \caption{Zhuo\cite{zhuo2011defocus}}
    \label{fig:fishrefined}
\end{subfigure}
\begin{subfigure}{0.16\linewidth}
    \includegraphics[trim=2cm 0 0 0, clip,width=1\linewidth]{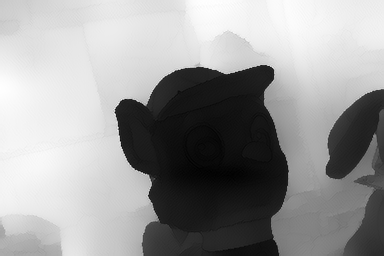}
    \caption{DMENet\cite{lee2019deep}}
    \label{fig:fishrefined}
\end{subfigure}
\begin{subfigure}{0.16\linewidth}
    \includegraphics[trim=2cm 0 0 0, clip,width=1\linewidth]{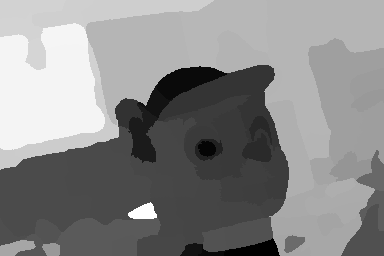}
    \caption{FDM\cite{chen2016fast} }
    \label{fig:fishrefined}
\end{subfigure}
\begin{subfigure}{0.16\linewidth}
    \includegraphics[trim=2cm 0 0 0, clip,width=1\linewidth]{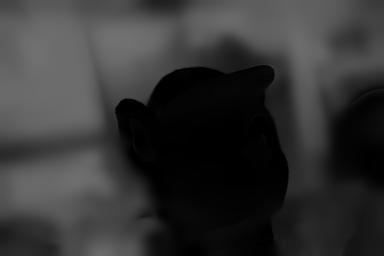}
    \caption{Ours }
    \label{fig:fishrefined}
\end{subfigure}

\caption{Blur map comparisons, a) Input image, b) Yi \cite{yi2016lbp} ($\beta = 0.25, s_1=7, s_2=11, s_3=15, T_{LBP}=0.025$), c) Zhuo\cite{zhuo2011defocus} ($ e=0.01, \sigma = 1, \lambda = 0.001, M = 10$), d) DMENet\cite{lee2019deep}, e) FDM\cite{chen2016fast} ($\sigma = 0.75$) , f) Proposed method.}
\label{fig:bird_comparison}
\end{figure*}

\subsection{Comparison with modern methods}

The bird image shown in the top row of Fig. \ref{fig:bird_comparison} presents different levels of defocus blur. Yi's algorithm was implemented using a threshold $ T_{LBP}=0.025$, for each level we use square local regions of size $s_1=7, s_2=11, s_3=15$ and finally we set the weight $\beta = 0.25$ for the multi-scale inference step. Yi's algorithm does not segment the bird which is completely in focus. Zhuo's algorithm is shown in column (c). We set the edge threshold as $e=0.01$ and the standard deviation as $\sigma = 1$ for the blur detection step. The interpolation step was performed using $\lambda = 0.001$ and a maximum blur level $M = 10$. Zhuo's method successfully identify that the bird is in focus. However a few details from the background are transferred to the defocus map after the interpolation step. Chen's method (FDM) uses Zhuo's approach to estimate the blur level on edges, but uses the SLIC algorithm \cite{achanta2012slic} for the interpolation. In our simulation, the SLIC parameters are set as: 200 super-pixels and a standard deviation $\sigma = 0.75$. As shown in the figure, for this particular image the focused region is not well defined. In comparison, the DMNet has a very good performance in capturing both background and foreground blurriness levels. The result produced by the proposed algorithm is shown in column (f). Our method can represent the different levels of blurriness in the bird image according to the \ac{CNN} prediction. Our refinement method reduces the number of details transferred from the original image to the defocus map, this can be appreciated more clearly in the background. 

Another comparison  example is shown in the last row of Fig. \ref{fig:bird_comparison}. We can see that Yi's algorithm performs better on this image capturing and segmenting focused pixels in the foreground. The methods of Zhuo and Chen have difficulty removing details in the foreground especially in the toy's eyes region. DMENet produces a sharp and accurate defocus map. Our method assigns the lowest blur level to the foreground allowing us to differentiate the focused region from the rest of the image.

%% file: 03a_Adaptive_enhancement.tex
% \begin{figure*}[!b]
% \centering
% \begin{subfigure}{0.32\textwidth}
%     \includegraphics[width=\textwidth]{images/enhancement/bird_original.png}
%     \caption{Input}
% \end{subfigure}
% \begin{subfigure}{0.32\textwidth}
%     \includegraphics[width=\textwidth]{images/enhancement/bird_bmap_refined_step16.png}
%     \caption{Refined blur map $M(n,m)$}
% \end{subfigure}
% \begin{subfigure}{0.32\textwidth}
%     \includegraphics[width=\textwidth]{images/enhancement/bird_gain.png}
%     \caption{Gain map $\lambda(n,m)$}
% \end{subfigure}
% \caption{Non linear transformation result. a) input image, b) Refined blur map (step size of 16 pixels), c) Gain map ($\alpha_1 = 30, \beta_1 = 0.15,\alpha_2 = 92, \beta_2 =0.4 $ and $\lambda_{max}=2$).}
% \label{fig:AdaptEnh}
% \end{figure*}

\subsection{Adaptive enhancement}

Unsharp masking (UM) \cite{unsharp_masking} is frequently used for image enhancement. The \ac{UM} algorithm can be described as \begin{equation}
    J(q) = I(q) + \lambda Z(q)
\label{eq:UM}
\end{equation}
where $\lambda$ is used to control the enhancement level, $Z(q) = I(q) - B(q)$,
 and $B$ is a low pass filtered version of $I$. For a fixed $\lambda$, UM performs a higher degree of enhancement in higher contrast or dynamic regions of the image. To raise the sharpness level of a region with low contrast, the parameter $\lambda$ has to be set to a relatively higher value, resulting in over enhancement in regions of high contrast. As such, an undesirable non-natural looking result may be produced. Besides, natural images may present different blur and contrast levels in different areas. Having a constant scale-factor $\lambda$ for the whole image could also lead to the sharpening of regions that are intentionally produced with a high level of blur, e.g., a highly smoothed background to produce a shallow depth of field. It is not desirable to sharpen those areas. Different methods have been proposed to address those problems, including using non-linear or edge-aware filters \cite{mitra1991new, ramponi1998cubic, ramponi1996nonlinear} and a pixel adaptive $\lambda$ that produces a spatially varying enhancement \cite{polesel2000image,ye2018blurriness}.

\begin{figure}[h!]
    \centering
    \includegraphics[width = 0.55 \linewidth]{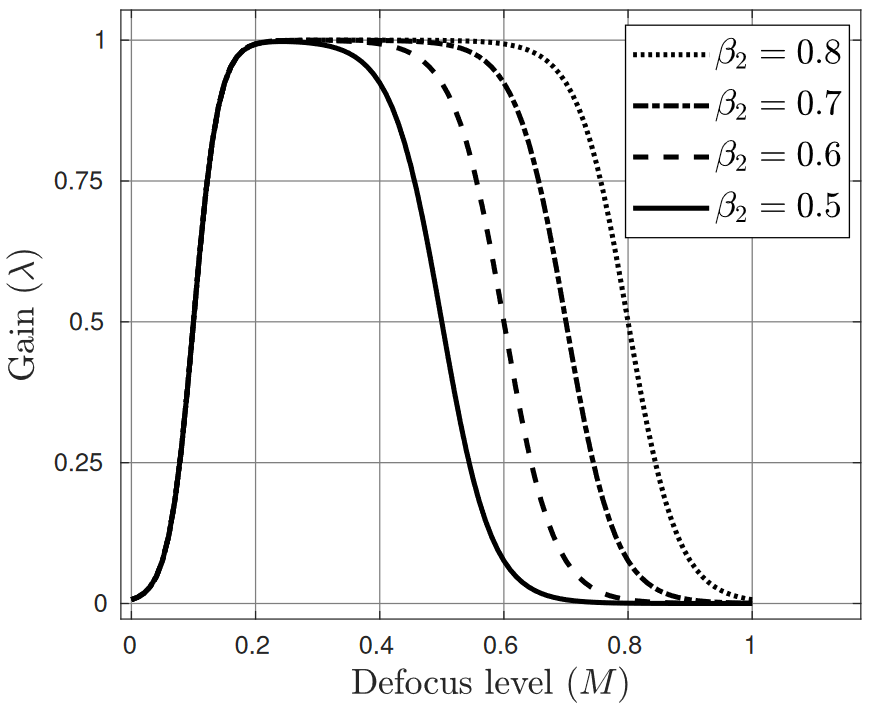}
    \caption{Non linear transformation of the defocus level. For easy visualization, the defocus level is normalized and the parameter settings are $\alpha_1=50$, $\beta_1=0.1$ and $\alpha_2=25$. }
    \label{fig:beta_distribution}
\end{figure}

We follow the same idea  of using a pixel adaptive $\lambda$ and propose an adaptive unsharp masking algorithm such that the parameter $\lambda$ for each pixel is a nonlinear function of the refined defocus blur map. The aim of the nonlinear function is to avoid the enhancement of sharp regions as well as in heavily blurred regions. We use the following nonlinear function:

\begin{equation}
    \lambda(q) =  \lambda_{max}\times \lambda_1(q) \times \lambda_2(q)
    \label{eq:beta}
\end{equation}
where $\lambda_{max}$ is the  desired maximum gain level,  $ \lambda_1$ and $ \lambda_2$ are two sigmoid functions: 

\begin{equation}
    \lambda_1(q) = \frac{1}{1+e^{-\alpha_1 \left (M(q)-\beta_1 \right)}}
\end{equation}

\begin{equation}
    \lambda_2(q) = 1-\frac{1}{1+e^{-\alpha_2 \left (M(q)-\beta_2 \right)}}
\end{equation}

The four parameters $\alpha_1$, $\beta_1$, $\alpha_2$ and $\beta_2$ are determined by the user to have control over the sharpening gain.  An example of controlling the gain is shown in Fig. \ref{fig:beta_distribution}. In this example, we set $\alpha_1=50$, $\beta_1=0.1$ and $\alpha_2=25$ are kept fixed while $\beta_2$ varies. The scale of the blur map is linearly normalized from the interval $[0,19]$ to $[0,1]$ for better visualization. $\beta_1$ and $\beta_2$ indicate the points where the gain is $0.5\times\lambda_{max}$ and $\alpha_1$ and $\alpha_2$ set the growth speed of the curve.

%%%% Test image
\begin{figure*}[t!]
\centering
    \begin{subfigure}{0.48\textwidth}
    \begin{tikzpicture}[
        zoomboxarray,
        zoomboxarray columns=2,
        zoomboxarray rows=2,
        %connect zoomboxes,
        zoombox paths/.append style={line width=0.75pt}
    ]
        \node [image node] { \includegraphics[trim=1cm 9cm 1cm 12cm, clip,width=0.48\textwidth]{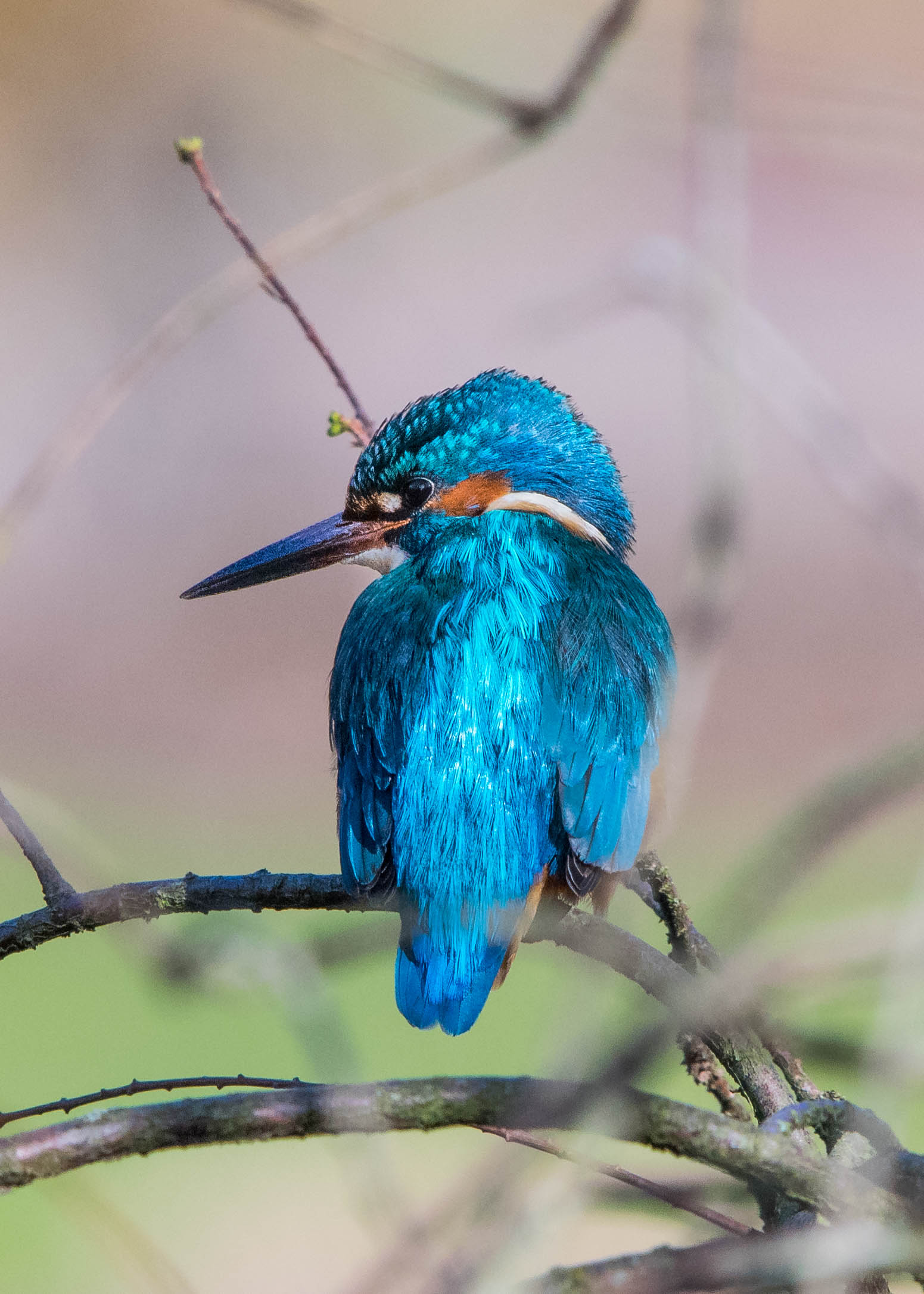} };
        \zoombox[color code = orange,magnification=4]{0.45,0.7} % background
        \zoombox[color code = green,magnification=5]{0.8,0.6} % chest
        \zoombox[color code = red,magnification=4]{0.25,0.25}% nest
        \zoombox[color code = blue,magnification=4]{0.4,0.5} % stick
    \end{tikzpicture}
        \caption{Input image}
        \label{fig:adaptive_enhancement_input}

    \end{subfigure}
    \begin{subfigure}{0.24\textwidth}
    \includegraphics[trim=1cm 9cm 1cm 12cm, clip,width=0.96\textwidth]{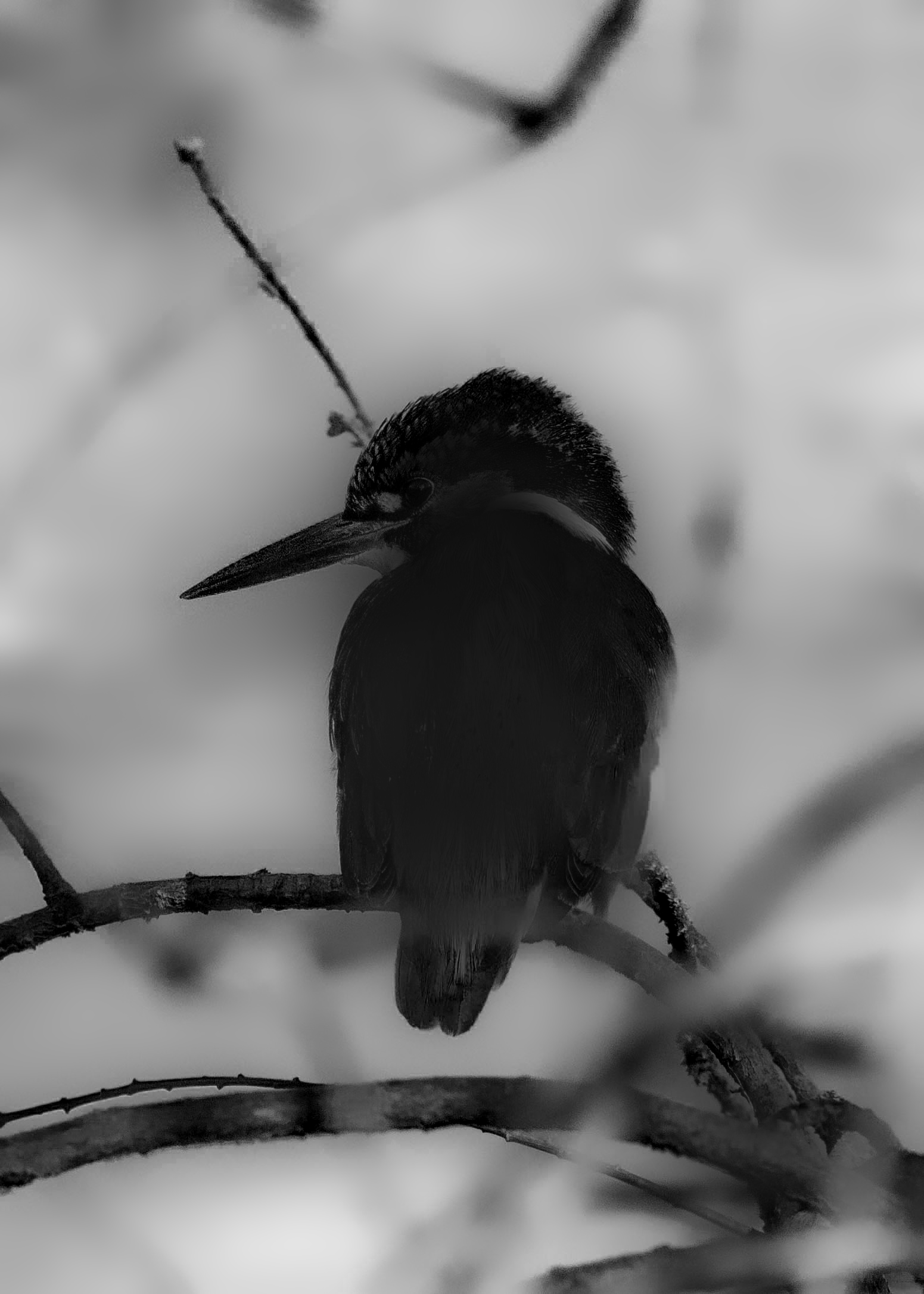}

        \caption{Refined blur map}
        \label{fig:adaptive_enhancement_bmap}
    \end{subfigure}
    \begin{subfigure}{0.24\textwidth}
    \includegraphics[trim=1cm 9cm 1cm 12cm, clip,width=0.96\textwidth]{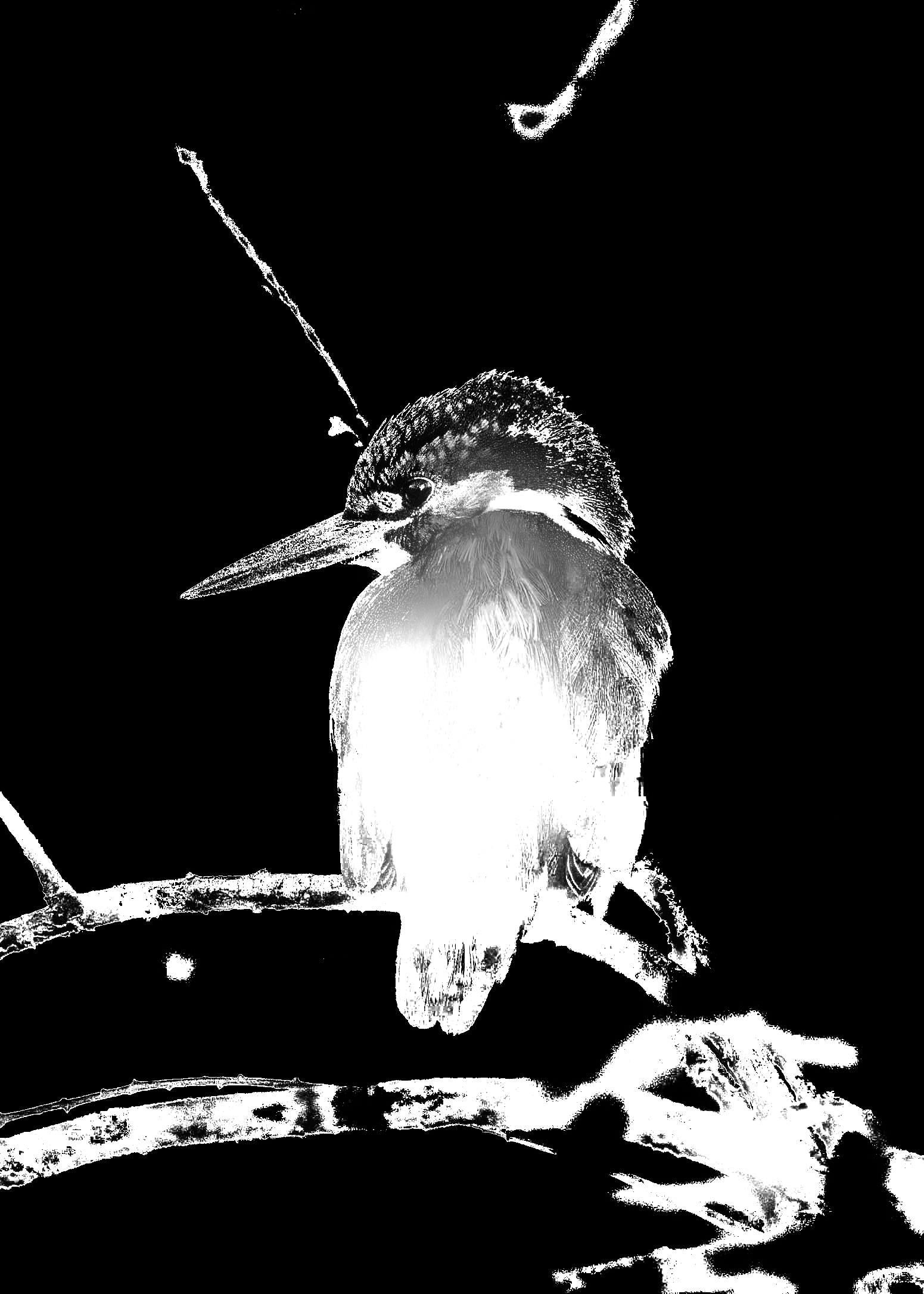}
        \caption{Gain map}
        \label{fig:adaptive_enhancement_gain}
    \end{subfigure}

    \begin{subfigure}{0.48\textwidth}
    \begin{tikzpicture}[
        zoomboxarray,
        zoomboxarray columns=2,
        zoomboxarray rows=2,
        %connect zoomboxes,
        zoombox paths/.append style={line width=0.75pt}
    ]
        \node [image node] { \includegraphics[trim=1cm 9cm 1cm 12cm, clip,width=0.48\textwidth]{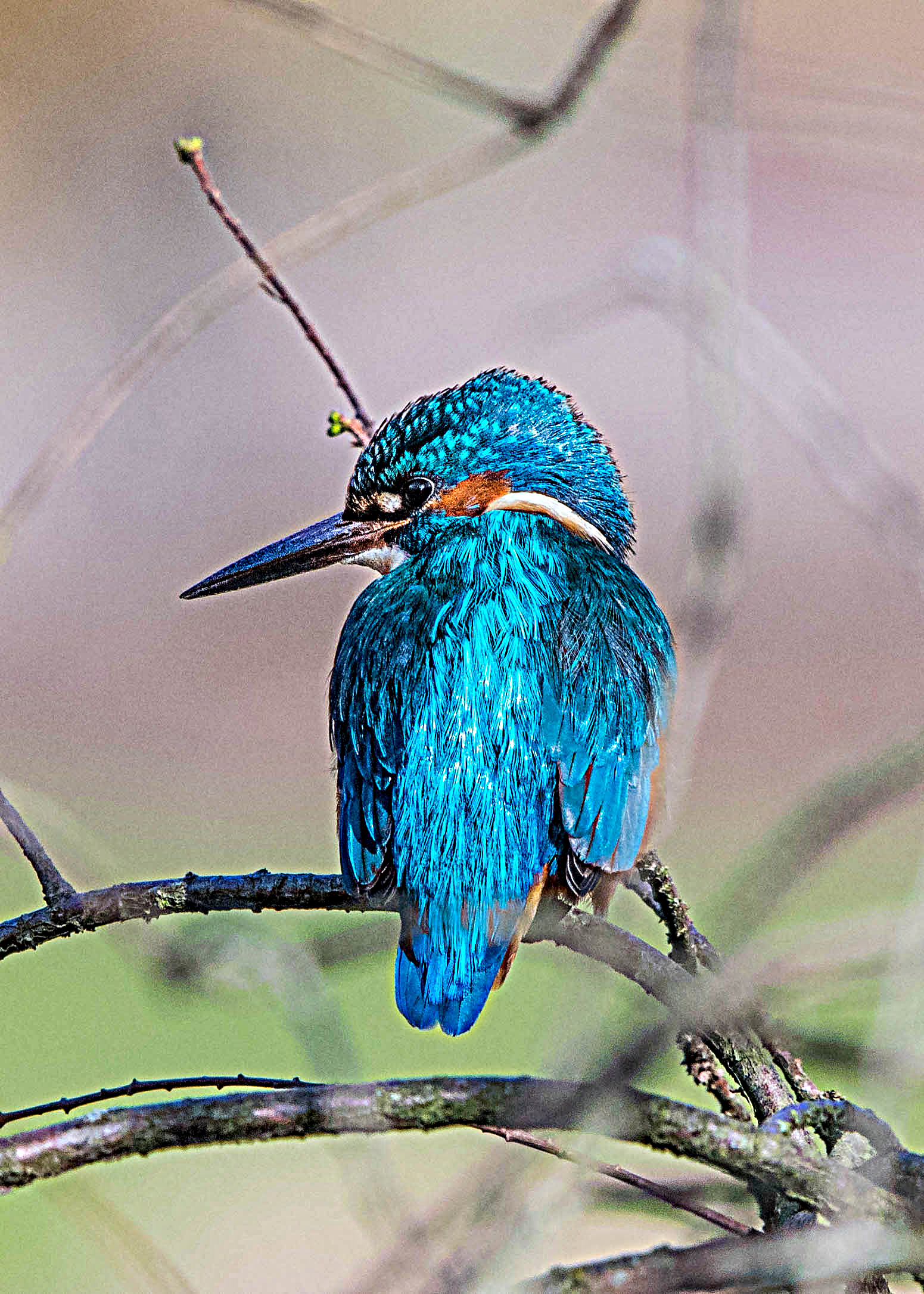} };
        \zoombox[color code = orange,magnification=4]{0.45,0.7} % background
        \zoombox[color code = green,magnification=5]{0.8,0.6} % chest
        \zoombox[color code = red,magnification=4]{0.25,0.25}% nest
        \zoombox[color code = blue,magnification=4]{0.4,0.5} % stick
    \end{tikzpicture}
        \caption{UM}
        \label{fig:adaptive_enhancement_UM}
    \end{subfigure}
        \begin{subfigure}{0.48\textwidth}
    \begin{tikzpicture}[
        zoomboxarray,
        zoomboxarray columns=2,
        zoomboxarray rows=2,
        %connect zoomboxes,
        zoombox paths/.append style={line width=0.75pt}
    ]
        \node [image node] { \includegraphics[trim=1cm 9cm 1cm 12cm, clip,width=0.48\textwidth]{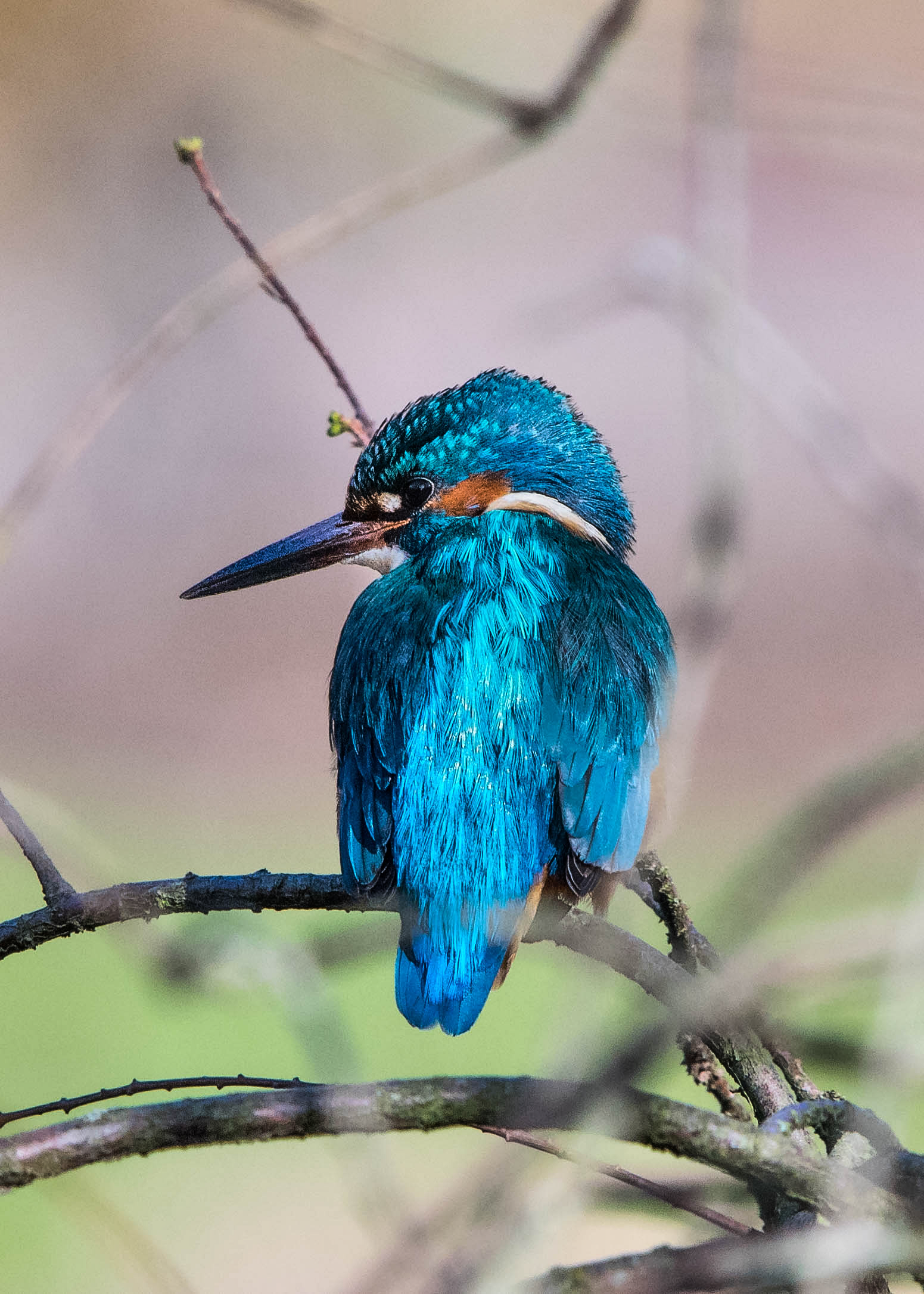} };
        \zoombox[color code = orange,magnification=4]{0.45,0.7} % background
        \zoombox[color code = green,magnification=5]{0.8,0.6} % chest
        \zoombox[color code = red,magnification=4]{0.25,0.25}% nest
        \zoombox[color code = blue,magnification=4]{0.4,0.5} % stick
    \end{tikzpicture}
        \caption{GUM}
        \label{fig:adaptive_enhancement_GUM}
    \end{subfigure}

        \begin{subfigure}{0.48\textwidth}
    \begin{tikzpicture}[
        zoomboxarray,
        zoomboxarray columns=2,
        zoomboxarray rows=2,
        %connect zoomboxes,
        zoombox paths/.append style={line width=0.75pt}
    ]
        \node [image node] { \includegraphics[trim=1cm 9cm 1cm 12cm, clip,width=0.48\textwidth]{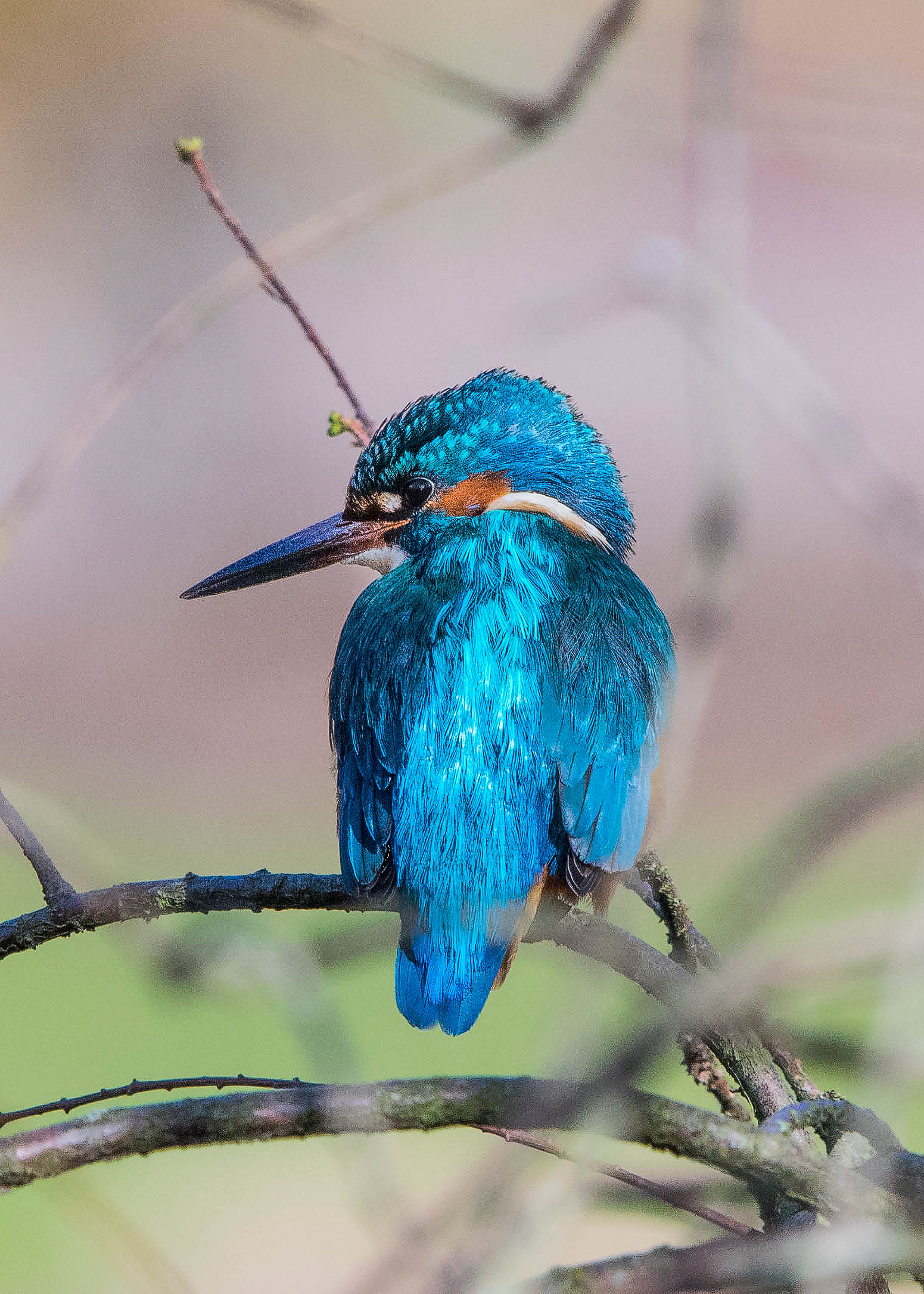} };
        \zoombox[color code = orange,magnification=4]{0.45,0.7} % background
        \zoombox[color code = green,magnification=5]{0.8,0.6} % chest
        \zoombox[color code = red,magnification=4]{0.25,0.25}% nest
        \zoombox[color code = blue,magnification=4]{0.4,0.5} % stick
    \end{tikzpicture}
        \caption{CAS}
        \label{fig:adaptive_enhancement_CAS}
    \end{subfigure}
        \begin{subfigure}{0.48\textwidth}
    \begin{tikzpicture}[
        zoomboxarray,
        zoomboxarray columns=2,
        zoomboxarray rows=2,
        %connect zoomboxes,
        zoombox paths/.append style={line width=0.75pt}
    ]
        \node [image node] { \includegraphics[trim=1cm 9cm 1cm 12cm, clip,width=0.48\textwidth]{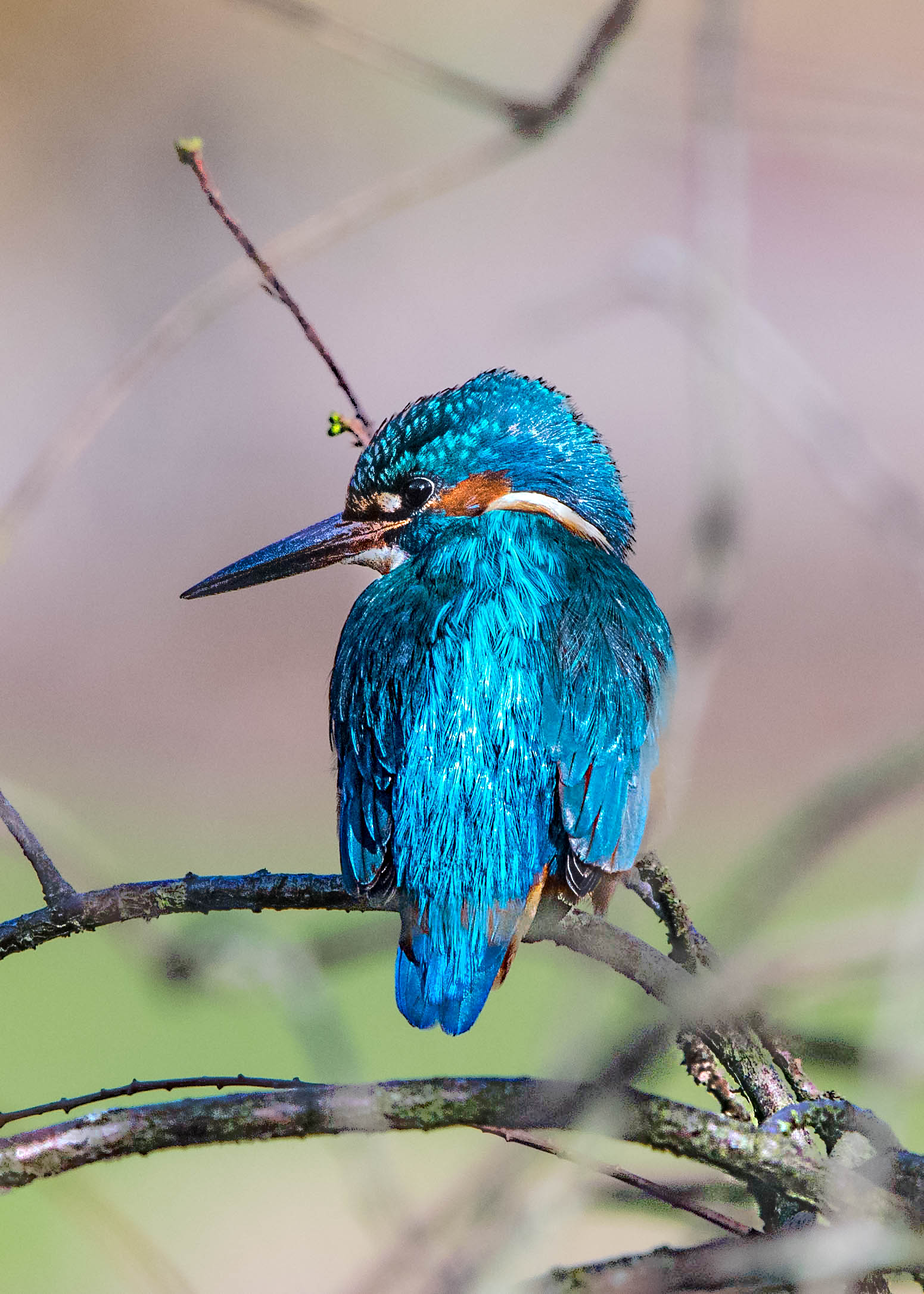} };
        \zoombox[color code = orange,magnification=4]{0.45,0.7} % background
        \zoombox[color code = green,magnification=5]{0.8,0.6} % chest
        \zoombox[color code = red,magnification=4]{0.25,0.25}% nest
        \zoombox[color code = blue,magnification=4]{0.4,0.5} % stick
    \end{tikzpicture}
        \caption{Proposed method}
        \label{fig:adaptive_enhancement_Ours}
    \end{subfigure}
    
\caption{Adaptive image enhancement comparison, a) Original image, b) Refined blur map (step size of 16 pixels), c) Gain map ($\alpha_1 = 30, \beta_1 = 0.15,\alpha_2 = 92, \beta_2 =0.4 $ and $\lambda_{max}=2$), d) UM algorithm $\lambda = 2$, e) GUM ($\kappa = 5\times10^{-4}, \lambda = 7$),  f) CAS ($\lambda = -0.125$), g) Proposed method ($\alpha_1 = 46, \beta_1 = 0.1,\alpha_2 = 183, \beta_2 =0.27 $ and $\lambda_{max}=2$).}
\label{fig:adaptive_enhancement}
\end{figure*}

In Fig. \ref{fig:adaptive_enhancement}, we present an example of obtaining the gain map through a non-linear transformation of the defocus map.  Fig. \ref{fig:adaptive_enhancement_input} shows the input image with different levels of blurriness. For example, the bird in the foreground is relatively sharp in comparison with the branches and the background which is heavily defocused. Enhancing this image is a real challenge. Fig.\ref{fig:adaptive_enhancement_bmap} and  Fig.\ref{fig:adaptive_enhancement_gain} show the defocus blur map and the gain map respectively. The gain map was calculated using the nonlinear transformation setting $\alpha_1 = 46, \beta_1 = 0.1,\alpha_2 = 183, \beta_2 =0.27$ and $\lambda_{max}=2$. We aim to adaptively enhance the image by preventing over-sharpening in focused regions and avoiding sharpening in blur sections. 

Fig.\ref{fig:adaptive_enhancement} also shows the comparison of our method with the classical unsharp masking (UM) and two newer methods contrast adaptive sharpening (CAS)\footnote{https://www.amd.com/en/technologies/radeon-software-fidelityfx} and Generalized Unsharp Masking (GUM) \cite{GUM_Dennis}.  We can see that applying UM with a fixed global gain $\lambda = 2$ produces an unpleasant sharpening in the background (green box) and over-sharpening of areas in focus such as the indicated by the orange, red and blur boxes. The CAS algorithm was implemented using a gain $\lambda = -0.125$. We can see in Fig. \ref{fig:adaptive_enhancement_CAS} that the algorithm performs very well in regions with medium and high defocus blur. However, it can not deal with the blur areas properly producing a noisy result(green and red boxes). The GUM algorithm (Fig. \ref{fig:adaptive_enhancement_GUM}) handles both background and foreground properly with a gain $\lambda = 7$ and a contrast factor $\kappa = 5\times10^{-4}$. But it produces strong halo artifacts when the gain increases. The result of applying the proposed method is shown in Fig. \ref{fig:adaptive_enhancement_Ours}. It does not sharpen the heavily blurred background (green box) and prevents over-sharpening of highly focused pixels by limiting the sharpening gain (orange, red and blue boxes).

\begin{table}[h!]
\centering
\caption{Image quality assessment of results in Fig.\ref{fig:adaptive_enhancement}. Best results are indicated in bold fonts.}
\label{tab:enhancement_metrics}
\begin{tabular}{llllll}
\toprule
           & Original & UM               & GUM         & CAS              & Proposed         \\
           \midrule
BRISQUE    & 20.83    & 44.41            & 24.89       & 55.19            & \textbf{20.61}   \\
IL-NIQE    & 26.72    & 25.331           & 26.0949     & 26.0814          & \textbf{24.4754} \\
NBIQA      & 66.244   & 66.4285          & 63.5012     & \textbf{76.4527} & 72.0607          \\
PSS        & 0.091    & 0.1161           & 0.0921      & 0.1407           & \textbf{0.0887}  \\
% BMPRI      & 24.1857  & \textbf{12.4072} & 18.9182     & 30.5073          & 26.0793          \\
BLIINDS-II & 42.5     & 40               & \textbf{37} & 52.5             & 43              \\
\bottomrule

\end{tabular}
\end{table}

 We perform the image quality assessment (IQA) of the images shown in Fig.\ref{fig:adaptive_enhancement} and present the results in Table \ref{tab:enhancement_metrics}. We use a series of no-reference (NR) IQA methods such as BRISQUE\cite{mittal2012no}, IL-NIQE\cite{zhang2015feature}, NBIQA\cite{ou2019novel}, PSS\cite{min2016blind} and BLIINDS-II\cite{saad2012blind} to objectively evaluate the performance of the proposed algorithm. These metrics take into account features from the spatial domain or transform domain to assess the quality and naturalness of an image. We should point out that these quality measurements use different scales. For example, while for the BRISQUE method a smaller score means a better image quality, for the NBIQA method a larger score means a better quality.    In Table \ref{tab:enhancement_metrics} we can see that our method achieves the best result on BRISQUE, IL-NIQUE and PSS, and second-best on NBIQA, This indicates that the proposed method produces a more natural-looking image enhancement than the other three methods. However, BLIINDS-II gives the second best score to UM algorithm leaving our method in 3rd position.  In fact, from Fig.\ref{fig:adaptive_enhancement} we can see that UM algorithm produces excessive halo artifacts due to enhancement in high contrast regions such as shown in the red box. The BLIINDS-II score for the UM result is thus debatable. Therefore, such quality metrics do not always generalize well as a full-reference metric. It is always advisable to perform a visual comparison when possible. Indeed, in image enhancement applications in which the user of the image is human, he/she will manipulate parameters of the algorithm to adjust the image until a satisfactory outcome is produced. This a very subjective process which cannot be replaced by a quality metric.

%% file: 03b_SDoF.tex
\begin{figure*}[h!]
\centering
\begin{subfigure}{0.24\linewidth}
    \includegraphics[width=1\linewidth]{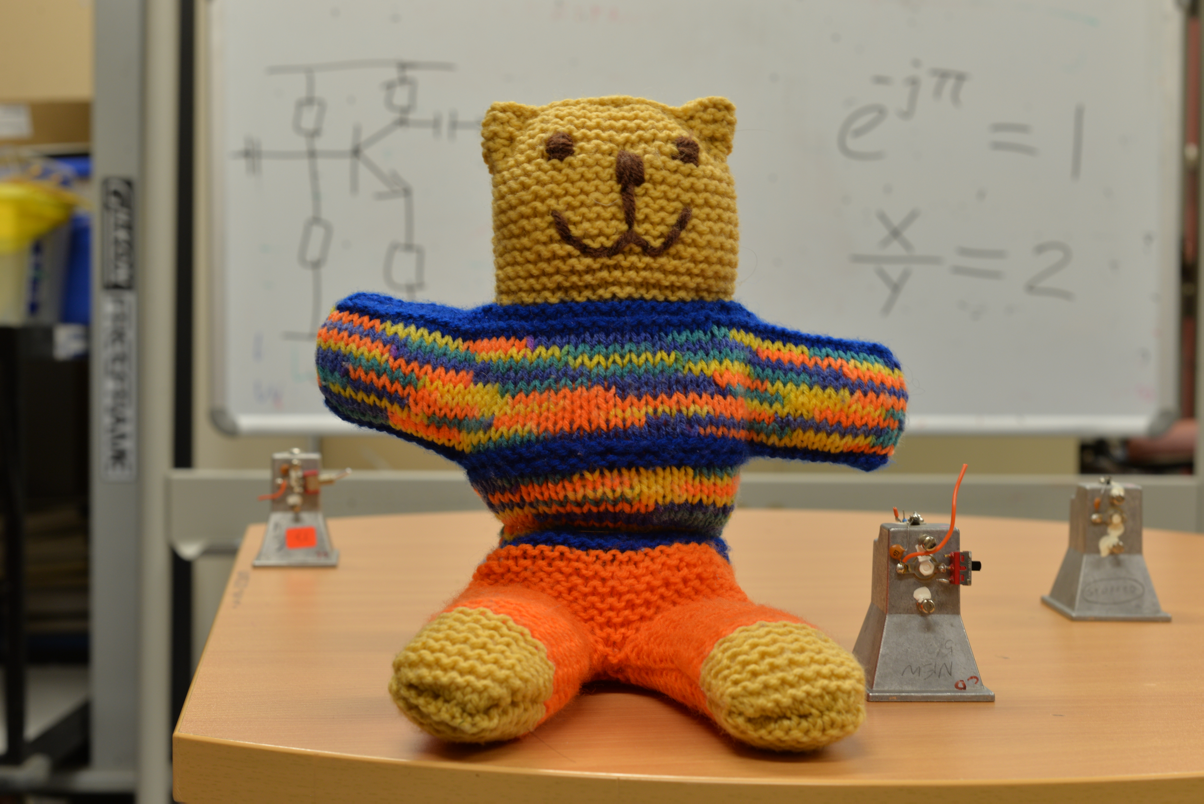}
\end{subfigure}
\begin{subfigure}{0.24\linewidth}
    \includegraphics[width=1\linewidth]{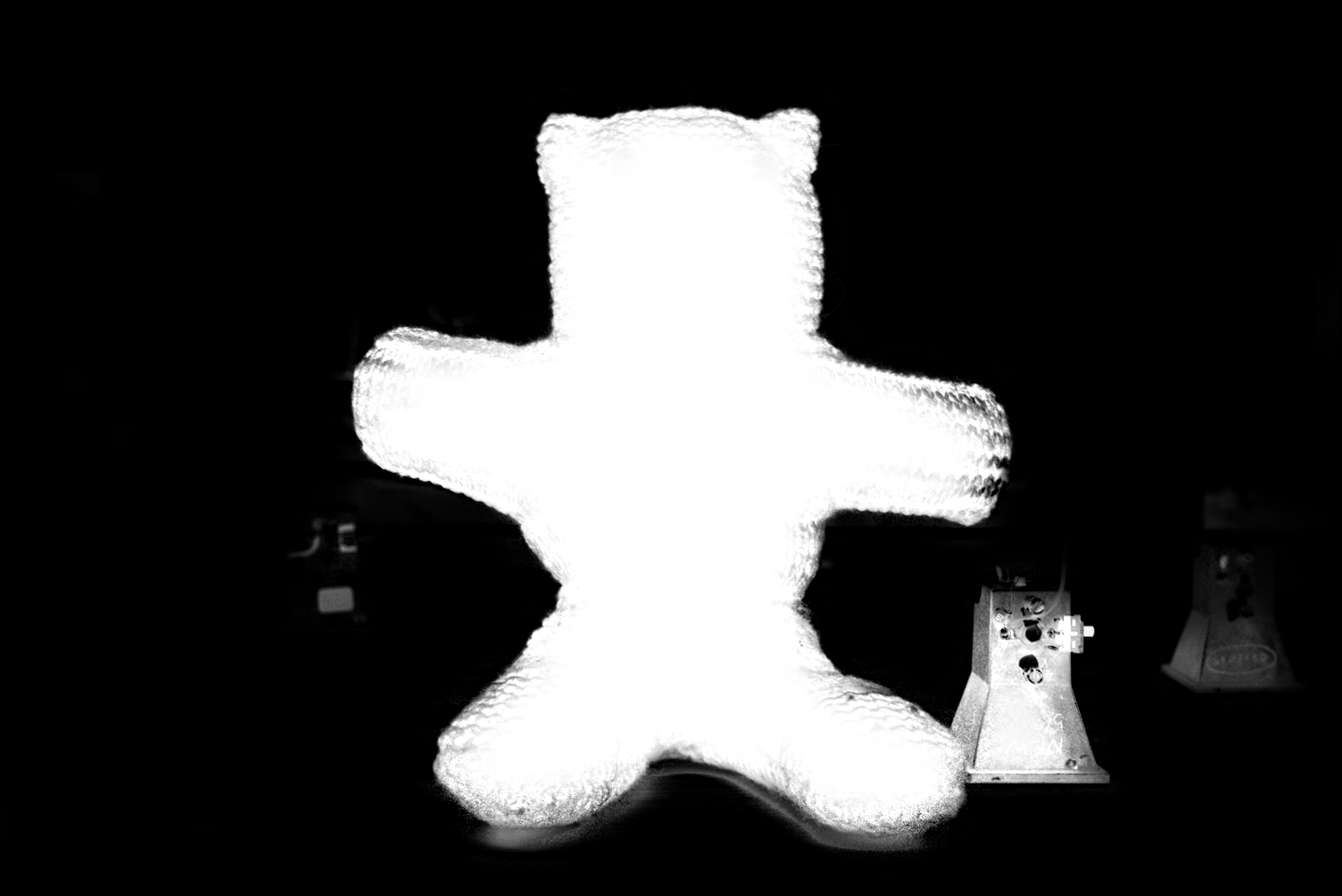}
\end{subfigure}
\begin{subfigure}{0.24\linewidth}
    \includegraphics[width=1\linewidth]{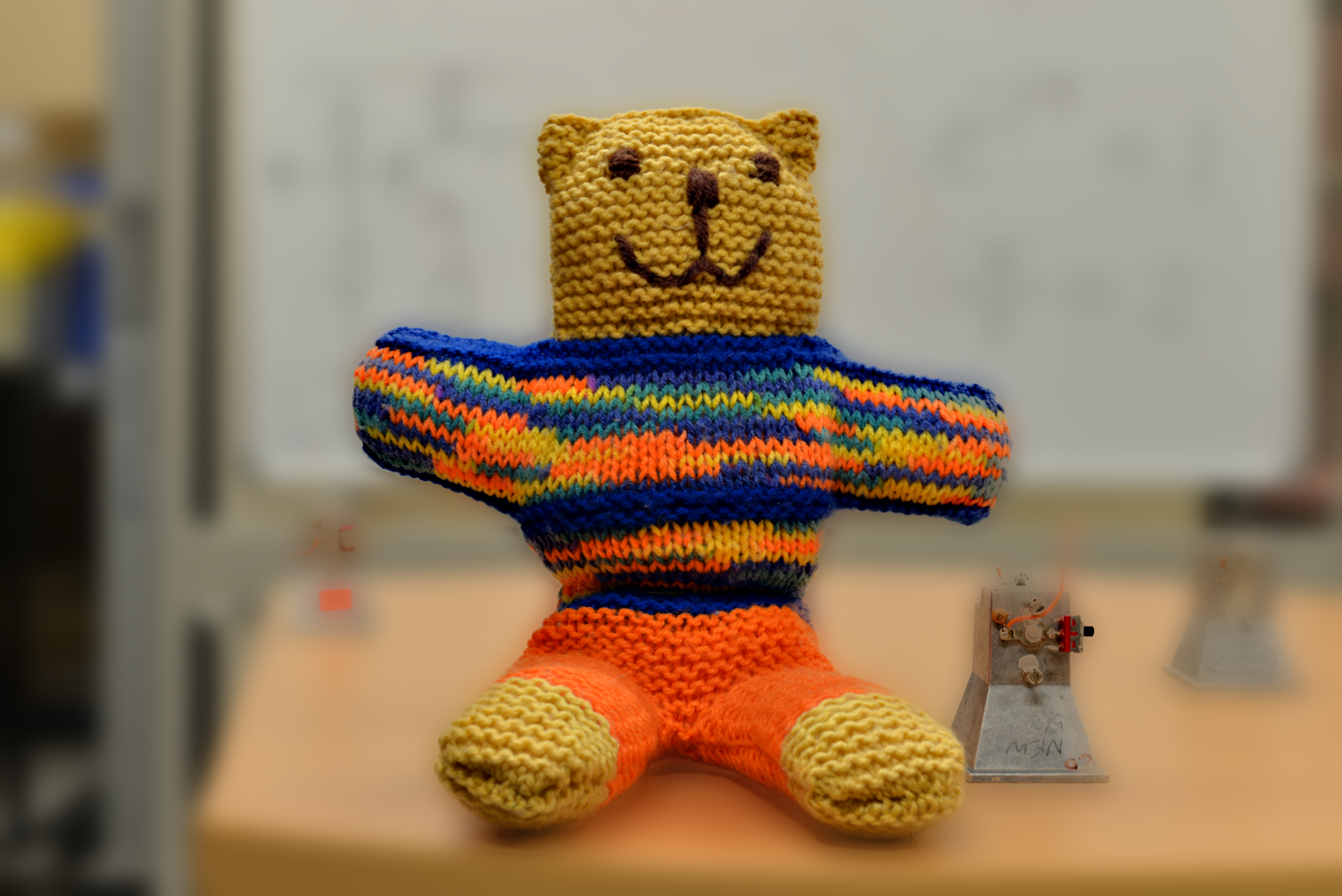}
\end{subfigure}
\begin{subfigure}{0.24\linewidth}
    \includegraphics[width=\columnwidth]{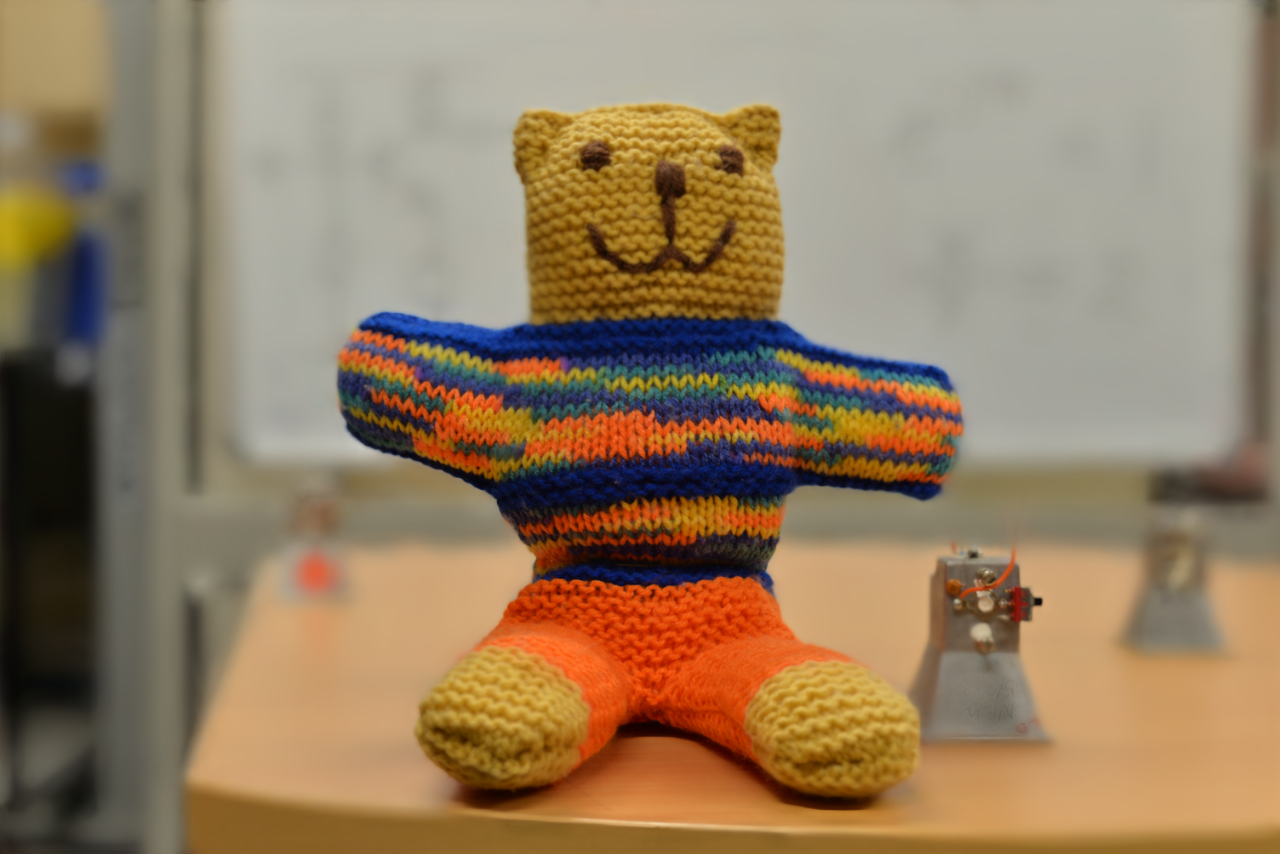}
\end{subfigure}

\begin{subfigure}{0.24\linewidth}
    \includegraphics[width=1\linewidth]{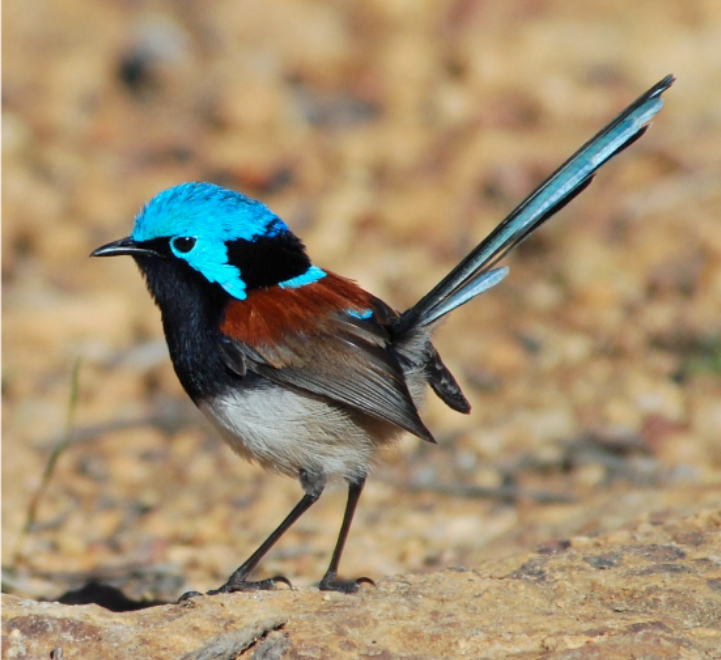}
\end{subfigure}
\begin{subfigure}{0.24\linewidth}
    \includegraphics[width=1\linewidth]{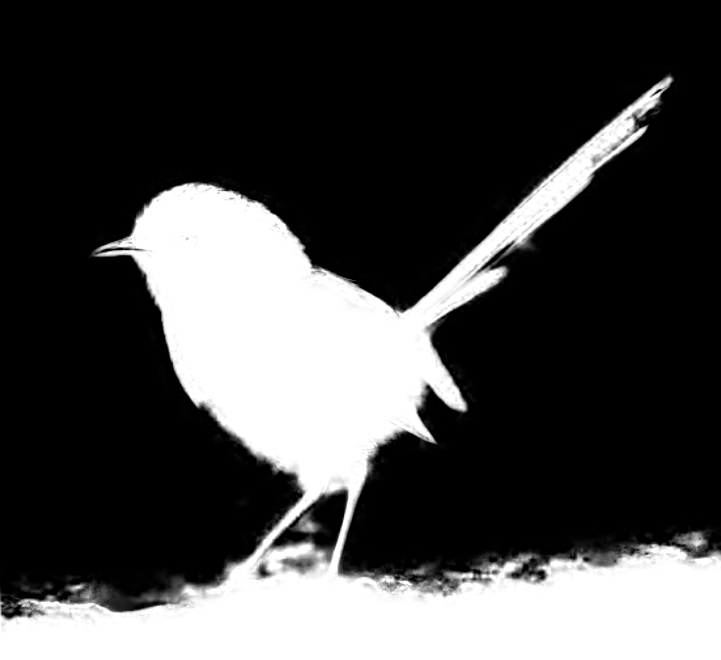}
\end{subfigure}
\begin{subfigure}{0.24\linewidth}
    \includegraphics[width=1\linewidth]{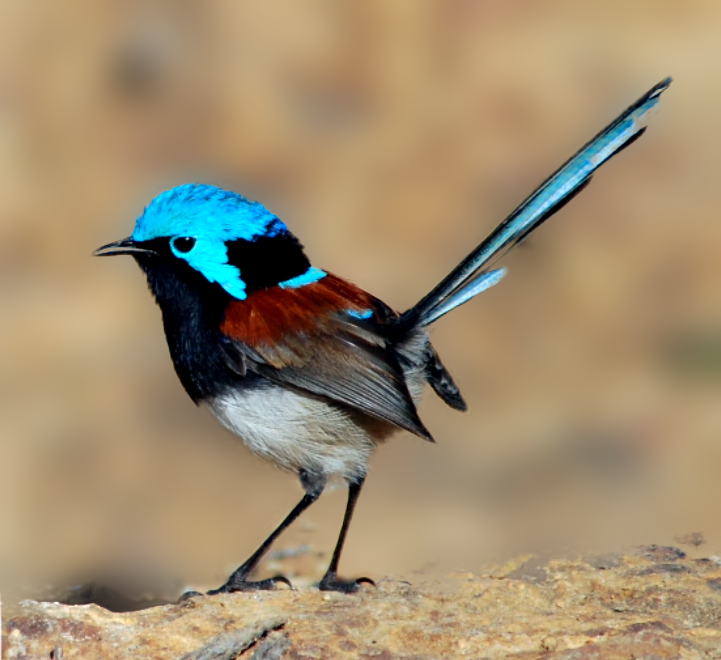}
\end{subfigure}
\begin{subfigure}{0.24\linewidth}
    \includegraphics[width=\columnwidth]{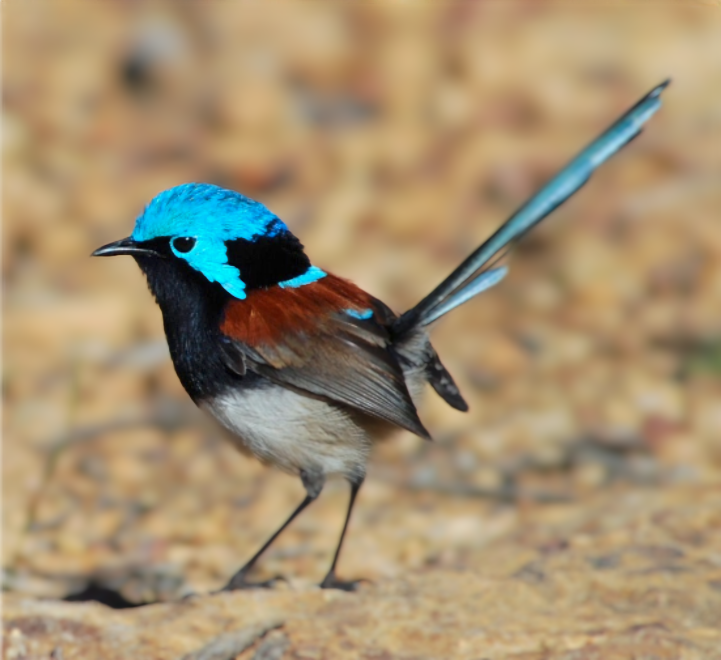}
\end{subfigure}

\begin{subfigure}{0.24\linewidth}
    \includegraphics[trim=0 0 0 80 cm,clip, width=1\linewidth]{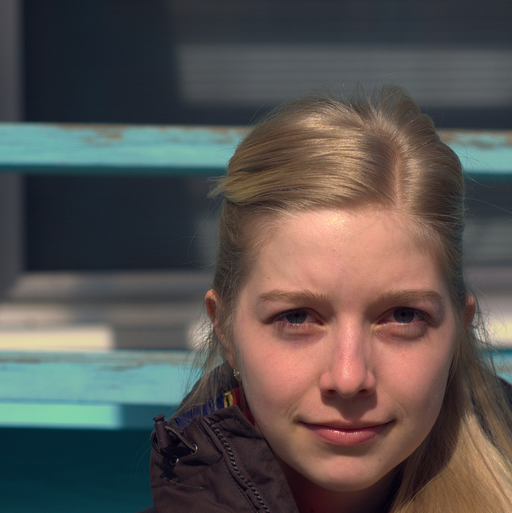}
        \caption{Input image}
\end{subfigure}
\begin{subfigure}{0.24\linewidth}
    \includegraphics[trim=0 0 0 80 cm,clip, width=1\linewidth]{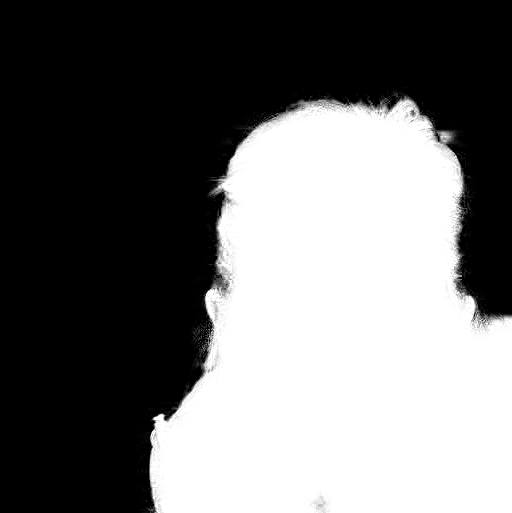}
        \caption{Weight map}
\end{subfigure}
\begin{subfigure}{0.24\linewidth}
    \includegraphics[trim=0 0 0 80 cm,clip, width=1\linewidth]{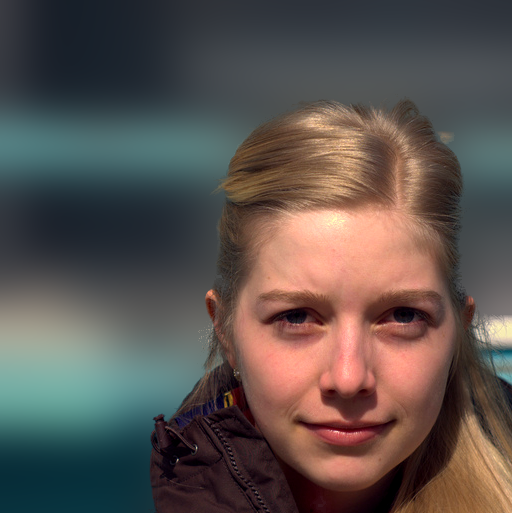}
        \caption{Proposed}
\end{subfigure}
\begin{subfigure}{0.24\linewidth}
    \includegraphics[trim=0 0 0 80 cm,clip, width=1\linewidth]{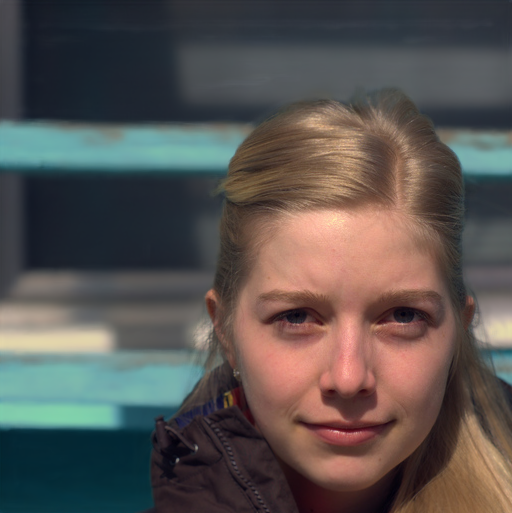}
        \caption{DeepLens \cite{wang2018deeplens}}

\end{subfigure}

% \begin{subfigure}{0.24\linewidth}
%     \includegraphics[width=\columnwidth]{images/defocus magnification/robert.png}
%     \caption{input image}
% \end{subfigure}
% \begin{subfigure}{0.24\linewidth}
%     \includegraphics[width=\columnwidth]{images/defocus magnification/robert_W_m0_40_m1_55_step_16.png}
%     \caption{Weight map}
% \end{subfigure}
% \begin{subfigure}{0.24\linewidth}
%     \includegraphics[width=\columnwidth]{images/defocus magnification/robert_SDoF_m0_40_m1_55_step_16.png}
%     \caption{Proposed}
% \end{subfigure}
% \begin{subfigure}{0.24\linewidth}
%     \includegraphics[width=\columnwidth]{images/defocus magnification/robert_sdof_deeplens.png}
%     \caption{DeepLens \cite{wang2018deeplens}}
% \end{subfigure}
\caption{SDoF effect example. a) Original image, b) Weight map, c) SDoF result, d) DeepLens result. }
\label{fig:SDoF_experiment1}
\end{figure*}

\subsection{Defocuss magnification}

In photography, the shallow depth-of-field (DoF) technique is used to make the main object stand out from the background by producing an image in which in-focus objects are with great details and contrast, while out-of-focus objects are greatly smoothed. In a camera, the DoF is controlled by a combination of three factors: the f-number, the focal length, and the distance of the camera from in-focus objects. Due to practical limitations such as a lens with a small diameter, it is not always possible to obtain the desired shallow DoF (SDoF) effect. Enhancement of an image by producing/increasing the SDoF effect is thus a practical image processing problem.

Computer vision algorithms were recently developed to achieve the shallow depth of field effect. For example, in DeepLens \cite{wang2018deeplens}, the depth information from a RGB-D sensor was used to train a CNN to create the SDoF effect from an image that has everything in focus. P. Sakurikar et al. \cite{sakurikar2019defocus} proposed an end-to-end GAN architecture that selectively increases the level of blurriness in the input image while keeping the edges of the in-focus object intact. Also, in \cite{google} an object segmentation based algorithm differentiates main objects from the those in the  background. The latter are blurred to create the SDoF effect.

A key idea in creating a SDoF effect from a single image is to perform a pixel adaptive weighted combination of a smoothed version denoted $B(p)$ and a sharpened version of the input image denoted $S(p)$, 
\begin{equation}
R(q)=W(q)S(q)+(1-W(q))B(q)\label{eq:sdof}
\end{equation}
We take this approach in this paper by using the defocus map to estimate the weights $W(q)$. The smoothed image $B$ is produced by using one of the edge-aware smoothing filters such as guided filter \cite{he2012guided} or one of its weighted versions \cite{li2015weighted, Sun2020}. The guided filter has two parameters: the radius of the neighbourhood $r$ and the smoothing parameter $\epsilon$. A bigger value of $r$ results in larger structures in the image being smoothed out, while a bigger value of $\epsilon$ leads to the overall greater degree of smoothness. To better control the smoothing result, we iterate the guided filter in a self-guidance mode. In our experiments presented in Fig. \ref{fig:SDoF_experiment1}, we use $r=33$, $\epsilon=128$, and 5 iterations. The sharpened image $S$ is produced by the classical unsharp masking algorithm with a fixed $\lambda=0.25$ for illustration purpose. 

The weight map ${W}$ is determined by the nonlinear mapping:
\begin{equation}
W(q)=1-\exp\left(-\left\lvert \frac{M(q)-10}{\sigma}\right\rvert ^{\gamma}\right)\label{eq:W}
\end{equation} 
where $\sigma$ and $\gamma$ are two parameters. 

The justification for the nonlinear transformation and the calculation of $\sigma$ and $\gamma$ can be explained as follows. When a pixel is within an in-focus area which corresponds to $M(q)\le c_{0}$ (e.g., $c_{0}=1$ is  user defined constant) the sharpened image should have greater weight such that the in-focus area is not blurred in the output image. Referring to equation (\ref{eq:sdof}), this requires that $W(q)=W_{0}\rightarrow1$. Similarly, when a pixel is within an out-of-focus area which corresponds to $M(q) \ge c_{1}$ (e.g., $c_{1}=7$ is also a user defined constant), the blurred image should have a greater weight such that in the output image the out-of-focus area is further blurred to create the SDoF effect. This requires $W(q)=W_{1}\rightarrow0$. With these considerations, the two parameters $\sigma$ and $\gamma$ can be determined. For example, we can set $W_{0}=0.999$ (for $M(q)=c_{0}$) and $W_{1}=0.001$ (for $M(q)=c_{1}$) and calculate the two parameters based on equation (\ref{eq:W}). In our experiments, the values of $c_{0}$ and $c_{1}$ were set depending on the image and the defocus levels present application.

Experimental results are shown in Fig. \ref{fig:SDoF_experiment1} which show that the proposed algorithm can effectively produce the SDoF effect. In the original images shown in Fig. \ref{fig:SDoF_experiment1}a the foreground is in focus while the background is slightly out of focus. After applying the non-linear transformation the weight map splits the image in two well-defined regions. Pixels in focus are with $W(q)\approx1$ while pixels out of focus are with $W(q)\approx0$ as shown in Fig. \ref{fig:SDoF_experiment1}b. The resulting SDoF image is shown in Fig. \ref{fig:SDoF_experiment1}c it is evident that the main subject remains sharp while the defocused background is further smoothed. The last column of this figure shows the results from using the DeepLens algorithm. We can see that the proposed algorithm is able to smooth the background to a greater degree.

%% file: 03c_Fusion.tex
\subsection{Multi-focus image fusion}

 Most of the lenses have an inherent reduced depth of field, so capturing an image where all the objects in the scene are properly focused is almost impossible. Multi-focus image fusion is a technique that combines two or more partially focused images to produce an image in which objects of interests are in focus. A common approach for multi-focus image fusion is through a weighted combination of partially in focus images where the weights define which pixels in the original image are in focus and which ones aren't. Multi focus image fusion is a widely studied topic and many algorithms have been proposed. A recent literature review  \cite{Liu2020a} divided the fusion algorithms into 4 main categories: transform domain, spatial domain, combined transform and spatial domains, and deep learning methods.

\begin{figure}[h!]
    \centering
    \includegraphics[width = 0.55\linewidth]{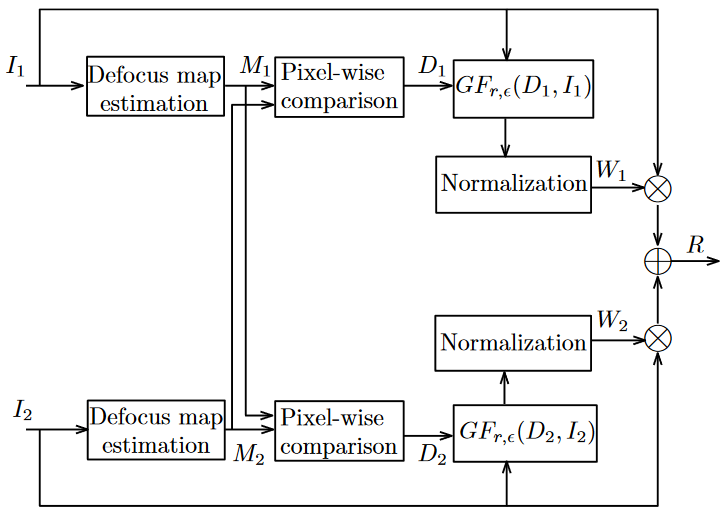}
    \caption{Multi focus image fusion block diagram.}
    \label{fig:fusion_block__diagram}
\end{figure}

In this paper, we propose a method to combine 2 or more images using the defocus map to recognize in-focus and out-of-focus regions on each input image. We generate a pixel-based weight map for each input image which allows us to blend them through a linear combination of the weighted images on a pixel level aiming to obtain an all in-focus image. Fig.\ref{fig:fusion_block__diagram} presents the block diagram for the proposed algorithm. The diagram only considers two input images for simplicity but the algorithm can handle multiple input images. The fusion of a set of $N$ images denoted $\{I_n\}_{n=1:N}$ can be described as:
\begin{equation}
    R(q)=\sum_{n=1}^{N} W_n(q)I_n(q) \label{eq:fusion}
\end{equation}
where the weights $W_n(q)$ are normalized such that $\sum_{n=1}^N W_n(q)=1$.

To determine the weight for each image  $I_{n}$, we first use the proposed CNN model to obtain blur map which is refined to produce $M_n$. We then generate the decision map as follows
\begin{equation}
        D_n(q) = \left\{\begin{matrix}
    0,  \quad \text{if} \quad M_n(q) \neq min(M_1(q), ... , M_n(q))
    \\ 
    1,  \quad \text{if} \quad M_n(q) = min(M_1(q), ... , M_n(q))
\end{matrix}\right.
\end{equation}
Thus the decision map $D_n(q)$ has a value of 1 only if the pixel of $I_{n}(q)$ presents the lowest level of blurriness among all $N$ input images. This map allows us to keep from each input image only those pixels that are most in focus.

To avoid halo artifacts around edges we use a guided filter \cite{he2012guided} to refine the edges of the decision map,

\begin{equation}
     \hat{W}_n = GF_{r, \epsilon}(D_n, I_n)
     \label{eq:refined_D_n}
\end{equation}
where we use $GF_{r,\epsilon}$ ($r,\epsilon$ are the two parameters) to denote the guided filtering process with $D_n$ being the image to be filtered and $I_n$ being the guidance image.  Since we applied the guided filter to each decision map, we must normalize each map to obtain the weights $W_n(q)$ as follows, 

\begin{equation}
    W_n(q) = \frac{\hat{W}_n(q)+\delta}{\sum_{n=1}^{N}(\hat{W}_n(q) + \delta)}
\end{equation}
$\delta$ represents a small positive regularization term to avoid the division by zero problem and to guaranty that there will be a contribution of each image to the result even when all the weights are zero. 

\begin{figure*}[h!]
\centering
\begin{subfigure}{0.16\linewidth}
    \includegraphics[width=\linewidth]{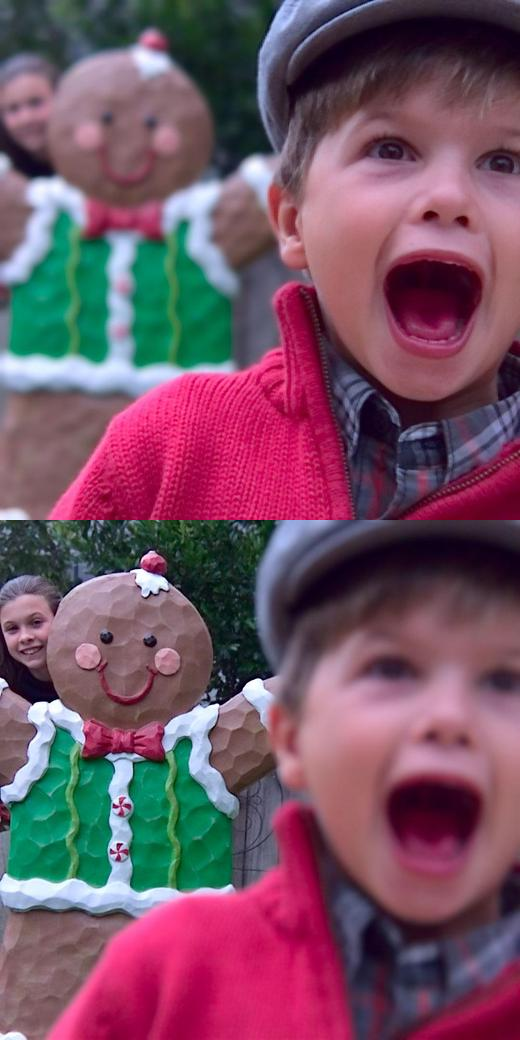}
    \caption{Inputs}
    \label{fig:MFF_weghts_inputs}
\end{subfigure}
\begin{subfigure}{0.16\linewidth}
    \includegraphics[width=\linewidth]{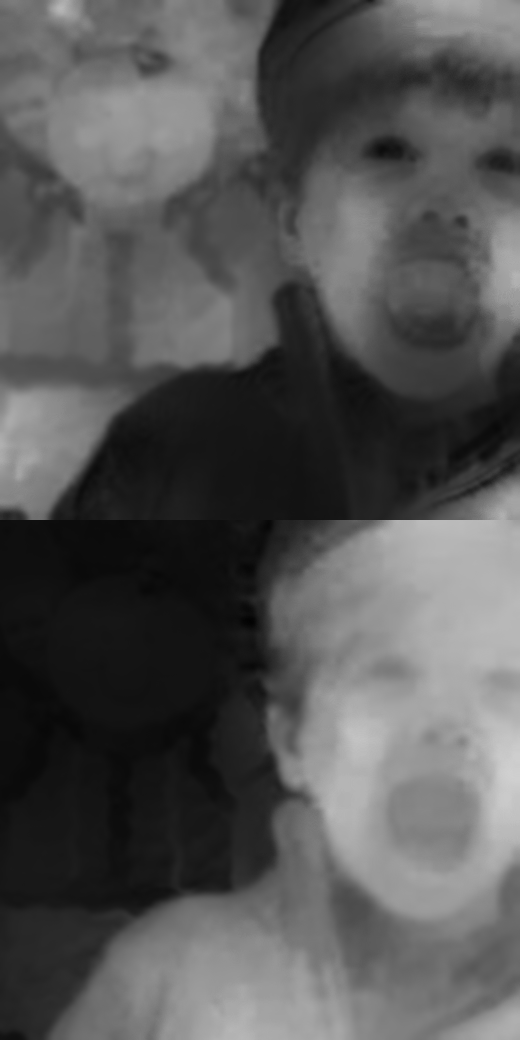}
    \caption{Defocus maps}
    \label{fig:MFF_weghts_dmap}
\end{subfigure}
\begin{subfigure}{0.16\linewidth}
    \includegraphics[width=\linewidth]{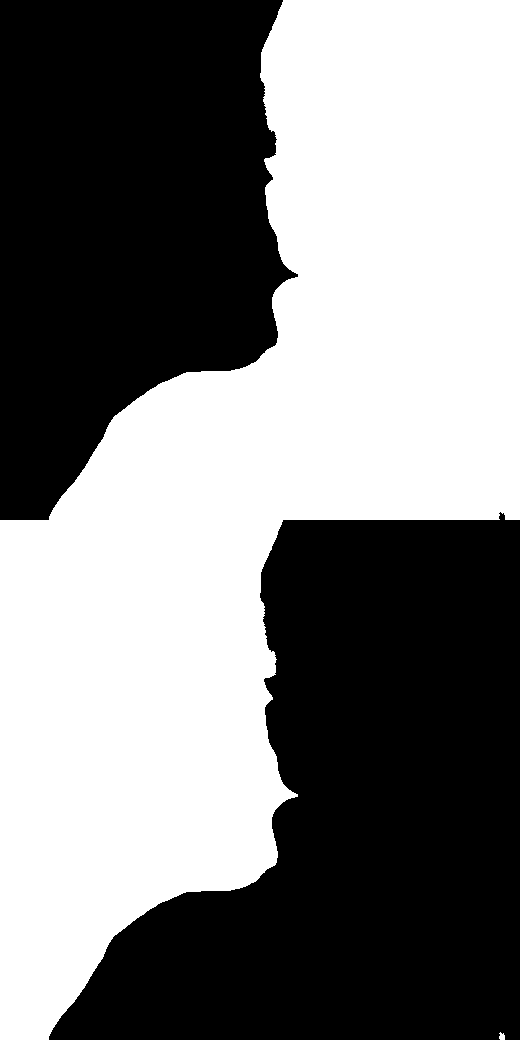}
    \caption{Decision maps}
    \label{fig:MFF_weghts_w}
\end{subfigure}
\begin{subfigure}{0.16\linewidth}
    \includegraphics[width=\linewidth]{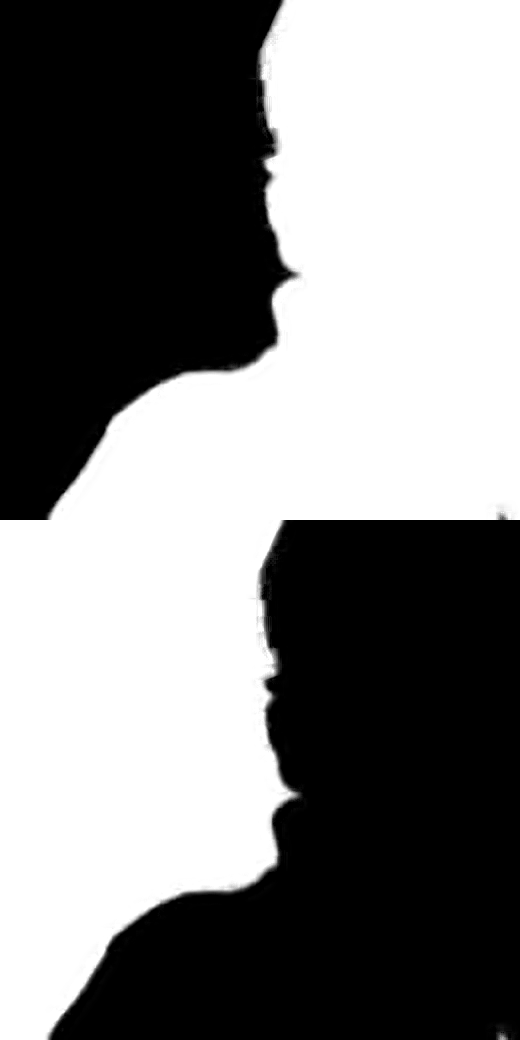}
    \caption{Weights}
    \label{fig:MFF_weghts_wr}
\end{subfigure}
\begin{subfigure}{0.32\linewidth}
    \includegraphics[width=\linewidth]{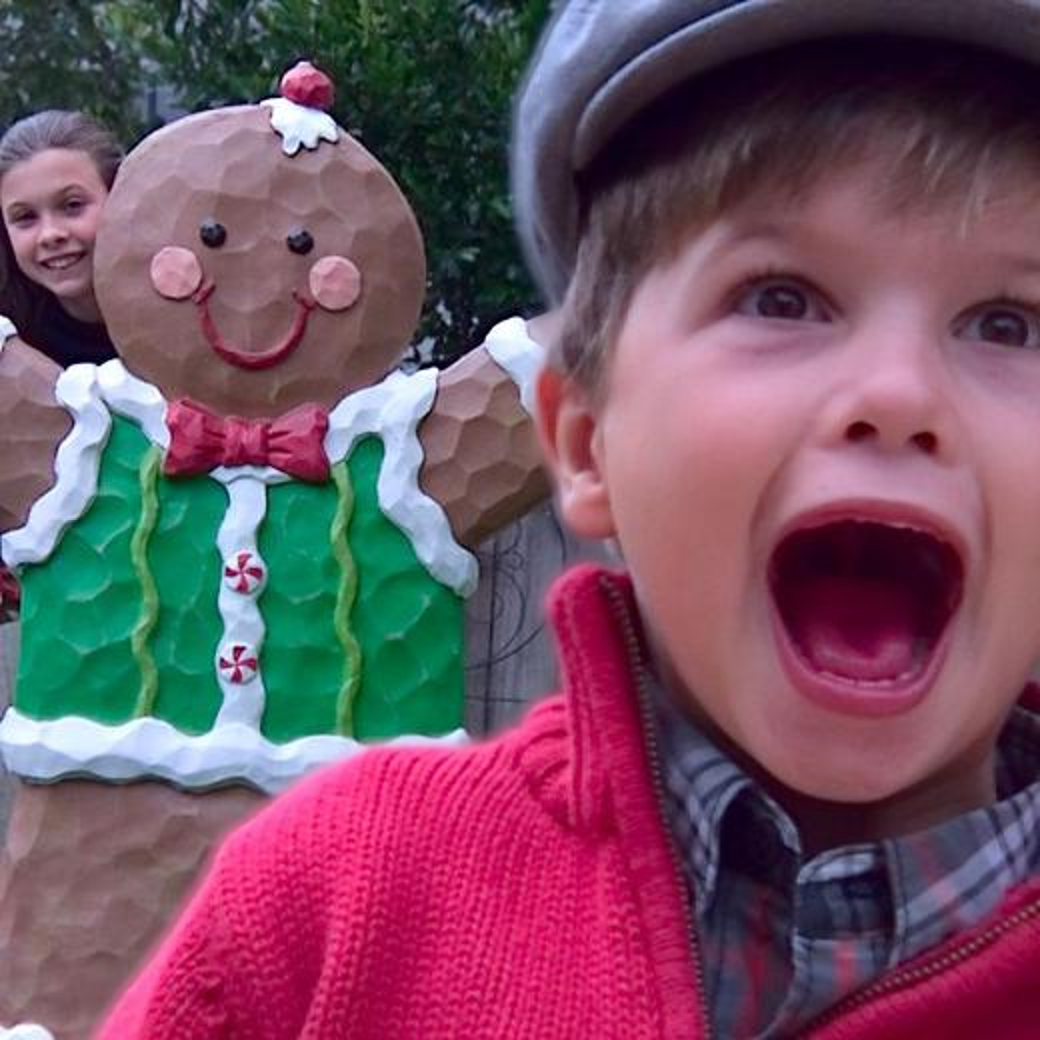}
    \caption{Fusion result}
    \label{fig:MFF_weghts_I}
\end{subfigure}
\caption{Multi focus image fusion. a) Input images, b) Defocus maps, c) Decision maps, d) Normalized weights ($r=7, \epsilon = 10^{-3} $), e) Fusion result.}
\label{fig:MFF_weghts}
\end{figure*}

To demonstrate the work-flow of the proposed method, we present the fusion process of two partially in-focus images in Fig.\ref{fig:MFF_weghts}. Input one has the foreground in focus while the background is out of focus. On the other hand, input two has the background in focus and the foreground out of focus. The defocus maps are presented in Fig. \ref{fig:MFF_weghts_dmap}. The resulting decision maps and the normalized weight maps are shown in Fig.\ref{fig:MFF_weghts_w} and \ref{fig:MFF_weghts}d, respectively. The fused result is shown in Fig.\ref{fig:MFF_weghts_I} which clearly shows that the resulting image has both foreground and background in focus. 

\begin{figure*}[t!]
\centering
\begin{subfigure}{0.13\linewidth}
        \begin{tikzpicture}[
        zoomboxarray,
        zoomboxarray columns=1,
        zoomboxarray rows=1,
        zoomboxes below,
        %connect zoomboxes,
        zoombox paths/.append style={line width=1pt}
    ]
        \node [image node] { \includegraphics[width=0.95\textwidth]{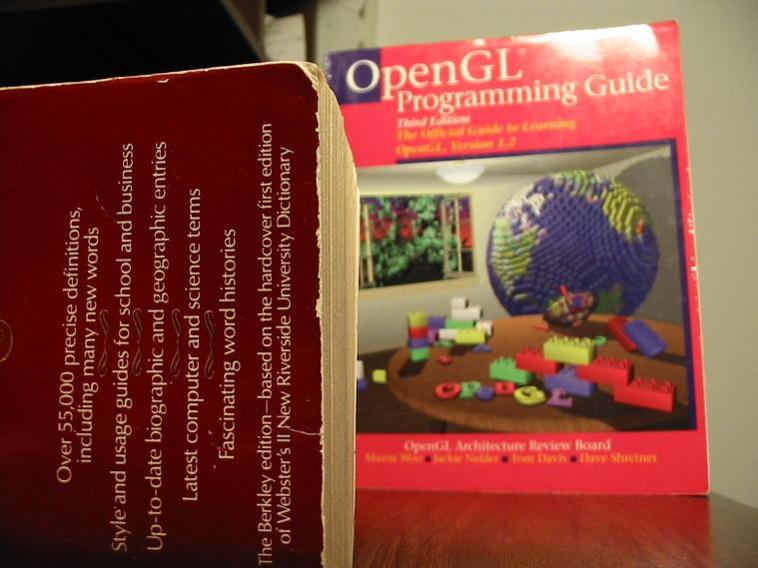} };
        \zoombox[color code = blue,magnification=4]{0.45,0.75} 

    \end{tikzpicture}
\end{subfigure}
\begin{subfigure}{0.13\linewidth}
    \begin{tikzpicture}[
        zoomboxarray,
        zoomboxarray columns=1,
        zoomboxarray rows=1,
        zoomboxes below,
        %connect zoomboxes,
        zoombox paths/.append style={line width=1pt}
    ]
        \node [image node] { \includegraphics[width=0.95\textwidth]{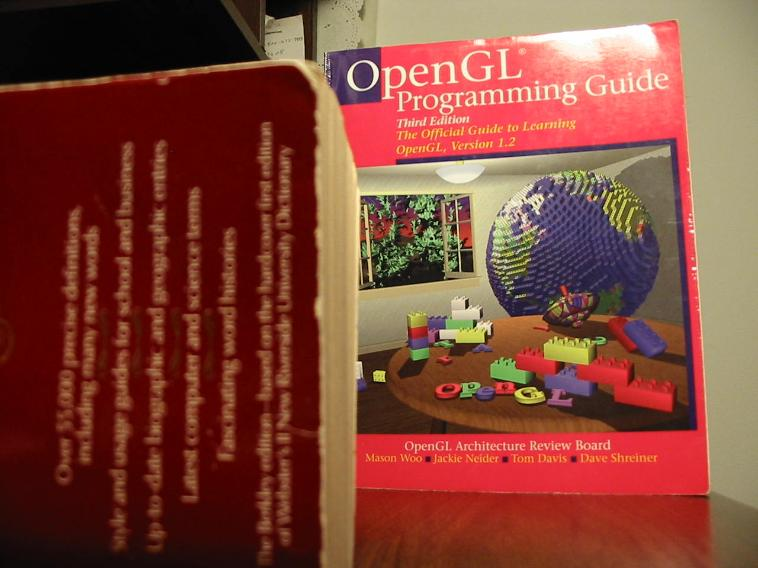} };
        \zoombox[color code = blue,magnification=4]{0.45,0.75} 
    \end{tikzpicture}
\end{subfigure}
\begin{subfigure}{0.13\linewidth}
        \begin{tikzpicture}[
        zoomboxarray,
        zoomboxarray columns=1,
        zoomboxarray rows=1,
        zoomboxes below,
        %connect zoomboxes,
        zoombox paths/.append style={line width=1pt}
    ]
        \node [image node] { \includegraphics[width=0.95\textwidth]{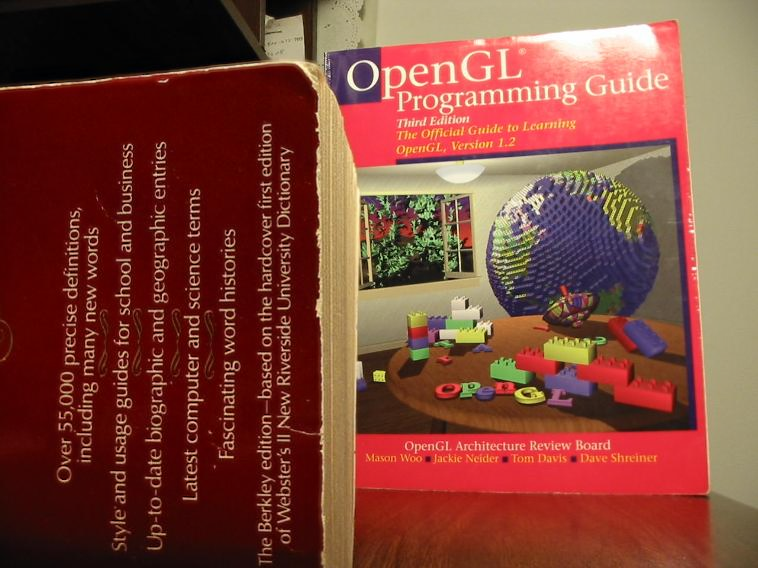} };
        \zoombox[color code = blue,magnification=4]{0.45,0.75} 
    \end{tikzpicture}
\end{subfigure}
\begin{subfigure}{0.13\linewidth}
        \begin{tikzpicture}[
        zoomboxarray,
        zoomboxarray columns=1,
        zoomboxarray rows=1,
        zoomboxes below,
        %connect zoomboxes,
        zoombox paths/.append style={line width=1pt}
    ]
        \node [image node] { \includegraphics[width=0.95\textwidth]{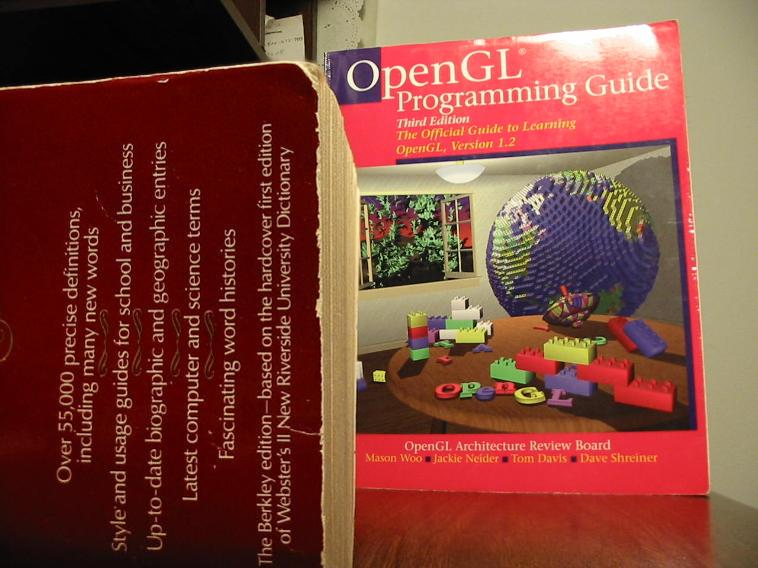} };
        \zoombox[color code = blue,magnification=4]{0.45,0.75} 
    \end{tikzpicture}
\end{subfigure}
\begin{subfigure}{0.13\linewidth}
        \begin{tikzpicture}[
        zoomboxarray,
        zoomboxarray columns=1,
        zoomboxarray rows=1,
        zoomboxes below,
        %connect zoomboxes,
        zoombox paths/.append style={line width=1pt}
    ]
        \node [image node] { \includegraphics[width=0.95\textwidth]{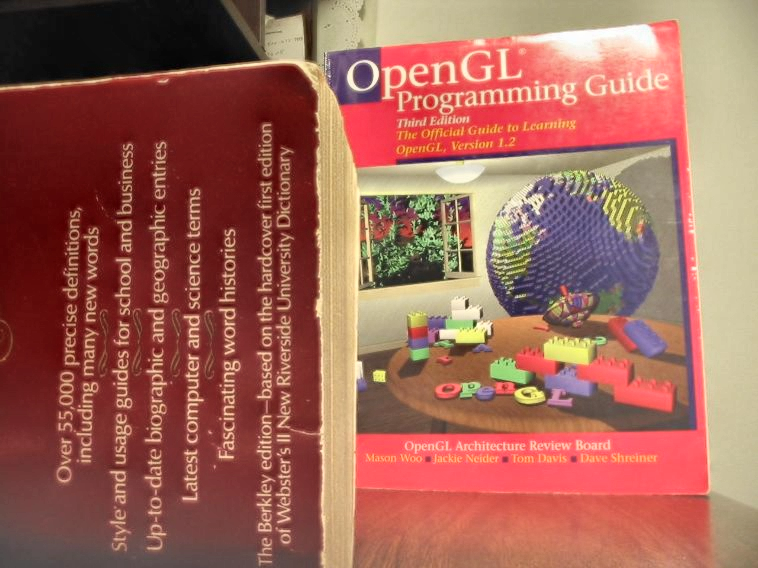} };
        \zoombox[color code = blue,magnification=4]{0.45,0.75} 
    \end{tikzpicture}
\end{subfigure}
\begin{subfigure}{0.13\linewidth}
        \begin{tikzpicture}[
        zoomboxarray,
        zoomboxarray columns=1,
        zoomboxarray rows=1,
        zoomboxes below,
        %connect zoomboxes,
        zoombox paths/.append style={line width=1pt}
    ]
        \node [image node] { \includegraphics[width=0.95\textwidth]{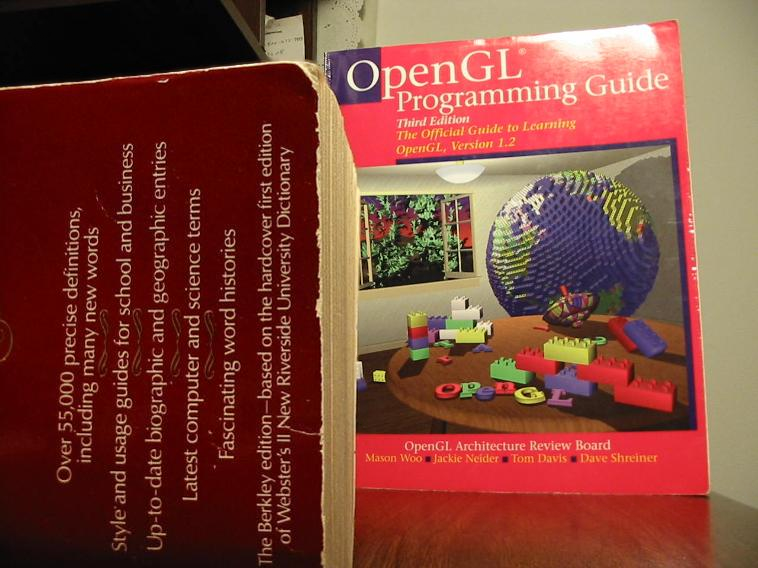} };
                \zoombox[color code = blue,magnification=4]{0.45,0.75} 
     \end{tikzpicture}
\end{subfigure}
\begin{subfigure}{0.13\linewidth}
        \begin{tikzpicture}[
        zoomboxarray,
        zoomboxarray columns=1,
        zoomboxarray rows=1,
        zoomboxes below,
        %connect zoomboxes,
        zoombox paths/.append style={line width=1pt}
    ]
        \node [image node] { \includegraphics[width=0.95\textwidth]{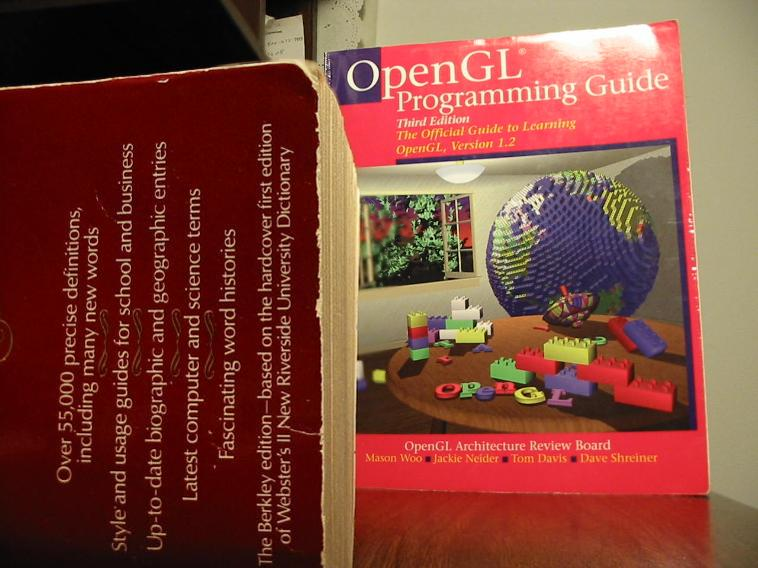} };
        \zoombox[color code = blue,magnification=4]{0.45,0.75} 
    \end{tikzpicture}
\end{subfigure}
\par\smallskip
\begin{subfigure}{0.13\linewidth}
        \begin{tikzpicture}[
        zoomboxarray,
        zoomboxarray columns=1,
        zoomboxarray rows=1,
        zoomboxes below,
        %connect zoomboxes,
        zoombox paths/.append style={line width=1pt}
    ]
        \node [image node] { \includegraphics[width=0.95\textwidth]{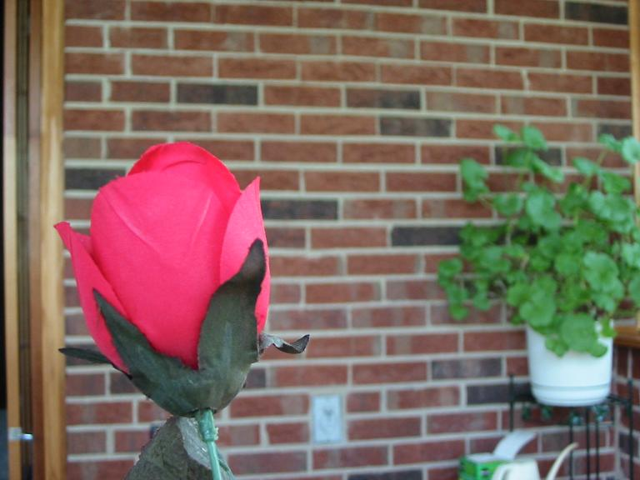} };
        \zoombox[color code = green,magnification=4]{0.35,0.15} 

    \end{tikzpicture}
    \caption{Input A}
\end{subfigure}
\begin{subfigure}{0.13\linewidth}
        \begin{tikzpicture}[
        zoomboxarray,
        zoomboxarray columns=1,
        zoomboxarray rows=1,
        zoomboxes below,
        %connect zoomboxes,
        zoombox paths/.append style={line width=1pt}
    ]
        \node [image node] { \includegraphics[width=0.95\textwidth]{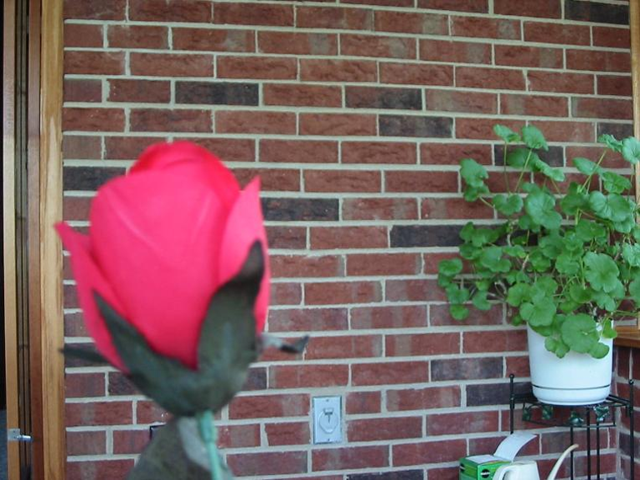} };
        \zoombox[color code = green,magnification=4]{0.35,0.15} 

    \end{tikzpicture}
    \caption{Input B}
\end{subfigure}
\begin{subfigure}{0.13\linewidth}
        \begin{tikzpicture}[
        zoomboxarray,
        zoomboxarray columns=1,
        zoomboxarray rows=1,
        zoomboxes below,
        %connect zoomboxes,
        zoombox paths/.append style={line width=1pt}
    ]
        \node [image node] { \includegraphics[width=0.95\textwidth]{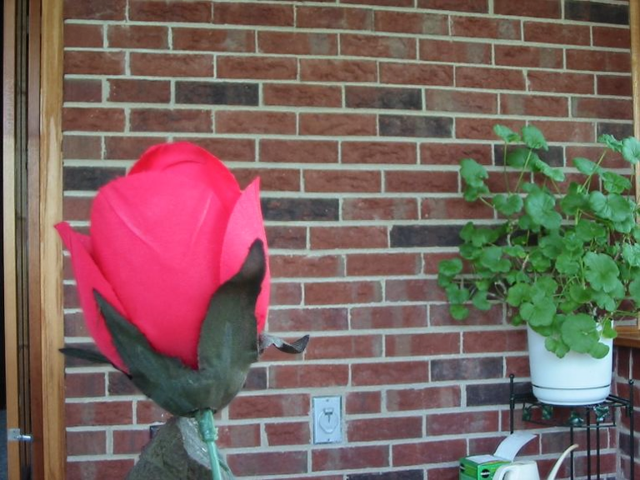} };
        \zoombox[color code = green,magnification=4]{0.35,0.15} 

    \end{tikzpicture}
    \caption{GFF \cite{li2013image}}
\end{subfigure}
\begin{subfigure}{0.13\linewidth}
        \begin{tikzpicture}[
        zoomboxarray,
        zoomboxarray columns=1,
        zoomboxarray rows=1,
        zoomboxes below,
        %connect zoomboxes,
        zoombox paths/.append style={line width=1pt}
    ]
        \node [image node] { \includegraphics[width=0.95\textwidth]{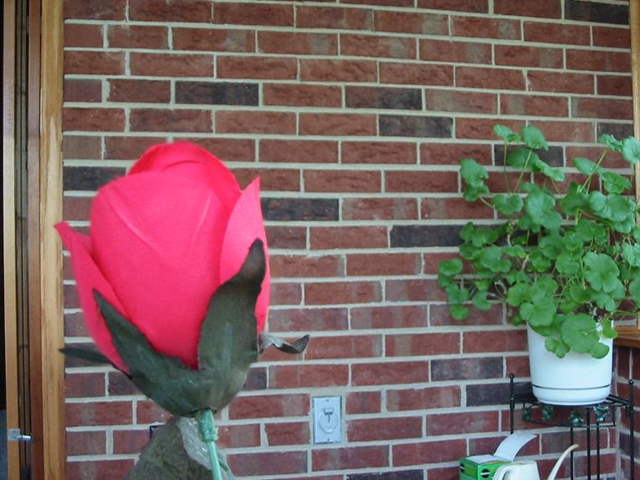} };
        \zoombox[color code = green,magnification=4]{0.35,0.15} 

    \end{tikzpicture}
    \caption{GFDF \cite{QIU201935GFDF}}
\end{subfigure}
\begin{subfigure}{0.13\linewidth}
        \begin{tikzpicture}[
        zoomboxarray,
        zoomboxarray columns=1,
        zoomboxarray rows=1,
        zoomboxes below,
        %connect zoomboxes,
        zoombox paths/.append style={line width=1pt}
    ]
        \node [image node] { \includegraphics[width=0.95\textwidth]{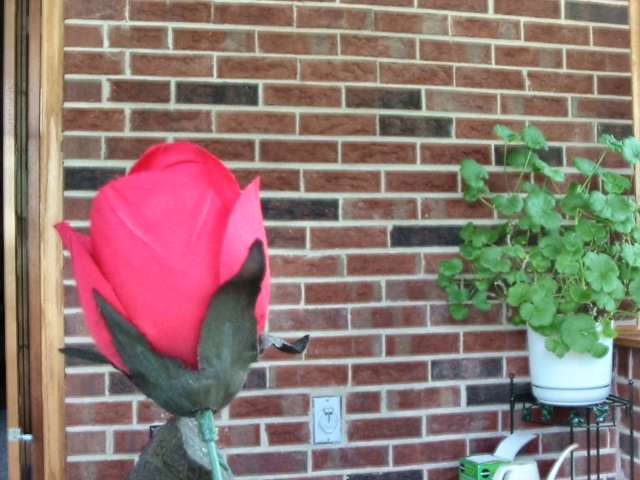} };
        \zoombox[color code = green,magnification=4]{0.35,0.15} 

    \end{tikzpicture}
    \caption{IFGD \cite{paul2016multIFGD}}
\end{subfigure}
\begin{subfigure}{0.13\linewidth}
        \begin{tikzpicture}[
        zoomboxarray,
        zoomboxarray columns=1,
        zoomboxarray rows=1,
        zoomboxes below,
        %connect zoomboxes,
        zoombox paths/.append style={line width=1pt}
    ]
        \node [image node] { \includegraphics[width=0.95\textwidth]{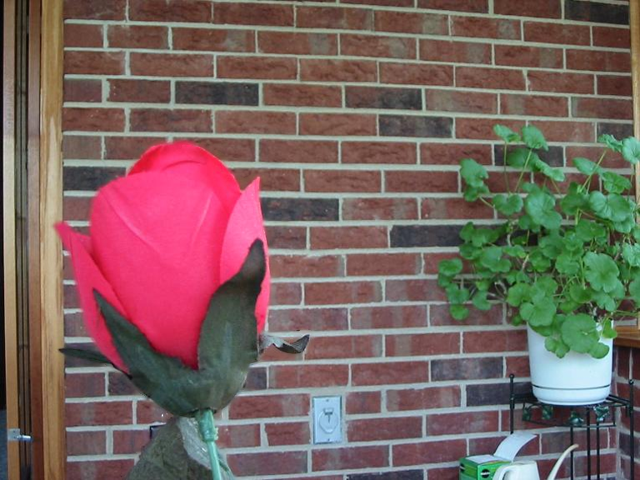} };
        \zoombox[color code = green,magnification=4]{0.35,0.15} 

    \end{tikzpicture}
            \caption{SESF\cite{ma2021sesf}}
\end{subfigure}
\begin{subfigure}{0.13\linewidth}
        \begin{tikzpicture}[
        zoomboxarray,
        zoomboxarray columns=1,
        zoomboxarray rows=1,
        zoomboxes below,
        %connect zoomboxes,
        zoombox paths/.append style={line width=1pt}
    ]
        \node [image node] { \includegraphics[width=0.95\textwidth]{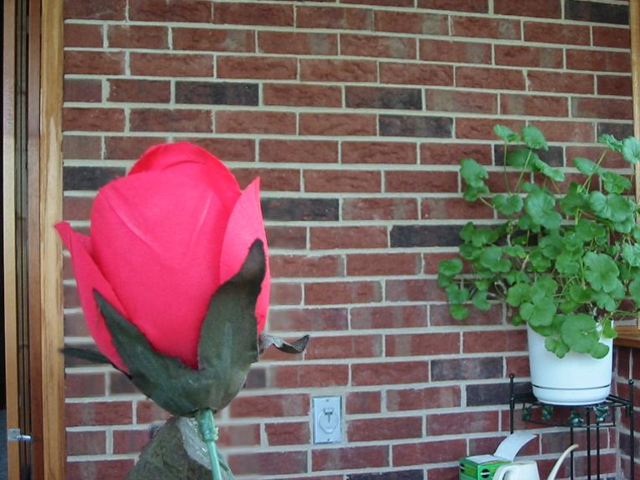} };
        \zoombox[color code = green,magnification=4]{0.35,0.15} 
    \end{tikzpicture}
            \caption{Proposed}
\end{subfigure}

\caption{Multi focus image fusion. a) Input image A, b) Input image B, 
c) GFF ($r_1=45, \epsilon_1=0.3, r_2=7, \epsilon_2 = 10^{-6} $),
d) GFDF($r =5, \epsilon = 0.3,w =7$), 
e) IFGD, 
f) SESF,
g) Ours ($step_{size} = 4, r=7, \epsilon = 10^{-3} $). }

\label{fig:mffbook}
\end{figure*}

Next, we present a comparison of the proposed method with four state-of-the-art methods: "Guided filter-based multi-focus image fusion through focus region detection" (GFDF) \cite{QIU201935GFDF}, "Image fusion with guided filtering" (GFF) \cite{li2013image}, "Multi-exposure and multi-focus image fusion in gradient domain" (IFGD) \cite{paul2016multIFGD} and a recent method called "SESF-Fuse: an unsupervised deep model for multi-focus image fusion" (SESF) \cite{ma2021sesf}. In Fig. \ref{fig:mffbook}  we present two fusion results, we applied these algorithms to two pairs of images from the Lytro dataset \cite{nejati2015multi_lytro}.

The first example is shown on the first row on Fig. \ref{fig:mffbook}, images on column a and b are the inputs to the fusion algorithm. Input A has the foreground in focus and input B has the background in focus. Both foreground and background contain text. The magnified boxes show that our algorithm fuses both images successfully. The resulting image,  shown in column f contains sharp text in both foreground and background. The result is comparable with GFF, GFDF and SESF in terms of sharpness and edge definition. Our method has better contrast than IFGD producing a more pleasant fusion effect and sharper text. A similar outcome can be seen in the second row on Fig.\ref{fig:mffbook}, this example uses two input images with a small defocus blur difference between them (columns a and b), we can see in column g that our algorithm can still fuse the images and obtain an all in focus result. The proposed method produces a similar result as GFF, GFDF and SESF but outperforms IFGD. 
Quantitative measurement is a complicated task due to the absence of ground truth. We use a set of standard quality metrics that have been used to evaluate the performance of the fusion methods: QG \cite{xydeas2000objectiveQG} is a gradient-based method that measures the edge information transferred from the input images to the result, QP \cite{zhao2007performanceQP} uses image phase congruence to measure the quality of the fused image, QY \cite{yang2008novelQy} employs the structural similarity metric (SSIM) \cite{Wang2004} between the inputs and fused image using local windows to evaluate at which level the fusion result contains all the complementary information from the inputs, QCB \cite{chen2009newQCB} considers contrast and saliency to evaluate the fusion result, and finally QFMI \cite{haghighat2014fastFMI} uses the mutual information to calculate the amount of information transferred to the fused image. Table \ref{tab:qualityMFF} shows the result of the objective assessment. The proposed method has high scores as GFDF and SESF but it is clearly better than  GFF and IFGD in these 3 images.

\input{table}

%% file: table.tex
\begin{table*}[h!]
\caption{Multi focus image fusion quality measurement. A bigger value indicates a better quality.}
\centering
\begin{tabular}{l l l l l l l }
\toprule
\textbf{Image}  \cite{nejati2015multi_lytro}                 & \textbf{Fusion Method} & \textbf{QG} \cite{xydeas2000objectiveQG}& \textbf{QP} \cite{zhao2007performanceQP} & \textbf{QY} \cite{yang2008novelQy}& \textbf{QCB} \cite{chen2009newQCB}& \textbf{QFMI}\cite{haghighat2014fastFMI} \\ 
\midrule
\multirow{4}{*}{\textbf{Cookie}} & \textbf{GFF}           & 0.68865     & 0.85206     & 0.96305     & 0.73283      & 0.50355       \\ 
                                 & \textbf{GFDF}          & 0.71176     & 0.86355     & 0.98848     & 0.79449      & 0.5963        \\ 
                                 & \textbf{GDIF}          & 0.63084     & 0.76759     & 0.85522     & 0.56627      & 0.40906       \\  
                                    & \textbf{SESF}                &  0.70739    &0.85783    &0.98619    &0.79372                &   0.58421      \\  
                                 & \textbf{Ours}          & 0.71277     & 0.85668     & 0.98892     & 0.79378      & 0.59566       \\ \midrule
\multirow{4}{*}{\textbf{Book}}   & \textbf{GFF}           & 0.61332     & 0.87643     & 0.90782     & 0.74186      & 0.51189       \\
                                 & \textbf{GFDF}          & 0.67992     & 0.8935      & 0.98554     & 0.80577      & 0.63095       \\  
                                 & \textbf{GDIF}          & 0.51231     & 0.76422     & 0.70457     & 0.53092      & 0.40471       \\  
                                 & \textbf{SESF}          & 0.67548    &0.89137    &0.98218    &0.80017    &0.62573       \\  
                                 & \textbf{Ours}          & 0.67762     & 0.89035     & 0.98691     & 0.80535      & 0.63139       \\ \midrule
\multirow{4}{*}{\textbf{Flower}} & \textbf{GFF}           & 0.65165     & 0.79573     & 0.96029     & 0.79185      & 0.53371       \\ 
                                 & \textbf{GFDF}          & 0.68329     & 0.79318     & 0.98302     & 0.83116      & 0.61103       \\  
                                 & \textbf{GDIF}          & 0.6425      & 0.76802     & 0.90415     & 0.6994       & 0.44966       \\  
                                 & \textbf{SESF}          & 0.68237    &0.79008   & 0.98262    &0.82927    &0.60822       \\  
                                 & \textbf{Ours}          & 0.68101     & 0.79275     & 0.98172     & 0.82869      & 0.60643       \\ \bottomrule
\end{tabular}

\label{tab:qualityMFF}
\end{table*}

%% file: main.bbl
\begin{thebibliography}{10}
\providecommand{\url}[1]{#1}
\csname url@samestyle\endcsname
\providecommand{\newblock}{\relax}
\providecommand{\bibinfo}[2]{#2}
\providecommand{\BIBentrySTDinterwordspacing}{\spaceskip=0pt\relax}
\providecommand{\BIBentryALTinterwordstretchfactor}{4}
\providecommand{\BIBentryALTinterwordspacing}{\spaceskip=\fontdimen2\font plus
\BIBentryALTinterwordstretchfactor\fontdimen3\font minus
  \fontdimen4\font\relax}
\providecommand{\BIBforeignlanguage}[2]{{%
\expandafter\ifx\csname l@#1\endcsname\relax
\typeout{** WARNING: IEEEtran.bst: No hyphenation pattern has been}%
\typeout{** loaded for the language `#1'. Using the pattern for}%
\typeout{** the default language instead.}%
\else
\language=\csname l@#1\endcsname
\fi
#2}}
\providecommand{\BIBdecl}{\relax}
\BIBdecl

\bibitem{Pentland1987}
A.~P. Pentland, ``A new sense for depth of field,'' \emph{IEEE Trans. Pattern
  Anal. Mach. Intell.}, no.~4, pp. 523--531, 1987.

\bibitem{Mather1996}
G.~Mather, ``Image blur as a pictorial depth cue,'' \emph{Proceedings of the
  Royal Society of London. Series B: Biological Sciences}, vol. 263, no. 1367,
  pp. 169--172, 1996.

\bibitem{xue2014novel}
F.~Xue and T.~Blu, ``A novel sure-based criterion for parametric psf
  estimation,'' \emph{IEEE Trans. Image Process.}, vol.~24, no.~2, pp.
  595--607, 2014.

\bibitem{chan2011single}
S.~H. Chan and T.~Q. Nguyen, ``Single image spatially variant out-of-focus blur
  removal,'' in \emph{Proc. International Conference on Image
  Processing}.\hskip 1em plus 0.5em minus 0.4em\relax IEEE, 2011, pp. 677--680.

\bibitem{cheong2015fast}
H.~Cheong, E.~Chae, E.~Lee, G.~Jo, and J.~Paik, ``Fast image restoration for
  spatially varying defocus blur of imaging sensor,'' \emph{Sensors}, vol.~15,
  no.~1, pp. 880--898, 2015.

\bibitem{Bae2007}
S.~Bae and F.~Durand, ``Defocus magnification,'' in \emph{Computer graphics
  forum}, vol.~26, no.~3.\hskip 1em plus 0.5em minus 0.4em\relax Wiley Online
  Library, 2007, pp. 571--579.

\bibitem{van2007robust}
I.~van Zyl~Marais and W.~H. Steyn, ``Robust defocus blur identification in the
  context of blind image quality assessment,'' \emph{Signal Processing: Image
  Communication}, vol.~22, no.~10, pp. 833--844, 2007.

\bibitem{al2015image}
F.~E. Al-Obaidi, ``Image quality assessment for defocused blur images,''
  \emph{American Journal of Signal Processing}, vol.~5, no.~3, pp. 51--55,
  2015.

\bibitem{Nasonov2019}
A.~Nasonov, A.~Krylov, and D.~Lyukov, ``{Image sharpening with blur map
  estimation using convolutional neural network},'' \emph{International
  Archives of the Photogrammetry, Remote Sensing and Spatial Information
  Sciences - ISPRS Archives}, vol.~42, no. 2/W12, pp. 161--166, 2019.

\bibitem{ye2018blurriness}
W.~Ye and K.-K. Ma, ``Blurriness-guided unsharp masking,'' \emph{IEEE Trans.
  Image Process.}, vol.~27, no.~9, pp. 4465--4477, 2018.

\bibitem{ziou2001depth}
D.~Ziou and F.~Deschenes, ``Depth from defocus estimation in spatial domain,''
  \emph{Computer vision and image understanding}, vol.~81, no.~2, pp. 143--165,
  2001.

\bibitem{lin2013absolute}
J.~Lin, X.~Ji, W.~Xu, and Q.~Dai, ``Absolute depth estimation from a single
  defocused image,'' \emph{IEEE Trans. Image Process.}, vol.~22, no.~11, pp.
  4545--4550, 2013.

\bibitem{shi2015break}
J.~Shi, X.~Tao, L.~Xu, and J.~Jia, ``Break ames room illusion: depth from
  general single images,'' \emph{ACM Trans. Graph.}, vol.~34, no.~6, pp. 1--11,
  2015.

\bibitem{Tang_2017_CVPR}
H.~Tang, S.~Cohen, B.~Price, S.~Schiller, and K.~N. Kutulakos, ``Depth from
  defocus in the wild,'' in \emph{Proc. IEEE Conf. CVPR}, 2017, pp. 2740--2748.

\bibitem{Levin2007}
A.~Levin, R.~Fergus, F.~Durand, and W.~T. Freeman, ``Image and depth from a
  conventional camera with a coded aperture,'' \emph{ACM Trans. Graph.},
  vol.~26, no.~3, pp. 70--es, 2007.

\bibitem{Vu2014}
D.~T. Vu, B.~Chidester, H.~Yang, M.~N. Do, and J.~Lu, ``Efficient hybrid
  tree-based stereo matching with applications to postcapture image
  refocusing,'' \emph{IEEE Trans. Image Process.}, vol.~23, no.~8, pp.
  3428--3442, 2014.

\bibitem{Zhou2009}
C.~Zhou, S.~Lin, and S.~K. Nayar, ``Coded aperture pairs for depth from defocus
  and defocus deblurring,'' \emph{International Journal of Computer Vision},
  vol.~93, no.~1, pp. 53--72, 2011.

\bibitem{tai2009single}
Y.-W. Tai and M.~S. Brown, ``Single image defocus map estimation using local
  contrast prior,'' in \emph{Proc. Int. Conf. Image Process., ICIP}.\hskip 1em
  plus 0.5em minus 0.4em\relax IEEE, 2009, pp. 1797--1800.

\bibitem{Zhang2016}
X.~Zhang, R.~Wang, X.~Jiang, W.~Wang, and W.~Gao, ``Spatially variant defocus
  blur map estimation and deblurring from a single image,'' \emph{Journal of
  Visual Communication and Image Representation}, vol.~35, pp. 257--264, 2016.

\bibitem{zhuo2011defocus}
S.~Zhuo and T.~Sim, ``Defocus map estimation from a single image,''
  \emph{Pattern Recognition}, vol.~44, no.~9, pp. 1852--1858, 2011.

\bibitem{Hu2006}
H.~Hu and G.~De~Haan, ``Low cost robust blur estimator,'' in \emph{Proc.
  International Conference on Image Processing}.\hskip 1em plus 0.5em minus
  0.4em\relax IEEE, 2006, pp. 617--620.

\bibitem{levin2007closed}
A.~Levin, D.~Lischinski, and Y.~Weiss, ``A closed-form solution to natural
  image matting,'' \emph{IEEE Trans. Pattern Anal. Mach. Intell.}, vol.~30,
  no.~2, pp. 228--242, 2007.

\bibitem{Chen2013}
Q.~Chen, D.~Li, and C.~K. Tang, ``{KNN matting},'' \emph{IEEE Trans. Pattern
  Anal. Mach. Intell.}, vol.~35, no.~9, pp. 2175--2188, 2013.

\bibitem{he2012guided}
K.~He, J.~Sun, and X.~Tang, ``Guided image filtering,'' \emph{IEEE Trans.
  Pattern Anal. Mach. Intell.}, vol.~35, no.~6, pp. 1397--1409, 2012.

\bibitem{Chakrabarti2010}
A.~Chakrabarti, T.~Zickler, and W.~T. Freeman, ``Analyzing spatially-varying
  blur,'' in \emph{Proc. Computer Society Conference on Computer Vision and
  Pattern Recognition}.\hskip 1em plus 0.5em minus 0.4em\relax IEEE, 2010, pp.
  2512--2519.

\bibitem{Zhu2013a}
X.~Zhu, S.~Cohen, S.~Schiller, and P.~Milanfar, ``Estimating spatially varying
  defocus blur from a single image,'' \emph{IEEE Trans. Image Process.},
  vol.~22, no.~12, pp. 4879--4891, 2013.

\bibitem{shi2015just}
J.~Shi, L.~Xu, and J.~Jia, ``Just noticeable defocus blur detection and
  estimation,'' in \emph{Proc. IEEE Conf. CVPR}, 2015, pp. 657--665.

\bibitem{AlexNet}
A.~Krizhevsky, I.~Sutskever, and G.~E. Hinton, ``Imagenet classification with
  deep convolutional neural networks,'' in \emph{Proc. Advances in neural
  information processing systems}, 2012, pp. 1097--1105.

\bibitem{sandler2018mobilenetv2}
M.~Sandler, A.~Howard, M.~Zhu, A.~Zhmoginov, and L.-C. Chen, ``Mobilenetv2:
  Inverted residuals and linear bottlenecks,'' in \emph{Proc. IEEE Conf. CVPR},
  2018, pp. 4510--4520.

\bibitem{ren2016faster}
S.~Ren, K.~He, R.~Girshick, and J.~Sun, ``Faster r-cnn: towards real-time
  object detection with region proposal networks,'' \emph{IEEE Trans. Pattern
  Anal. Mach. Intell.}, vol.~39, no.~6, pp. 1137--1149, 2016.

\bibitem{lateef2019survey}
F.~Lateef and Y.~Ruichek, ``Survey on semantic segmentation using deep learning
  techniques,'' \emph{Neurocomputing}, vol. 338, pp. 321--348, 2019.

\bibitem{jin2019flexible}
Z.~Jin, M.~Z. Iqbal, D.~Bobkov, W.~Zou, X.~Li, and E.~Steinbach, ``A flexible
  deep cnn framework for image restoration,'' \emph{IEEE Trans. Multimedia},
  vol.~22, no.~4, pp. 1055--1068, 2019.

\bibitem{anwar2020diving}
S.~Anwar and C.~Li, ``Diving deeper into underwater image enhancement: A
  survey,'' \emph{Signal Processing: Image Communication}, vol.~89, p. 115978,
  2020.

\bibitem{yang2019deep}
W.~Yang, X.~Zhang, Y.~Tian, W.~Wang, J.-H. Xue, and Q.~Liao, ``Deep learning
  for single image super-resolution: A brief review,'' \emph{IEEE Trans.
  Multimedia}, vol.~21, no.~12, pp. 3106--3121, 2019.

\bibitem{Nasonova2015}
A.~Nasonova and A.~Krylov, ``Deblurred images post-processing by poisson
  warping,'' \emph{IEEE Signal Processing Letters}, vol.~22, no.~4, pp.
  417--420, 2014.

\bibitem{Ma2018a}
K.~Ma, H.~Fu, T.~Liu, Z.~Wang, and D.~Tao, ``{Deep Blur Mapping: Exploiting
  High-Level Semantics by Deep Neural Networks},'' \emph{IEEE Trans. Image
  Process.}, vol.~27, no.~10, pp. 5155--5166, 2018.

\bibitem{zhao2019defocus}
W.~Zhao, F.~Zhao, D.~Wang, and H.~Lu, ``Defocus blur detection via multi-stream
  bottom-top-bottom network,'' \emph{IEEE Trans. Pattern Anal. Mach. Intell.},
  vol.~42, no.~8, pp. 1884--1897, 2019.

\bibitem{jiang2020multianet}
Z.~Jiang, X.~Xu, C.~Zhang, and C.~Zhu, ``Multianet: a multi-attention network
  for defocus blur detection,'' in \emph{Proc. International Workshop on
  Multimedia Signal Processing (MMSP)}.\hskip 1em plus 0.5em minus 0.4em\relax
  IEEE, 2020, pp. 1--6.

\bibitem{tang2020br}
C.~{Tang}, X.~{Liu}, S.~{An}, and P.~{Wang}, ``Br$^2$net: Defocus blur
  detection via a bidirectional channel attention residual refining network,''
  \emph{IEEE Trans. Multimedia}, vol.~23, pp. 624--635, 2021.

\bibitem{zhai2021global}
Y.~Zhai, J.~Wang, J.~Deng, G.~Yue, W.~Zhang, and C.~Tang, ``Global context
  guided hierarchically residual feature refinement network for defocus blur
  detection,'' \emph{Signal Processing}, p. 107996, 2021.

\bibitem{lee2019deep}
J.~Lee, S.~Lee, S.~Cho, and S.~Lee, ``Deep defocus map estimation using domain
  adaptation,'' in \emph{Proc. IEEE Conf. CVPR}, 2019, pp. 12\,222--12\,230.

\bibitem{bohra2020texturetomtf}
M.~Bohra, S.~Maheshwari, and V.~Gandhi, ``Texturetomtf: predicting spatial
  frequency response in the wild,'' \emph{Signal, Image and Video Processing},
  pp. 1--8, 2020.

\bibitem{ying2020patches}
Z.~Ying, H.~Niu, P.~Gupta, D.~Mahajan, D.~Ghadiyaram, and A.~Bovik, ``From
  patches to pictures (paq-2-piq): Mapping the perceptual space of picture
  quality,'' in \emph{Proc. IEEE Conf. CVPR}, 2020, pp. 3575--3585.

\bibitem{Sakamoto1984}
T.~Sakamoto, ``Model for spherical aberration in a single radial gradient-rod
  lens,'' \emph{Applied Optics}, vol.~23, no.~11, pp. 1707--1710, 1984.

\bibitem{Tai2009a}
Y.-W. Tai and M.~S. Brown, ``Single image defocus map estimation using local
  contrast prior,'' in \emph{Proc. Int. Conf. Image Process., ICIP}.\hskip 1em
  plus 0.5em minus 0.4em\relax IEEE, 2009, pp. 1797--1800.

\bibitem{Favaro2005}
P.~Favaro and S.~Soatto, ``A geometric approach to shape from defocus,''
  \emph{IEEE Trans. Pattern Anal. Mach. Intell.}, vol.~27, no.~3, pp. 406--417,
  2005.

\bibitem{Kubota2005}
A.~Kubota and K.~Aizawa, ``Reconstructing arbitrarily focused images from two
  differently focused images using linear filters,'' \emph{IEEE Trans. Image
  Process.}, vol.~14, no.~11, pp. 1848--1859, 2005.

\bibitem{Agustsson_2017_CVPR_Workshops}
E.~Agustsson and R.~Timofte, ``Ntire 2017 challenge on single image
  super-resolution: Dataset and study,'' in \emph{Proc. IEEE Conf. CVPR}, 2017,
  pp. 126--135.

\bibitem{NASNet}
B.~Zoph, V.~Vasudevan, J.~Shlens, and Q.~V. Le, ``Learning transferable
  architectures for scalable image recognition,'' in \emph{Proc. IEEE Conf.
  CVPR}, 2018, pp. 8697--8710.

\bibitem{DenseNet}
G.~Huang, Z.~Liu, L.~Van Der~Maaten, and K.~Q. Weinberger, ``Densely connected
  convolutional networks,'' in \emph{Proc. IEEE Conf. CVPR}, 2017, pp.
  4700--4708.

\bibitem{ma2018shufflenet}
N.~Ma, X.~Zhang, H.-T. Zheng, and J.~Sun, ``Shufflenet v2: Practical guidelines
  for efficient cnn architecture design,'' in \emph{Proc. European conference
  on computer vision (ECCV)}, 2018, pp. 116--131.

\bibitem{tan2019efficientnet}
M.~Tan and Q.~V. Le, ``Efficientnet: Rethinking model scaling for convolutional
  neural networks,'' \emph{arXiv preprint arXiv:1905.11946}, 2019.

\bibitem{Adam}
D.~P. Kingma and J.~Ba, ``Adam: A method for stochastic optimization,''
  \emph{arXiv preprint arXiv:1412.6980}, 2014.

\bibitem{li2014weighted}
Z.~Li, J.~Zheng, Z.~Zhu, W.~Yao, and S.~Wu, ``Weighted guided image
  filtering,'' \emph{IEEE Trans. Image Process.}, vol.~24, no.~1, pp. 120--129,
  2014.

\bibitem{guo2018mutually}
X.~Guo, Y.~Li, J.~Ma, and H.~Ling, ``Mutually guided image filtering,''
  \emph{IEEE Trans. Pattern Anal. Mach. Intell.}, vol.~42, no.~3, pp. 694--707,
  2018.

\bibitem{yi2016lbp}
X.~Yi and M.~Eramian, ``Lbp-based segmentation of defocus blur,'' \emph{IEEE
  Trans. Image Process.}, vol.~25, no.~4, pp. 1626--1638, 2016.

\bibitem{chen2016fast}
D.-J. Chen, H.-T. Chen, and L.-W. Chang, ``Fast defocus map estimation,'' in
  \emph{Proc. Int. Conf. Image Process., ICIP}.\hskip 1em plus 0.5em minus
  0.4em\relax IEEE, 2016, pp. 3962--3966.

\bibitem{variance_lapla}
J.~L. {Pech-Pacheco}, G.~{Cristobal}, J.~{Chamorro-Martinez}, and
  J.~{Fernandez-Valdivia}, ``Diatom autofocusing in brightfield microscopy: a
  comparative study,'' in \emph{Proceedings 15th International Conference on
  Pattern Recognition. ICPR-2000}, vol.~3, 2000, pp. 314--317 vol.3.

\bibitem{achanta2012slic}
R.~Achanta, A.~Shaji, K.~Smith, A.~Lucchi, P.~Fua, and S.~S{\"u}sstrunk, ``Slic
  superpixels compared to state-of-the-art superpixel methods,'' \emph{IEEE
  Trans. Pattern Anal. Mach. Intell.}, vol.~34, no.~11, pp. 2274--2282, 2012.

\bibitem{unsharp_masking}
G.~Ramponi, ``A cubic unsharp masking technique for contrast enhancement,''
  \emph{Signal Processing}, vol.~67, no.~2, pp. 211--222, 1998.

\bibitem{mitra1991new}
S.~K. Mitra, H.~Li, I.-S. Lin, and T.-H. Yu, ``A new class of nonlinear filters
  for image enhancement,'' in \emph{Proc. IEEE Int. Conf. Acoustics, Speech,
  and Signal Processing}, 1991, pp. 2525--2526.

\bibitem{ramponi1998cubic}
G.~Ramponi, ``A cubic unsharp masking technique for contrast enhancement,''
  \emph{Signal Processing}, vol.~67, no.~2, pp. 211--222, 1998.

\bibitem{ramponi1996nonlinear}
G.~Ramponi, N.~K. Strobel, S.~K. Mitra, and T.-H. Yu, ``Nonlinear unsharp
  masking methods for image contrast enhancement,'' \emph{Journal of Electronic
  Imaging}, vol.~5, no.~3, pp. 353--367, 1996.

\bibitem{polesel2000image}
A.~Polesel, G.~Ramponi, and V.~J. Mathews, ``Image enhancement via adaptive
  unsharp masking,'' \emph{IEEE Trans. Image Process.}, vol.~9, no.~3, pp.
  505--510, 2000.

\bibitem{GUM_Dennis}
G.~{Deng}, ``A generalized unsharp masking algorithm,'' \emph{IEEE Trans. Image
  Process.}, vol.~20, no.~5, pp. 1249--1261, 2011.

\bibitem{mittal2012no}
A.~Mittal, A.~K. Moorthy, and A.~C. Bovik, ``No-reference image quality
  assessment in the spatial domain,'' \emph{IEEE Trans. Image Process.},
  vol.~21, no.~12, pp. 4695--4708, 2012.

\bibitem{zhang2015feature}
L.~Zhang, L.~Zhang, and A.~C. Bovik, ``A feature-enriched completely blind
  image quality evaluator,'' \emph{IEEE Trans. Image Process.}, vol.~24, no.~8,
  pp. 2579--2591, 2015.

\bibitem{ou2019novel}
F.-Z. Ou, Y.-G. Wang, and G.~Zhu, ``A novel blind image quality assessment
  method based on refined natural scene statistics,'' in \emph{Proc. Int. Conf.
  Image Process., ICIP}.\hskip 1em plus 0.5em minus 0.4em\relax IEEE, 2019, pp.
  1004--1008.

\bibitem{min2016blind}
X.~Min, G.~Zhai, K.~Gu, Y.~Fang, X.~Yang, X.~Wu, J.~Zhou, and X.~Liu, ``Blind
  quality assessment of compressed images via pseudo structural similarity,''
  in \emph{IEEE Int. Conf. Multimed. Expo}.\hskip 1em plus 0.5em minus
  0.4em\relax IEEE, 2016, pp. 1--6.

\bibitem{saad2012blind}
M.~A. Saad, A.~C. Bovik, and C.~Charrier, ``Blind image quality assessment: A
  natural scene statistics approach in the dct domain,'' \emph{IEEE Trans.
  Image Process.}, vol.~21, no.~8, pp. 3339--3352, 2012.

\bibitem{wang2018deeplens}
L.~Wang, X.~Shen, J.~Zhang, O.~Wang, Z.~Lin, C.-Y. Hsieh, S.~Kong, and H.~Lu,
  ``Deeplens: shallow depth of field from a single image,'' \emph{arXiv
  preprint arXiv:1810.08100}, 2018.

\bibitem{sakurikar2019defocus}
P.~Sakurikar, I.~Mehta, and P.~Narayanan, ``Defocus magnification using
  conditional adversarial networks,'' in \emph{2019 IEEE Winter Conference on
  Applications of Computer Vision (WACV)}.\hskip 1em plus 0.5em minus
  0.4em\relax IEEE, 2019, pp. 1337--1346.

\bibitem{google}
N.~Wadhwa, R.~Garg, D.~E. Jacobs, B.~E. Feldman, N.~Kanazawa, R.~Carroll,
  Y.~Movshovitz-Attias, J.~T. Barron, Y.~Pritch, and M.~Levoy, ``Synthetic
  depth-of-field with a single-camera mobile phone,'' \emph{ACM Trans. Graph.},
  vol.~37, no.~4, pp. 1--13, 2018.

\bibitem{li2015weighted}
Z.~Li, J.~Zheng, Z.~Zhu, W.~Yao, and S.~Wu, ``Weighted guided image
  filtering,'' \emph{IEEE Trans. Image Process.}, vol.~24, no.~1, pp. 120--129,
  2014.

\bibitem{Sun2020}
\BIBentryALTinterwordspacing
Z.~Sun, B.~Han, J.~Li, J.~Zhang, and X.~Gao, ``Weighted guided image filtering
  with steering kernel,'' \emph{IEEE Trans. Image Process.}, vol.~29, pp.
  500--508, 2020. [Online]. Available:
  \url{https://doi.org/10.1109/tip.2019.2928631}
\BIBentrySTDinterwordspacing

\bibitem{Liu2020a}
\BIBentryALTinterwordspacing
Y.~Liu, L.~Wang, J.~Cheng, C.~Li, and X.~Chen, ``{Multi-focus image fusion: A
  Survey of the state of the art},'' \emph{Inform. Fusion}, vol.~64, pp.
  71--91, 2020. [Online]. Available:
  \url{http://www.sciencedirect.com/science/article/pii/S1566253520303109}
\BIBentrySTDinterwordspacing

\bibitem{QIU201935GFDF}
X.~Qiu, M.~Li, L.~Zhang, and X.~Yuan, ``Guided filter-based multi-focus image
  fusion through focus region detection,'' \emph{Signal Processing: Image
  Communication}, pp. 35--46, 2019.

\bibitem{li2013image}
S.~Li, X.~Kang, and J.~Hu, ``Image fusion with guided filtering,'' \emph{IEEE
  Trans. Image Process.}, vol.~22, no.~7, pp. 2864--2875, 2013.

\bibitem{paul2016multIFGD}
S.~Paul, I.~S. Sevcenco, and P.~Agathoklis, ``Multi-exposure and multi-focus
  image fusion in gradient domain,'' \emph{Journal of Circuits, Systems and
  Computers}, vol.~25, no.~10, pp. 1--18, 2016.

\bibitem{ma2021sesf}
B.~Ma, Y.~Zhu, X.~Yin, X.~Ban, H.~Huang, and M.~Mukeshimana, ``Sesf-fuse: An
  unsupervised deep model for multi-focus image fusion,'' \emph{Neural
  Computing and Applications}, vol.~33, no.~11, pp. 5793--5804, 2021.

\bibitem{nejati2015multi_lytro}
M.~Nejati, S.~Samavi, and S.~Shirani, ``Multi-focus image fusion using
  dictionary-based sparse representation,'' \emph{Inform. Fusion}, vol.~25, pp.
  72--84, 2015.

\bibitem{xydeas2000objectiveQG}
C.~S. Xydeas and V.~S. Petrovic, ``Objective pixel-level image fusion
  performance measure,'' in \emph{Proc. Sensor Fusion: Architectures,
  Algorithms, and Applications IV}, vol. 4051.\hskip 1em plus 0.5em minus
  0.4em\relax International Society for Optics and Photonics, 2000, pp. 89--98.

\bibitem{zhao2007performanceQP}
J.~Zhao, R.~Laganiere, and Z.~Liu, ``Performance assessment of combinative
  pixel-level image fusion based on an absolute feature measurement,''
  \emph{Int. J. Innov. Comput. I.}, vol.~3, no.~6, pp. 1433--1447, 2007.

\bibitem{yang2008novelQy}
C.~Yang, J.-Q. Zhang, X.-R. Wang, and X.~Liu, ``A novel similarity based
  quality metric for image fusion,'' \emph{Inform. Fusion}, vol.~9, no.~2, pp.
  156--160, 2008.

\bibitem{Wang2004}
Z.~Wang, A.~C. Bovik, H.~R. Sheikh, and E.~P. Simoncelli, ``{Image quality
  assessment: From error visibility to structural similarity},'' \emph{{IEEE
  Trans. Image Process.}}, vol.~13, no.~4, pp. 600--612, 2004.

\bibitem{chen2009newQCB}
Y.~Chen and R.~S. Blum, ``A new automated quality assessment algorithm for
  image fusion,'' \emph{Image and Vision Computing}, vol.~27, no.~10, pp.
  1421--1432, 2009.

\bibitem{haghighat2014fastFMI}
M.~Haghighat and M.~A. Razian, ``Fast-fmi: non-reference image fusion metric,''
  in \emph{Proc. AICT}.\hskip 1em plus 0.5em minus 0.4em\relax IEEE, 2014, pp.
  1--3.

\end{thebibliography}
